\documentclass[12pt]{article}

\usepackage[utf8]{inputenc}
\usepackage{geometry}
\usepackage{setspace}
\usepackage{natbib}
\usepackage{amsmath}
\usepackage{amsthm}
\usepackage{amssymb}
\usepackage{amsfonts}
\usepackage{listings}
\usepackage{tcolorbox} 
\usepackage{soul}
\usepackage{placeins} 
\usepackage{booktabs}
\usepackage{graphicx}   
\usepackage[bf]{caption}
\usepackage{subcaption}
\usepackage{multirow} 
\usepackage{makecell}
\usepackage{datetime}
\usepackage{rotating}   
\usepackage{siunitx}    
\usepackage{pdflscape} 
\usepackage{stackengine,scalerel} 
\usepackage{hyperref}

\newdateformat{monthyeardate}{\monthname[\THEMONTH] \THEYEAR}
\hypersetup{colorlinks=true,allcolors=black}

\definecolor{blue}{RGB}{51, 51, 153}
\definecolor{orange}{RGB}{255, 166, 0}
\definecolor{green}{RGB}{51, 153, 51}
\definecolor{red}{RGB}{153, 51, 51}

\title{
    \textbf{The Pay and Non-Pay Content of Job Ads}\thanks{
        Acknowledgements: The views below are those of the authors and do not necessarily reflect the position of the Federal Reserve Bank of New York or the Federal Reserve System. We thank Statistics Norway and the Norwegian Public Employment Agency (NAV) for providing us access to their data. For helpful feedback and suggestions, we are thankful to Christoph Heidrich, Lisa Kahn, Gizem Kosar, Thomas Le Barbanchon, Steve Machin, Alan Manning, Ryan Michaels, Chris Moser, Suresh Naidu, Barbara Petrongolo, Giorgio Topa, as well as conference and seminar participants at the FAIR Midway Conference 2022 (Troms\o), the 2023 Nordic Meeting on Register Data and Economic Modelling (Oslo), the 2023 EEA-ESEM Meetings (Barcelona), the 3rd CRC Labor Workshop (Mannheim), the Federal Reserve System Applied Micro Conference (Chicago), the 2024 CEPR Annual Symposium in Labour Economics (London), the 2024 NBER SI Labor Studies/Personnel Economics Meeting (Cambridge), the NHH/FAIR CELE Meeting, the 2025 Econometric Society World Congress (Seoul), and the 2026 ASSA/AEA Annual Meeting (Philadelphia). This project has received funding from the European Union's Horizon Europe Research and Innovation Programme under Grant Agreement No. 101043127 and the Norwegian Research Council under Grant Agreement No. 275123.
    }
}

\author{
    Richard Audoly\footnote{
        Federal Reserve Bank of New York. 
        E-mail: \href{mailto:richard.audoly@ny.frb.org}{richard.audoly@ny.frb.org}
    } 
    \and
    Manudeep Bhuller\footnote{
        University of Oslo, Statistics Norway, CEPR, IZA, CESifo. 
        E-mail: \href{mailto:manudeep.bhuller@econ.uio.no}{manudeep.bhuller@econ.uio.no}
    } 
    \and
    Tore Adam Reiremo\footnote{
        Norwegian School of Economics. 
        E-mail: \href{mailto:tore.reiremo@nhh.no}{tore.reiremo@nhh.no}
    }
}

\begin{document}

\date{}

\maketitle

\begin{center}
First Version: July 2024 \\
This Version: \monthyeardate\today
\end{center}

\begin{abstract}

    
    How informative are job ads about the actual pay and non-pay attributes offered by employers? Using a comprehensive database of job ads posted by Norwegian employers, we develop a methodology to systematically classify the pay and non-pay job attributes advertised in vacancy texts. About 60\% of job ads provide pay-related information and nearly all ads feature information on non-pay attributes. We link these advertised attributes to the employers posting the ads and validate this information against revealed-preference measures of employer quality, realized attributes, and choices from a survey experiment. All three strategies confirm that job ads provide reliable signals of employer quality. We then incorporate the detailed job attributes in a monopsony framework and quantify their contribution to labor market inequality.
\end{abstract}

\vfill



\thispagestyle{empty}
\clearpage

\onehalfspacing

\setcounter{page}{1}
\section{Introduction}
\label{sec:introduction}

There is a long tradition in labor economics of seeing jobs as more than pay \citep[e.g.,][]{brown1980equalizing,duncan1983adam,rosen1986theory}. Phrases such as ``compensating differentials,'' ``amenities,'' or ``workplace differentiation'' all capture the notion that jobs are multidimensional bundles of attributes that workers trade off against pay when considering job opportunities with alternative employers (see, e.g., \cite{mas2025nonwage} for a recent overview). Such trade-offs can be substantial: recent experimental evidence shows that workers at a US call center are willing to give up 8\% of their salary for the option to work from home \citep{mas2017valuing}, while undergraduates at a top US university would accept a 5.1\% salary cut for jobs that offer the possibility to work flexible hours \citep{wiswall2018preference}.


What is unclear from this evidence, however, is how workers gather information on the pay and amenity attributes of a job while exploring their options in the labor market. The content of job postings represents one important search channel through which workers can gather information about these attributes.\footnote{Recently, \cite{carrillo2023matching} provide survey evidence from Germany showing that 55.3\% of firms used job postings in their latest hire, while 88.1\% of workers used job postings in their search process. Notably, however, job postings are not exclusive of other search methods, such as social networks. \cite{carrillo2023matching} find that firms and workers use, respectively, 1.9 and 2.3 channels on average.} But as employers ultimately decide on the content of publicly advertised job postings, it remains an open question whether such content reflects the actual pay and amenities associated with different employers.

In this paper, we study in detail the information content of job ads. Our first objective is to provide a detailed description of the pay and non-pay content advertised by employers in their vacancies. Our second objective is to analyze how employers of varying quality, as characterized by several alternative measures, systematically advertise different types of pay and non-pay attributes in their job ads. Our third and final objective is to separately quantify the contribution of advertised amenities to labor market inequality, providing empirical content to the amenity index typically featured in structural models of the labor market.

To characterize the distribution of job attributes advertised in job ads, we use the near-universe of publicly posted vacancies in Norway. A key contribution is to systematically extract a broad range of pay and non-pay job attributes from these texts. Using natural language processing tools, we identify nearly fifty commonly advertised pay and non-pay attributes, such as ``competitive pay,'' ``flexible work hours,'' and ``nice work environment.'' We build a large collection of unique natural language expressions associated with each attribute that we make publicly available.\footnote{The full list is available at: \url{https://zenodo.org/records/14973628}.}  

We find that about 60\% of job ads provide pay-related information, with 30\% revealing explicit salary information, such as an actual salary figure, a salary bracket, or referring to a collective bargaining agreement. Nearly all ads also feature non-pay amenities, such as contract duration, irregular hours, shift work, flexible work hours, workplace attributes, task-related attributes, or other minor perks. These results indicate that employers advertise a broad range of pay and non-pay attributes when attracting workers.


The key strength of our data set is that each ad can be directly traced to the establishment posting the job. Using this feature of the data, we analyze the information that workers can infer from job ads with three complementary empirical designs. First, we analyze how advertised job attributes correlate with several revealed-preference measures of employer quality. Second, we link ads to hires and document the concordance between advertised and realized attributes, for the subset of attributes with clear counterparts in the register data. Third, we present evidence from a survey experiment fielded by \cite{BhullerAEA2024} in which respondents state their preferences over real-world job ads rather than hypothetical jobs.

Across measures of employer quality, we find that high-quality employers are more likely to advertise attributes typically seen as positive, such as more generous pensions, flexible working hours, and permanent jobs, while low-quality employers tend to advertise negative attributes, such as shift-work and temporary jobs. This holds even for less tangible job attributes: high-quality employers also more often advertise a professional environment with ``good colleagues'' and work involving ``challenging tasks.'' Similarly, we show that job ad attributes meaningfully predict employer quality, with more than 60\% of the variation explained across quality measures. This result holds after removing the variation accounted for by industry, occupation, and location controls: job ad attributes still explain 15--20\% of the remaining dispersion in employer quality. Finally, using the linked ad-hire data we construct, we confirm that advertised and realized job attributes generally coincide. 

The survey experiment adds further evidence on the reliability of job ads as an information channel. We estimate the implied willingness-to-pay for the job attributes extracted from the ads. We find that our estimates have sensible signs and, for the attributes documented in prior work, plausible magnitudes, indicating that respondents rely on this information in making their choices. Moreover, based solely on the ad text, respondents are also more likely to state that they prefer job ads posted by higher-quality employers, as measured with a revealed-preference approach using information for the population at large.

Having documented that job ads contain useful information about employer quality, we quantify the contribution of advertised attributes to employer-level inequality in the labor market. We incorporate these attributes into an otherwise standard static monopsony model in which workers have heterogeneous preferences over alternative employers \citep{card2018firms,lamadon2022imperfect}. With amenity information derived from the ads, we can distinguish between advertised and intrinsic amenities, where the latter capture employer characteristics that are not explicitly mentioned in job ads but still valued by workers.

Our estimates suggest that the distinction between advertised and intrinsic amenities is useful for analyzing the employer-level drivers of labor market inequality. We find that the advertised component of non-pay utility augments pay differentials, while the intrinsic component represents a compensating differential. Our model simulations show that, all else equal, wages would decrease by 2.3 log points in the absence of advertised non-pay attributes and increase by 3.8 log points without intrinsic amenities. Employer-level pay dispersion increases by 13\% when all employers provide the same advertised amenities, and decreases by 25\% when intrinsic amenities are equalized. More broadly, our framework helps connect findings in the empirical literature focusing on specific non-pay attributes \citep[e.g.,][]{mas2017valuing,maestas2023value} with the structural labor literature that collapses amenities into a scalar index \citep[e.g.,][]{hall2018wage,morchio2024gender}.

We see our approach as the first attempt at quantifying the breadth and quality of information present in the texts of job ads. We show that job openings contain a broad range of attributes relevant to workers, and that their content provides reliable signals of employer quality. These findings have implications for a broad class of job search models, such as models with a multi-dimensional job ladder \citep{hwang1998hedonic,jarosch2023searching}. We leave it to future research to shed light on employers' endogenous choices over how much and what type of content to include in their job ads, and how these choices relate to their recruitment objectives.

Our results also have implications for the design of policies related to the information contained in job ads. There is a current policy drive to improve pay transparency within organizations in many countries, and several US jurisdictions are also implementing policies to promote pay transparency in job openings \cite[][]{cullen2024pay}.\footnote{In January 2021, Colorado enacted a transparency law requiring online job postings to include information about the expected salary, while in November 2022, New York City mandated employers to include a salary bracket in job ads, followed by Washington State in January 2023 \citep[see][]{audoly2025employers}.} Our findings on the prevalence of pay-related information in job ads, and on which non-pay attributes more attractive employers choose to advertise, can help inform the design of such policies.

\paragraph{Related literature}
Several recent studies have used vacancy texts as a source of information on the type of jobs advertised by employers. A large portion of this literature focuses on skill requirements. \cite{marinescu2020opening} study the role of job titles in accounting for the number of applicants per vacancy. \cite{deming2018skill}, \cite{atalay2020evolution}, and \cite{deming2020earnings} study variation in demand for specific skills across local labor markets and over time.\footnote{Recent contributions have focused specifically on demand for computer science and AI skills \citep[see, e.g.,][]{alekseeva2021demand,acemoglu2022artificial,braxton2023technological,contractor2022effect}.}

We contribute to this literature by considering a broad set of pay and non-pay attributes advertised by employers in job ads. Several recent contributions focus on a single attribute, such as flexible work arrangements \citep{adams2023flexible}, remote work \citep{hansen2023}, or on-the-job training \citep{adams2023job}. Our approach encompasses these attributes but also captures what employers communicate about compensation and workplace quality. \cite{sockin2022} similarly uses text analysis to retrieve amenity information from employer reviews posted by workers. We see our results as complementary, as vacancy texts reflect the information advertised by employers rather than worker assessments.\footnote{In other recent related work, \cite{arold2024} and \cite{lagos2024} use text analysis to retrieve information on workplace amenities from collective bargaining agreement texts, while \cite{caldwell2025bargaining}, \cite{Humlum2025}, \cite{gupta2025workplace}, and \cite{folke2022} design large-scale surveys to collect information on specific workplace amenities, which these authors are able to link to administrative data.}

A large body of work has emphasized that non-pay attributes are important determinants of how much workers value alternative employers. One strand elicits workers' preferences over specific attributes using quasi-experimental designs \citep[see, e.g.,][]{barbanchon2021gender} or survey experiments \citep[see, e.g.,][]{maestas2023value}.\footnote{Notable examples include workplace flexibility \citep{mas2017valuing, wiswall2018preference, drakeetal2023}, job security \citep{datta2019}, commuting time \citep{barbanchon2021gender,butikofer2025parenthood}, working from home \citep{nagleretal2022, lewandowski2022}, and shift work \citep{desierewalter2023}. See \cite{mas2025nonwage} for a recent overview.} The other strand recovers structural estimates of employers' composite amenity value by modeling workers' job choices in markets with heterogeneous firms. In this approach, amenity values are identified by reconciling workers' choices with their pay \citep{rosen1986theory}, using employer size in static monopsony models \citep[e.g.,][]{card2018firms,lamadon2022imperfect} or mobility flows in dynamic search models \citep[e.g.,][]{sorkin2018ranking,taber2020estimation,morchio2024gender}.

Our paper contributes to both strands of the literature. We quantify the willingness-to-pay for non-pay attributes mentioned in job ads using a survey experiment, and use information on advertised attributes in a structural monopsony framework. Further, we relate survey responses to common revealed-preference measures of employer quality, building on \citet{sorkin2018ranking} and \citet{bagger2019empirical}. By distinguishing advertised and intrinsic amenities, we can test which amenities represent augmenting or compensating differentials. Our framework therefore helps connect these two important strands of the literature.\footnote{\cite{morchio2024gender} also relate their estimates of the amenity values of employers to some of the workplace characteristics observed in their data, such as ``workplace hazards'' and ``working hours flexibility''.}
 
We also contribute to the expanding body of work aimed at unpacking search frictions in the labor market. \cite{jager2022worker} elicit workers' beliefs about their reemployment wages in the event of job loss and compare them to actual outcomes. \cite{horton2021cheap} and \cite{belot2022wage} design experiments to study how job seekers respond to wage variation in job vacancies. Our evidence also relates to recent work on how posted wages affect recruitment outcomes \citep[e.g.,][]{faberman2018}.\footnote{Recent studies have examined the wage elasticity of vacancy duration \citep{mueller2024}, applications \citep{azar2022,banfi2019highwage}, and hires \citep{bassier2022, hirsch2022}.} We add to these studies by focusing on job ad texts as a source of information on both pay and non-pay attributes, and by relating these attributes to revealed-preference measures of employer quality.

Finally, our work relates to a broader literature on the role of advertising, which emphasizes the distinction between its informational and persuasive functions.\footnote{See \cite{bagwell2007economic} for a review.} Isolating these motives empirically requires either experimental variation \citep{bertrand2010s} or data on advertising effort combined with a structural model \citep{hastings2017sales}. While we see our work as a first step toward linking realized job attributes to those advertised in job ads, we leave it for future research to devise similar research designs for the labor market.

\paragraph{Outline} 
Section \ref{sec:data} describes our data sources. Section \ref{sec:advertised_attributes} explains how we retrieve information on the attributes advertised by employers in vacancy texts and provide evidence on their prevalence in our data.  Section \ref{sec:learning_value} provides our evidence on the information content of job ads, where we link the information in job ads to revealed-preference measures of employer quality, realized attributes in employment contracts, and responses from a survey experiment. Section \ref{sec:static_model} provides our evidence on the contribution of amenities to labor market inequality from a monopsony framework with detailed job attributes. Section \ref{sec:conclusion} concludes.

\section{Data and Institutional Context}
\label{sec:data}

\subsection{Vacancy Data}
\label{sec:vacancy_data}

Our data include almost 2.15 million job ads covering the near universe of publicly posted vacancies in Norway between 2015 and 2024.\footnote{In an earlier version of this paper, we used data on Norwegian job ads going back to 2002.} These data are maintained by the Norwegian Public Employment Agency (NAV). The employment agency collects information about vacancies from several sources, including online job boards and newspapers, as well as direct job opening reports from employers.\footnote{In accordance with the Labor Market Act §7, Norwegian employers are required by law to report publicly posted job openings to the agency, which maintains a comprehensive database of publicly posted vacancy ads with the stated goal of providing job-seekers with current information on suitable job opportunities.} The share of online postings has increased gradually over the past two decades, and about 90\% of job postings recorded in the agency's database since 2015 were retrieved from various online job boards.\footnote{\cite{bhuller2023} study the consequences on labor market matching of increased online job search and recruitment triggered by a roll-out of broadband internet across Norway during the early 2000s.} The remaining job ads were either (i) scanned or transcribed from newspapers by caseworkers in the employment agency, or (ii) enclosed in the notifications of job openings sent by employers directly to the agency. The ability to observe virtually all publicly posted vacancy ads in the economy with full-text corpora is an important advantage of our setting, which differentiates it from the existing literature using vacancy texts that often relies on information from selected online job portals.

In addition to the job title and actual text of each job opening, our data contain the following structured information about each ad: unique establishment identifiers, the dates when the ad was registered and filled or removed (i.e., vacancy duration), the number of job openings per vacancy, and some additional information about job characteristics submitted by employers. The establishment identifiers are central to our analysis. They allow us to link job openings to matched employer-employee administrative data and to compare the information we extract from the text to actual outcomes at the establishment level.\footnote{
    Notably, around 9\% of job ads in our data were posted through recruitment or temporary employment agencies, and we drop these from most of our main analysis, as we are unable to link such ads to the actual establishment where the job is placed. Further, in around 12\% of job ads, the posting employer has for various reasons decided not to disclose the establishment name in the publicly posted information, but we do have the corresponding establishment identifier in our data as the agency maintains this information and could share this with us for research purposes (according to the Labor Market Act §7-4).
    } 
Each job ad in our data set has an occupational code based on the 4-digit ISCO classification. This occupational code is assigned by caseworkers based on the job title and textual information on job descriptions and skill requirements stated by the employers in the vacancy, and is a novel feature of our data. Besides the structured information, the text of a typical job ad contains an average of 310 words. As we describe in Section \ref{sec:advertised_attributes}, we use tools from the natural language processing literature to retrieve information on job attributes from these texts.

As in many other contexts \citep[see, e.g.,][]{card2024ads}, Norwegian employers are required to follow anti-discrimination laws (Working Environment Act §3; Equality and Anti-Discrimination Act §29a), which limit the listing of explicit preferences, e.g., based on gender, age, or ethnicity, in job ads. Moreover, public-sector state employers are subject to the ``qualification principle'' (the State Employees Act §3), which requires that the best-qualified applicant be appointed. This implies that the skill requirements listed in state-sector job ads have binding implications for how candidates are evaluated. In other respects, Norwegian employers may design and describe job ads as they see fit. Importantly, there is no general legal requirement to disclose information on salaries or workplace amenities in job ads.



\subsection{Matched Employer-Employee Data} 
\label{sec:matched employer-employee data}

We have access to Norwegian administrative matched employer-employee data covering several decades. From January 2015, these data are available as monthly files, where each record has information on the individual's employer, as well as their pre-tax earnings, hours of work, employment start and end date. Individuals can have several recorded spells in any given month if they receive earnings from more than one employer. This data set has additional information on several background characteristics of employees (gender, education) and employers (location, industry). Notably, our data set also includes information on certain non-pecuniary job attributes, such as whether the employment spell involves shift work, and taxable in-kind benefits, which we use in our analysis for comparisons to some of the attributes retrieved from job ads. Since 2021, the data set has additional job attributes (e.g., contract type) for each employment spell that employers must report to Statistics Norway.

We use information from the matched employer-employee data set at several places in our analysis. First, we infer revealed-preference measures of employer quality in a structural approach, where we use information on measures of average hourly wages, employment size and worker flows across employers. For this purpose, we aggregate all employment spells to the annual level for each worker and retain observations corresponding to the main employer, defined as the establishment with the largest annual earnings. Using information on annual earnings and annual contracted hours of work, we calculate the average hourly wage for each worker in their main job in each year. Second, we use detailed monthly information on hiring dates for workers in each employment spell to relate posted job ads to potential hires. Moreover, we draw on information from the matched employer-employee data set to define comparable worker and employer characteristics for our various analyses.

\subsection{Job Postings and Recruitment}
\label{sec:posting hiring}

An underlying assumption in our analysis is that job postings are an important source through which employers recruit workers and where workers learn about available jobs. In order to assess the importance of job postings for recruitment, we use information on (i) establishment identifiers available both in vacancy and matched employer-employee data sets, and (ii) the dates of job ad postings and hires. Based on this information, we find that 85.3\% of all publicly posted non-staffing agency job ads can be associated to at least one new hire in the establishment within the following six months after the posting date. This applies to 92.8\% for ads posted by public sector state employers, 83.5\% of ads by non-state public sector employers (e.g., local municipalities, public schools, public hospitals), and 84.2\% of ads by private sector employers. Conversely, 48.6\% of all hires can be related to a posted ad within the previous six months; this applies to 87.0\% of hires by public sector state employers, 70.4\% of hires by non-state public employers, and 39.9\% of private sector hires. Notably, employers may use other recruitment channels, such as social networks, or directly solicit applications, and thus hire workers without publicly posting a job ad.\footnote{While all employers have an obligation to report the vacancies they publicly post to the Norwegian employment agency, it is up to them to decide whether or not to post a public vacancy for the purposes of recruitment. However, recruitment by public-sector state employers is regulated by state legislation, which requires as a general rule that all job openings are publicly advertised (the State Employees Act §4).} These shares reported above nonetheless suggest that job ads are an important recruitment channel in the Norwegian labor market.

To further assess the associations between job postings and recruitment in our data, we perform a series of descriptive event studies around the month of each job ad posting, and follow hiring outcomes at the establishment level around such events. This descriptive evidence can help us evaluate the extent of ``excess'' hiring associated with job postings, as opposed to hires that may reflect other recruitment channels. We show in Appendix Figure \ref{fig:hiring_pattern} that there is a clear increase in the average number of hires in the six months following an ad posting, peaking about two to three months after the posting month, while hiring outcomes are stable prior to the posting month. In total, the number of hired workers increases by 0.43 workers in the six month period after an ad is posted relative to the months before. When focusing on hires within narrowly defined 4-digit occupations, i.e., the posted occupation in the job ad and the occupation associated with the hire are identical, we find an ``excess'' hiring of about 0.2 workers in the months following the ad posting. This suggests that postings and recruitment are closely related, as one may expect.

Further, in some of our analyses, we use micro-level information on workers hired in the different job postings. While the above suggests there is a tight relationship between postings and hires, our data sets do not provide direct links between the hired workers and the job postings at the ad-hire level. In a similar vein as the event study setup described above, we thus construct a dataset of hires linked to job ads based on probabilistic linkages, within narrowly defined establishment-occupation cells and posting-hiring periods. We use this restrictive approach to avoid measurement error in the linking of hires to job ads, with the downside that we do not capture all hires that could be related to job postings.\footnote{For instance, the posted occupation in the job ad and the occupation associated with the hire may not necessarily be identical, even though the worker is clearly hired as a result of the job posting. As noted in Section \ref{sec:vacancy_data}, the occupational code in our job ad data is assigned by caseworkers in the employment agency based on the job title and other textual information stated by the employers in the vacancy, which may be different from the occupational code filed by the employer in the administrative records at the time of hiring.} Specifically, for each job ad we look for hires in the six months after the ad is posted. We restrict links to establishment-occupation cells with a single hire in the following six months to reduce the likelihood of linking an ad to a worker hired independently of the posting. We further exclude ads that were posted on the same date in the same establishment-occupation cell. We also exclude recall hires having a prior spell with the employer, as they are less likely to be hired through postings. Finally, we avoid linking a single job spell to multiple vacancies by removing it if the spell is already linked to a job ad, i.e., we link ads to hires without replacement. In the following, we refer to the resulting dataset as the ad-hire sample.

\subsection{Survey Experiment}
\label{sec:survey data}

Besides information on job ads and matched employer-employee data, we use anonymized data from a survey experiment performed by \citet{BhullerAEA2024}. Notably, the survey used hypothetical choice experiments to elicit willingness to pay (WTPs) for workplace amenities, and further included components where respondents were asked to rank real job ads, where the choice sets respondents faced were experimentally designed for the purposes of identification. The survey experiment was pre-registered with trial number \textit{AEARCTR-0013426} in the AEA RCT registry. In recent work, \citet{andresen2026aea} use data from this survey to study the role of pay beliefs on job choices, using experimentally designed features of the survey. We instead use information on respondents' rankings of real job ads, and relate these rankings to revealed-preference measures of employer quality, which we estimate using the matched employer-employee data for the population at large. As in \citet{andresen2026aea}, we provide estimates of respondents' WTPs for specific workplace amenities.

We use data from the survey waves that were targeted at students enrolled at the University of Oslo, which is a large public university in Norway. Around 2,800 students were invited to participate in the survey, of whom 1,059 gave their informed consent and 1,040 completed one of the relevant modules, which gave a response rate of about 37\% (see details in \citet{andresen2026aea}). Each respondent faced up to 40 experimentally designed hypothetical choice scenarios, where the respondents were asked to state their preferred alternative across the two options that were made available to them in each scenario. Half of the choice scenarios featured choices between purely hypothetical jobs as in \citet{wiswall2018preference}. Each alternative listed the monthly salary and a set of posted amenities that were randomly sampled using the methodology developed by \citet{drake22}, and the respondents were explicitly told that the alternatives were otherwise identical in all aspects. This approach facilitates consistent estimation of respondents' WTPs for randomized workplace amenities.

Importantly, for the purposes of our study, the survey also included experimentally designed components featuring around 1,200 real job ads that were posted by around 500 unique employers in Oslo or the surrounding labor market regions over a six-month period from June to December 2023. These job ads were used in experimentally designed choice scenarios as part of the survey, where the respondents were asked to rank between two job ads in each scenario, as illustrated in Appendix Figure \ref{fig:experiment_scenario}. In a given scenario, the job ads were randomly selected from the relevant pool of posted ads within the same sector and broad occupation group. The respondents were allowed to read the full texts of each job posting, and could thus learn about amenities and other job attributes associated with each option, before stating their preferred alternative. To ease comparisons, all job ads featured full-time jobs. Notably, none of the job ads used in the survey experiment explicitly stated the compensation level as part of the actual ad text. These features imply that comparisons of respondents' rankings of job ads to revealed-preference measures of employer quality derived from market-level data are informative about what job seekers can potentially learn about employer quality solely from job ad texts. Using our matched employer-employee data across all employers in the Norwegian labor market, we are able to construct measures of employer quality for most of the employers included in the survey for the relevant period. 

\subsection{Sample Selection}
\label{sec:sample_selection}

For most of our analysis we focus on the years 2021-2024, which allows us to leverage the richness of our administrative data and relate our findings to evidence from the survey experiment that was fielded during the later part of this period. We exclude 2020 from the analysis as recruitment and hiring were heavily disrupted in Norway at the onset of the Covid-19 pandemic. From 2021, job postings and unemployment rates had largely returned to their pre-pandemic levels. We also provide additional results from a longer period covering the years 2015-2024, using only information from job ads and administrative data sources.

\begin{table}[h!]
    \caption{Overview of Sample Selection.}\vspace{-1em}
   \label{tab:sample_selection}
    \begin{center}
        \scalebox{.7}{\def\sym#1{\ifmmode^{#1}\else\(^{#1}\)\fi}
\begin{tabular}{lS[table-format=9.0, group-separator={,}, group-minimum-digits = 4]lS[table-format=9.0, group-separator={,}, group-minimum-digits = 4]lS[table-format=9.0, group-separator={,}, group-minimum-digits = 4]lS[table-format=9.0, group-separator={,}, group-minimum-digits = 4]l} \hline\midrule
&&&&& \multicolumn{4}{c}{Workers in Included Establishments} \\ \cmidrule(lr){6-9} 
\multicolumn{1}{c}{} & \multicolumn{2}{c}{Job Ads} & \multicolumn{2}{c}{Establishments} & \multicolumn{2}{c}{Workers} & \multicolumn{2}{c}{Worker-Years} \\\multicolumn{1}{c}{} & \multicolumn{2}{c}{(1)} & \multicolumn{2}{c}{(2)} & \multicolumn{2}{c}{(3)} & \multicolumn{2}{c}{(4)}  \\ 
\midrule 
\textbf{All Observations} & 1031687 & (100\%) & 264084 & (100\%) & 3000154 & (100\%) & 12220469 & (100\%) \\
\multicolumn{9}{l}{\textbf{Main Analysis}} \\
\quad Text Analysis Sample & 943721 & (91.5\%) & 117896 & (44.6\%) & 2792164 & (93.1\%) & 10985594 & (89.9\%) \\
\quad Employer Quality Sample & 675832 & (65.5\%) & 70046 & (26.5\%) & 2210151 & (73.7\%) & 8253567 & (67.5\%) \\
\quad Static Monopsony Sample & 390722 & (37.9\%) & 12973 & (4.9\%) & 1688064 & (56.3\%) & 5902421 & (48.3\%) \\
\multicolumn{9}{l}{\textbf{Additional Analysis}} \\
\quad Linked Ad-Hire Sample & 110913 & (10.8\%) & 47877 & (18.1\%) & 2036676 & (67.9\%) & 6999923 & (57.3\%) \\
\quad Survey Experiment Sample & 1028 & (0.1\%) & 417 & (0.2\%) & 96529 & (3.2\%) & 214123 & (1.8\%) \\
\midrule\hline
\end{tabular}
}
    \end{center} 
    \vspace{-.5em}
    {\footnotesize \textit{Notes:} 
    This table documents the sample sizes for different sample restrictions imposed for the 2021-2024 period. The first row shows the total number of observations in our data on job ads (Column 1) and matched employer-employee data (Columns 2-4), prior to the various sample selection steps. The ``Text Analysis Sample'' refers to the set of job ads with non-missing information on establishment identifiers, occupational codes, and vacancy texts, and with at least one section written in Norwegian. The ``Employer Quality Sample'' refers to the sample we retain by restricting the sample to workers aged 20 to 60 (inclusive) and the strongly connected set of employers used in the estimation of employer values, while removing establishments registered as recruitment or temporary employment agencies.    
    The ``Static Monopsony Sample'' refers to a subset of large employers that have at least 30 stayers and 15 movers during the 2021-2024 period. The ``Linked Ad-Hire Sample'' refers to a restricted set of job ads that could be uniquely linked to a new hire at the establishment-4-digit occupation level within the six-month period since job ad posting, as described in Section \ref{sec:posting hiring}. The ``Survey Experiment Sample'' refers to the subset of job ads used in the survey experiment, as described in Section \ref{sec:survey data}, that could be linked to employer quality measures taken from the ``Employer Quality Sample''. Appendix Table \ref{tab:sample_characteristics_2021_2024} documents the characteristics of employers and workers for each of these samples, while Appendix Figure \ref{fig:balance_attributes} documents the detected job ad attributes for the three ``Main Analysis'' samples. Appendix Table \ref{tab:sample_selection_2015_2024} documents the corresponding sample sizes for the 2015-2024 period.} 
\end{table}

We use different samples at various stages of our analysis. Table \ref{tab:sample_selection} summarizes these samples and shows how the number of observations differ across samples used in our baseline analysis for the 2021-2024 period.\footnote{Appendix Table \ref{tab:sample_selection_2015_2024} shows the corresponding sample restrictions for the 2015-2024.} As shown in the first row, prior to additional sample restrictions, our initial matched employer-employee data has about 264,000 establishments covering about 3 million unique workers and 12.2 million worker-year observations. During this period, we observe about 1 million job ads in the employment agency's database.\footnote{An ad is on average associated with 1.6 job openings, so these ads cover 1.6 million job openings.} 

In our main analysis, we use either data on job ads or job ads combined with administrative employment data. As shown in the second row of Table \ref{tab:sample_selection}, we retain around 92\% of job ads with non-missing information on establishment and posted occupation, as well as non-missing vacancy texts with at least one section written in Norwegian. Note, however, that less than half of the establishments in our initial sample publicly posted at least one job ad in the 2021-2024 period. Meanwhile, the posting establishments tend to be substantially larger, and for our text analysis of job ads, we still retain establishments covering around 90\% of workers and worker-year observations, as shown in Columns (3)-(4). In Appendix Table \ref{tab:sample_characteristics_2021_2024}, we provide a comparison of how sample compositions differ across samples. Notably, employers considered in our text analysis of job ads are quite similar in terms of average worker characteristics to the initial sample drawn from the matched employer-employee data, yet these employers are larger in size and more likely to belong in the public sector.

Next, we consider the sample used for estimating revealed-preference measures of employer quality. In this sample, we include only prime-aged workers, i.e., aged 20--60, and further, we restrict the estimation to employers that are in the set of strongly connected establishments, i.e., the set of establishments across which we observe two-way job mobility flows during the 2021-2024 period. Finally, we also remove establishments registered as recruitment or temporary employment agencies.\footnote{Around 90.1\% of job ads classified as ``staffing agency'' ads in our data were posted by temporary employment agencies (e.g., Manpower), while recruitment agencies account for the remaining share.} With these restrictions, we exclude more than 70\% of the establishments in our initial sample, but still retain establishments covering about 74\% (67\%) of all workers (worker-year observations). Comparing the average characteristics of employers considered in the text analysis and the employer quality measures in Appendix Table \ref{tab:sample_characteristics_2021_2024}, we find that the two samples are broadly similar in terms of industry and worker composition, while employers used in the latter sample are larger.

Further, we consider a more restricted set of large employers that we use to estimate a static monopsony model of the labor market. This sample includes only large employers that have at least 30 stayers and 15 movers during the 2021-2024 period. Despite covering only 5\% of establishments in our initial sample, this sample nonetheless retains employers that account for 38\% of job ads and around 56\% (48\%) of all workers (worker-year observations). Given the sample restrictions, these employers are also substantially larger than the average employer in our initial sample, having a higher share of college graduates and higher average hourly wage, and these employers are more likely to belong in the public sector.

To assess how well job attributes advertised in job ads correspond to realized job characteristics observed in administrative data, we further use a restrictive sample of job ads where we can create unique links between each job ad and a newly hired worker, as described more closely in Section \ref{sec:posting hiring}. This sample covers 11\% of all job ads and 18\% of establishments, which altogether account for 68\% (57\%) of workers (worker-year observations) in our initial data. Despite the restrictive nature of the ad-hire links, this sample is fairly similar in terms of overall worker and establishment characteristics to the sample used in our text analysis, as shown in Columns (2) and (5) of Appendix Table \ref{tab:sample_characteristics_2021_2024}. Notably, however, the characteristics of newly hired workers differ from the average characteristics of workers in these establishments. The newly hired workers are younger, more likely to be female, and have lower wages; see Appendix Table \ref{tab:characteristics_linked_and_survey}. This reflects that the linking of ads to hire naturally captures workers overrepresented in labor market flows, as compared to the average worker.

Finally, we use data from the survey experiment performed by \citet{BhullerAEA2024}. For the purposes of this study, we are able to link measures of employer quality to 1,028 job ads that were posted by 417 employers. Despite the small number of employers featured in the survey experiment, as shown in the numbers reported in the final row of Table \ref{tab:sample_selection}, these employers account for around 3\% of workers and 2\% of worker-year observations in our initial sample. This reflects that employers posting job ads in Oslo and the surrounding labor market regions are larger. As shown in Appendix Table \ref{tab:sample_characteristics_2021_2024}, these employers have a larger share of female and college graduated workers and pay higher wages, and are more likely to belong in the public sector. Meanwhile, the composition of survey respondents departs significantly from the average characteristics of workers in our data; see Appendix Table \ref{tab:characteristics_linked_and_survey}. As mentioned in Section \ref{sec:survey data}, the survey targeted university students, and thus survey respondents are in their early 20s, almost 70\% are female, and less than 50\% had completed a college (bachelor's) degree at the time of interview. Notably, however, nearly all survey respondents report having at least one prior employment spell, receiving average hourly wages that are about 20\% lower than for the newly hired workers in our linked ad-hire sample.

\section{Pay and Non-Pay Attributes in Vacancy Texts}
\label{sec:advertised_attributes}

This section describes how we retrieve the job attributes publicly advertised by employers in job ads. 
These attributes are extracted from vacancy texts using tools developed in natural language processing. 
To the best of our knowledge, we are among the first to systematically extract a comprehensive set of pay and amenity attributes from the texts of job ads.

\subsection{Extracting Advertised Job Attributes}
\label{sec:extraction}

Consider the sample vacancy text for an IT-consultant position in Figure \ref{fig:sample_ad}. This text contains a variety of information on the type of tasks associated with the position, the required skills and experience necessary to perform these tasks, and the pay and non-pay attributes of the job.
As this example makes clear, these attributes cover many different aspects of the position, ranging from the duration of the contract and regular hours (``full time, permanent position'') to characteristics of the workplace (``good working environment'') and information about the level of pay (``competitive conditions'').
Our goal is to systematically extract this information from the open text of all vacancies in the database.

\begin{figure} 
    \centering
    \caption{Sample Vacancy Text: IT-consultant.}\vspace{-1em}
    \label{fig:sample_ad}
    \include{figures/sample_ads/it_consultant.tex}
       {\footnotesize \textit{Notes:} Translation from Norwegian by the authors. In Appendix Figures \ref{fig:sample_ads_teacher}-\ref{fig:sample_ads_engineer}, we provide additional examples of representative vacancy texts covering a teaching substitute and a civil engineer, respectively.} 
\end{figure}

The key difficulty with extracting job attributes from the raw texts of job ads is that a single attribute can be expressed in many distinct ways in natural language. 
Besides, while prior research in this area has centered on specific workplace amenities, we do not have a definitive list of the job attributes advertised by employers in vacancy texts.
We therefore proceed in three broad steps to extract these attributes from all job postings: (i) we pin down a list of attributes to extract, (ii) we ascribe a set of expressions to each of these attributes, and (iii) we label the entire corpus of vacancy texts based on these expressions.

\paragraph{Step 1: List of job attributes to include}
Our choice of which pay and non-pay job attributes to include in the analysis is based partly on information that is commonly stated in ads, partly on our knowledge of the particular context of the Norwegian labor market, and partly on attributes that are considered important in the existing literature. 

We identify commonly mentioned attributes in publicly posted vacancies using two distinct strategies: (i) human recognition on a subset of job ads, and (ii) identifying commonly used phrases in the ``we offer'' sections of vacancy texts.
We first directed several research assistants to read through the texts of 1,200 randomly chosen job ads in the vacancy corpus. The research assistants were asked to make a list of the commonly advertised attributes that are valuable to workers, to take note of the corresponding expressions associated with these attributes, and to indicate the presence of these attributes in each job posting.

As a second source of information on commonly advertised attributes, we isolate and extract common phrases from the ``we offer'' sections of job ads. We do this in two steps: First, we identify and extract all structured lists from the vacancy corpus.\footnote{These lists are found by searching for consecutive sentences starting with a hyphen or bullet point and by searching for the HTML tags used to generate lists in online vacancies. By design, this approach identifies attributes that can be organised in structured lists. In the later steps, we search for the common phrases identified in structured lists across the full corpus of job ads, including ads that were not posted online or lacked structured lists, as well as search for alternative similar phrases identified using text analysis tools.} As an example, in the sample vacancy text in Figure \ref{fig:sample_ad}, we retrieve three distinct lists: ``Tasks,'' ``Required qualifications,'' and ``We can offer.'' We find that 50\% of the vacancy texts in our sample feature one or more such lists. Second, we isolate the ``we offer'' lists by applying unsupervised topic modeling \citep{blei2003latent} to the collection of lists retrieved in the previous step. Further methodological details are provided in Appendix \ref{app:text_analysis}. The most common expressions in the ``we offer'' lists constitute the second source of information on commonly advertised job attributes that we are able to identify.

Based on the two approaches described above, we are able to identify 47 distinct job attributes, as well as an initial set of expressions associated with each attribute. For the purpose of illustration, we summarize the 47 attributes in ten broad categories within classes of pay and non-pay job attributes, as shown in Table \ref{tab:broad_categories}. Taken together, these attributes cover a variety of different job characteristics advertised in job ads, such as information on career opportunities, convenient/inconvenient hours, and the quality of the workplace. The full list of job attributes with descriptions is provided in Appendix Tables \ref{tab:attributes_pay_explained}-\ref{tab:attributes_nonpay2_explained}.

\begin{table}
    \caption{Main Categories of Job Attributes.}\vspace{-1em}
    \label{tab:broad_categories}
    \vspace{-0.5em}
    \begin{center}
        \begin{tabular}{@{}lll@{}}
    \toprule
    \textbf{Main Category}    & \textbf{Examples}                            & \\ 
    \midrule
    \textbf{A. Pay Attributes}       &                                          & \\
    \quad -- Compensation Scheme     & Compensation level; Collective agreement pay; Incentive pay       & \\
    \quad -- Financial Attributes    & Insurance scheme; Pension scheme; Mortgage possibility         & \\
    \quad -- Career Opportunities    & On-the-job training; Career opportunities & \\
    \textbf{B. Non-Pay Attributes}  &                                             & \\
    \quad -- Hours of Work           & Full-time; Part-time; Full-time/part-time choice          & \\
    \quad -- Convenient Hours        & Regular daytime work; Possibility to work flexible hours  & \\
    \quad -- Inconvenient Hours      & Shift-work; Weekend/evening/nights; On-call employment         & \\
    \quad -- Contract Duration       & Permanent job; Temporary job                       & \\
    \quad -- Workplace Attributes    & Social environment; Good colleagues; Remote work            & \\
    \quad -- Task-Related Attributes & Interesting tasks; Challenging tasks       & \\
    \quad -- Other Minor Perks       & Central location; Company vehicle; Company gym       & \\ 
    \bottomrule
\end{tabular}

    \end{center} 
    \vspace{-.5em}
    {\footnotesize \textit{Notes:} This table documents broad categories of attributes detected in vacancies, with notable examples of individual attributes. The full list of job attributes with descriptions is provided in Appendix Tables \ref{tab:attributes_pay_explained}-\ref{tab:attributes_nonpay2_explained}.}
\end{table}

\paragraph{Step 2: Expressions corresponding to each job attribute}
Each job attribute retrieved in Step 1 is associated with an initial set of expressions. For instance, the attribute ``flexible working hours,'' is associated with the target expressions ``flexible working hours'' and ``flexible work time arrangements.''\footnote{The corresponding expressions in Norwegian are ``fleksibel arbeidstid'' and ``fleksible arbeidstidsordninger,'' which denote working hour schemes that offer workers the possibility to choose their own schedule.} In Step 2, we seek to enrich this initial set by including other common ways of referring to the same attributes. 

To expand on the set of possible expressions in our analysis, we rely on the Continuous Bag-of-Words (CBOW) Word2Vec algorithm \citep{mikolov2013word2vec}\footnote{See \cite{atalay2020evolution} for an application of this algorithm to vacancy job titles.}. This approach measures the degree of similarity between phrases by exploiting that words with similar meanings tend to occur in similar contexts.
As an example, information about flexible working hours is commonly given with information about whether the spell is full-time or part-time in our corpus.
We train the CBOW model using our collection of job ads, thus associating words that are similar in the context of job ads but not necessarily in other contexts. 
We then loop through the full collection of job ads and use the trained model to store phrases of one, two, or three consecutive words most similar to the phrases in our original lists from Step 1. Finally, we remove expressions that are wrongly classified as similar by the model. An example of a removed phrase is ``flexible job adaptation,'' which is classified as similar to ``flexible working hours'' by the model, but does not imply flexibility in one's schedule.\footnote{Notably, as there are two variants of written Norwegian (Bokmål and Nynorsk), we translate all phrases from Bokmål to Nynorsk, and append phrases that are not already captured by our dictionary.} 
This discussion highlights the main advantage of our approach. 
We are able to identify a large number of attributes that can be described by many different expressions, but we still retain control over which specific expressions are included. 

In total, our dictionary consists of 1,772 unique expressions, and each of the 47 attributes is associated with around 39 expressions on average, although some attributes have many fewer associated expressions than others. 
We have made this full list of expressions available in an online repository for other researchers to use.\footnote{This repository can be accessed at this link: \url{https://zenodo.org/records/14973628}.} 
Differences in the number of associated phrases reflect both the number of ways an attribute is presented in vacancy texts and the specificity of an attribute. 
For instance, the attribute ``inclusive work-life scheme'' is usually discussed using specific terms in Norwegian, which is why the model identifies only a small number of expressions for this attribute. Conversely, the attribute ``shift work'' is associated with many alternative expressions since there are many ways jobs can involve shift work.

\paragraph{Step 3: Apply to all vacancies}
We use our 1,772 unique expressions to generate job attribute indicators for all ads in the data.
A job ad is defined as advertising a given attribute if any of the expressions associated with the attribute is found in the ad text.

\subsection{Validation}\label{sec:validation}

A potential drawback of our approach is that it might fail to detect attributes implied by more complex phrases. Consider the sentence ``we offer flexibility in starting date as well as in working hours.'' As there is some text between ``flexibility'' and ``working hours,'' our dictionary approach would not detect flexibility in working hours in this case. More generally, we can expect that a job posting does offer a specific attribute when our procedure identifies it (high precision rate), but we can be less certain that a job posting does not contain that same attribute when it is not identified by our procedure (low sensitivity rate). Another potential concern is that our approach does not differentiate between affirmations and negations when searching for specific expressions. In practice, this concern is limited, as negations related to advertised attributes are rare.\footnote{We examined the prevalence of negations by searching for 22 negation cues (e.g., ``not,'' ``no,'' ``absence,'' ``never'') within the six words preceding each detected attribute phrase. At most 0.8\% of detected attribute mentions indicated a potential negation.}

\begin{table}[h!] 
    \caption{Validation of Detected Job Attributes.}\vspace{-1em} 
    \label{tab:attribute_distribution}
    \begin{center}
        \scalebox{.80}{\def\sym#1{\ifmmode^{#1}\else\(^{#1}\)\fi}

\begin{tabular}{lcccccc} \hline\midrule 
\multicolumn{1}{c}{} & \multicolumn{1}{c}{All Ads} & \multicolumn{5}{c}{Job Ads in the Validation Sample} \\ \cmidrule(lr){2-2} \cmidrule(lr){3-7} 
\shortstack{} & \shortstack{Text\\Analysis} & \shortstack{Text\\Analysis} & \shortstack{Manual\\Recognition} & \shortstack{Success\\Rate} & \shortstack{Precision\\Rate} & \shortstack{Sensitivity\\Rate} \\ 
 & (1) & (2) & (3) & (4) & (5) & (6) \\ 
\midrule 
\textbf{A. Pay Attributes}&\textbf{70.8}&\textbf{79.2}&\textbf{73.2}&\textbf{89.0}&\textbf{89.3}&\textbf{96.6} \\  
\quad –– Compensation Scheme&55.9&64.8&61.8&92.5&91.9&96.4 \\  
\quad –– Financial Attributes&32.4&37.8&34.2&96.5&90.7&100.0 \\  
\quad –– Career Opportunities&43.6&48.8&41.8&79.0&71.3&83.2 \\  
\textbf{B. Non-Pay Attributes}&\textbf{96.5}&\textbf{98.0}&\textbf{97.0}&\textbf{97.5}&\textbf{98.2}&\textbf{99.2} \\  
\quad –– Hours of Work&77.0&76.0&71.8&92.2&92.1&97.6 \\  
\quad –– Convenient Hours&20.5&15.8&15.8&94.0&81.0&81.0 \\  
\quad –– Inconvenient Hours&27.0&28.0&22.8&92.8&77.7&95.6 \\  
\quad –– Contract Duration&65.0&67.2&66.8&88.0&90.7&91.4 \\  
\quad –– Workplace Attributes&50.7&55.0&51.0&82.0&80.0&86.3 \\  
\quad –– Task-Related Attributes&55.9&65.0&46.5&72.0&64.2&89.8 \\  
\quad –– Minor Perks&39.8&46.0&18.5&66.5&33.7&83.8 \\  
Any Observed Attribute&97.0&98.2&98.0&98.2&99.0&99.2 \\  
\midrule 
Average Number of Attributes&6.11&6.64&5.56&–&–&– \\  
Number of Job Ads&860,467&400&400&400&–&– \\  
\midrule\hline 
\end{tabular}}
    \end{center}
     {\footnotesize \textit{Notes}: This table compares prevalence rates for ten broad categories of job attributes measured using our text analysis approach and manual recognition. A broad category of job attribute is considered present if at least one of the underlying distinct job attributes is detected. Column (1) is for the full sample of job ads with non-missing vacancy text posted 2015--2019, while Columns (2)-(6) consider a random sample of 400 job ads used in the validation. Columns (1)-(3) report prevalence rates, while Columns (4)-(6) provide summary statistics that compare prevalence rates from text analysis and manual recognition. Success rate is defined as the share of job ads where the text analysis yields the same result as manual recognition. Precision is the rate of agreement between the two methods given that an attribute was detected in our text analysis, while, sensitivity is the rate of agreement given that an attribute was detected in the manual recognition. Precision is a measure of how many detected attributes are false positives, and sensitivity is a measure of how good our method is at recovering true attributes. See Appendix Table \ref{tab:validation_nonpay} for results from validation exercises for each of the 47 underlying job attributes that contribute to the ten broad categories shown here.}
\end{table}

To check the importance of the above concerns in our setting, we performed an additional round of manual recognition. We selected another random sample of 400 job ads posted during the five-year period from 2015 to 2019, and directed another group of research assistants to manually classify job attributes in this sample. We deliberately recruited a distinct group of assistants from the ones who contributed to the initial manual classification (Step 1 of Section \ref{sec:extraction}). We gave each research assistant our full list of 47 attributes and a few sample expressions for each attribute, and tasked them to search for each of these attributes in the full text of each job ad. We could then systematically compare the attributes retrieved with our automated procedure to those detected in the manual human recognition.

Table \ref{tab:attribute_distribution} shows the results of this comparison for ten broad categories of job attributes. Column (1) reports the prevalence rates measured based on our text analysis for each of these categories in the sample of job ads from 2015--2019. Column (2) reports the corresponding rates for the random sample of 400 job ads, also based on our text analysis procedure, while Column (3) reports the prevalence rates from manual recognition for this random sample. Finally, Columns (4)-(6) provide rates of success, precision, and sensitivity by comparing  prevalence rates for each category across text analysis and manual recognition.

Across the ten broad categories of job attributes, we find that our procedure performs well in terms of success, precision, and sensitivity rates. Success and precision rates are above or around 80\% for most categories. The notable exceptions are ``task-related attributes'', where we find a success rate of 72\% and precision of 64.2\%, and ``minor perks'', where we find a success rate of 66.5\% and precision of 33.7\%. The former is driven by differences in detection rates for whether the job ``involves leadership responsibilities'' and ``work involves travelling,'' while the latter is driven by ``central location,'' which is more frequently detected in our text analysis.\footnote{See Appendix Table \ref{tab:validation_nonpay}, which shows the results of the same validation exercise for each of the 47 underlying attributes contributing to the ten broad categories in Table \ref{tab:attribute_distribution}.} Overall, high levels of precision indicate that the attributes recovered by text analysis reflect the actual content of the attributes well. Sensitivity is always equal to or larger than precision, above 80\% for all attributes.

\FloatBarrier

\subsection{The Prevalence of Advertised Job Attributes}
\label{sec:prevalence}

\begin{figure}[h!]
    \caption{The Prevalence of Job Attributes Advertised in Vacancy Texts.}\vspace{-1em}
    \label{fig:share_attributes}
    \begin{center}
        \includegraphics[width=.72\textwidth]{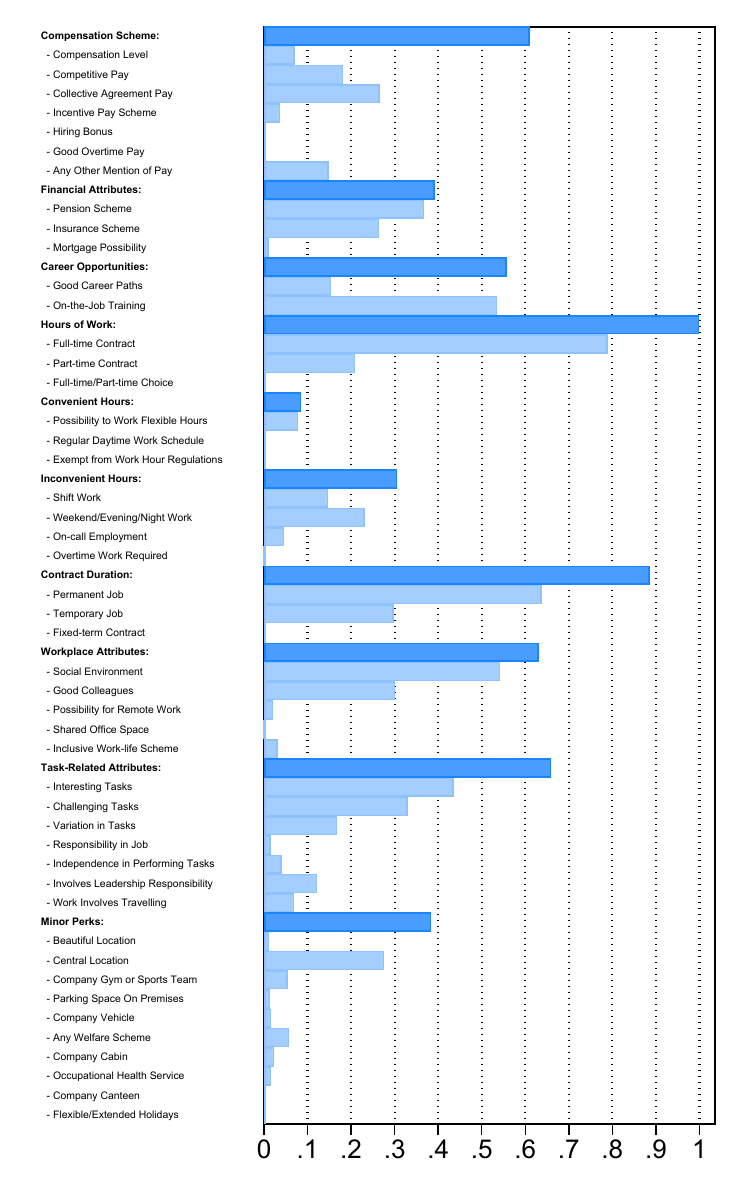}
    \end{center}
    \vspace{-2em}
    {\footnotesize \textit{Notes:} This figure documents the prevalence of job attributes detected in job ads posted in Norway between 2021 and 2024 [N=943,721]. The light blue bars show the share of ads detected with each distinct job attribute. The dark blue bars show the share of ads detected with at least one attribute within ten broad categories. Appendix Figure \ref{fig:share_attributes_2015_2024} shows the corresponding results for the 2015-2024 period as a whole.} 
\end{figure}

We now use the detailed advertised job attributes to highlight several salient descriptive statistics. Figure \ref{fig:share_attributes} reports the share of job ads posted between 2021 and 2024 advertising each of the 47 attributes, while Appendix Figure \ref{fig:share_attributes_2015_2024} shows the corresponding results for the 2015-2024 period as a whole. About 60\% of job ads provide some compensation-related information. However, explicit information about the actual level of pay is scarce. Less than 10\% of ads mention a salary number or bracket (``compensation level'') and slightly more than 25\% of ads indicate that pay is set by a collective wage bargaining agreement (``collective agreement pay'').\footnote{By comparison, \cite{batra2023online} find that 13\% of job ads have salary information in US data.} Taken together, we find that 30\% of job ads feature explicit information about the actual pay level or CBAs (e.g., pay scales or floors).

\begin{figure}
    \begin{center}
    \caption{Explained Variation in Publicly Advertised Job Attributes.}
    \label{fig:explained_variation}
    \vspace{-0.5em}\includegraphics[width=.85\textwidth]{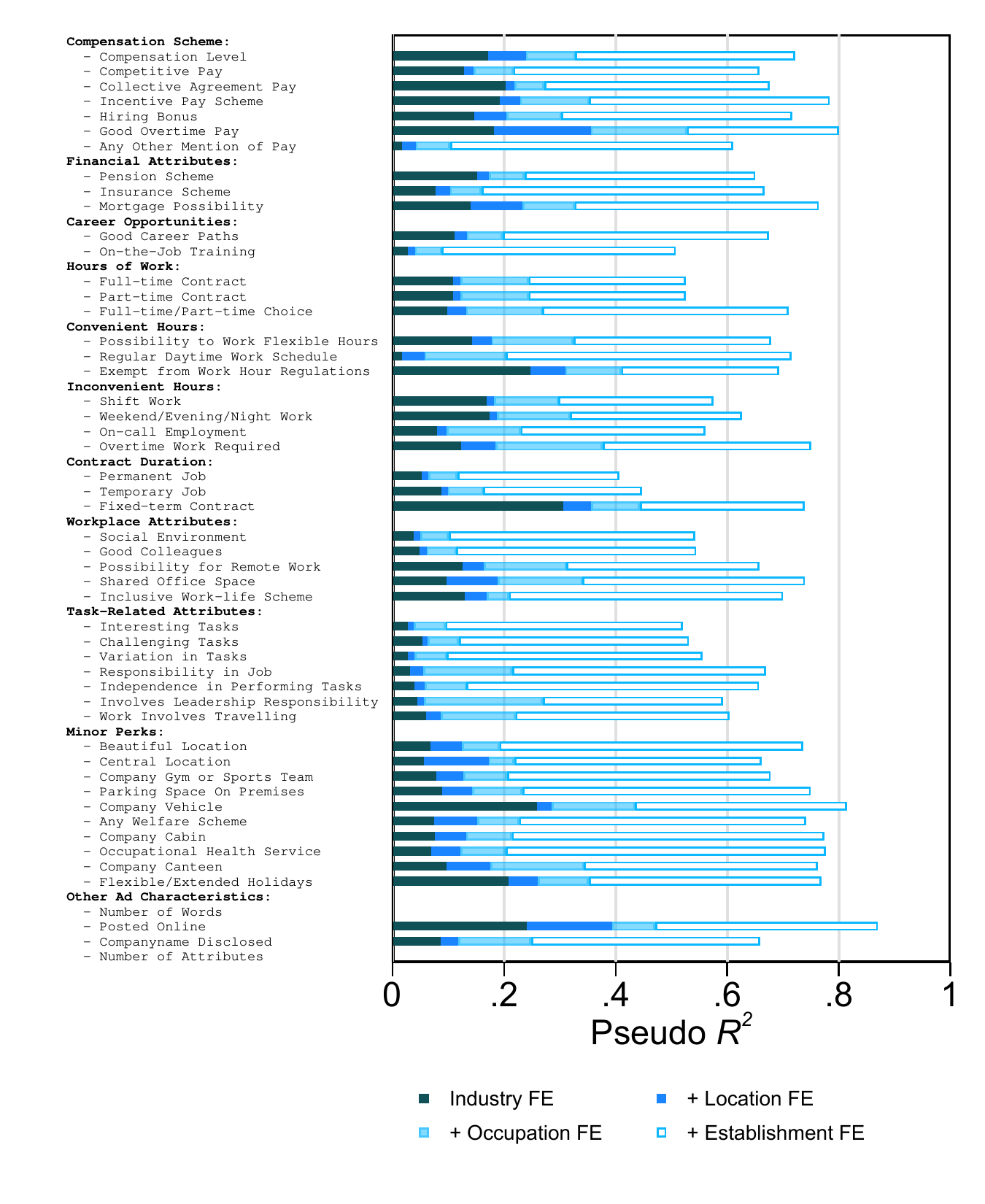}
    \end{center} 
    \vspace{-3em}
    {\footnotesize \textit{Notes:} This figure shows pseudo $R^2$ \citep{mcfadden1974} from separate logistic regressions of binary job attributes detected in job ads posted in Norway between 2021 and 2024 [N=748,870] on fixed effects denoting unique combinations of 2-digit industries (88 groups), location indicators (10 groups), 2-digit occupations (44 groups), and 57,656 establishments. We start by including industry fixed effects, continue by including industry$\times$location fixed effects, and so on. The last set of regressions controls for 133,964 unique combinations. The regression excludes 194,851 ads with a unique combination of establishment, occupation, location, and industry indicators, since these observations are perfectly predicted by the last set of regressions by construction. Location indicators group 422 municipalities into 10 groups based on the number of workers, so municipalities assigned to the same group have a similar number of workers, with a specific indicator for job postings in Oslo. Appendix Figure \ref{fig:explained_variation_shapley} shows the corresponding Shapley Value decomposition, while Appendix Figure \ref{fig:explained_variation_pseudo_all} shows the corresponding results for the 2015-2024 period as a whole.}
\end{figure}

Beyond information on the more tangible attributes of the job, such as pay, hours of work and contract duration, Figure \ref{fig:share_attributes} also shows that employers frequently advertise more subjective aspects of the working environment, such as characteristics of the workplace (``workplace attributes'') or some appreciation of the type of tasks involved (``task-related attributes''). Around 35--40\% of ads describe these tasks as ``challenging'' or ``interesting.'' Similarly, more than 50\% of vacancies feature some language about the quality of the ``social environment'' at work and 30\% advertise ``good colleagues.''

Next, we analyze to what extent differences in advertised attributes are explained by observable characteristics of the job ad, such as the industry of the posting establishment or the occupation associated with the job title stated in the job ad. To describe the variation in advertised job attributes that can be explained by these characteristics, we estimate a series of logistic regressions for the probability that a given attribute is advertised in the text of a vacancy with several sets of fixed effects: industries, locations, occupations, and establishments. Figure \ref{fig:explained_variation} reports the estimated pseudo-$R^2$, a summary measure of goodness of fit for binary outcomes \citep{mcfadden1974}, from logistic regressions estimated separately for each attribute.\footnote{As an alternative, we also estimated linear probability models using OLS with the same set of fixed effects separately for each attribute (results available upon request). However, as all of our pay and non-pay attributes are binary, one could be concerned about interpreting standard goodness of fit from such regressions, which restricts the range of the coefficient of determination $R^2$ \citep{cox1992}.} This measure therefore captures the variation in the content of job ads within industries, locations, occupations, and establishments.

A possibility is that the attributes advertised for a given position are largely determined by its industry or occupation, so the pay and non-pay attributes mentioned in vacancy texts directly reflect the type of job advertised. For example, it can be expected that working as a hospital nurse involves shift-work, in which case there is no value-added to mentioning this information in the corresponding job description. The data instead suggest that there is variation in advertised job attributes for very similar positions. Figure \ref{fig:explained_variation} shows that industries, occupations, and location fixed-effects explain at most 30--40\% of the variation in the attributes advertised by employers across the 47 categories we retrieve. By contrast, when we add establishment fixed-effects, we find that the explained variation jumps up to 50--70\% across attributes, which suggests that advertised attributes are in large part correlated across vacancies for establishments observed with several job ads. These patterns are confirmed in a Shapley value decomposition (Appendix Figure \ref{fig:explained_variation_shapley}), and also hold when we extend the analysis to the years 2015-2024 (Appendix Figure \ref{fig:explained_variation_pseudo_all}). These patterns are also consistent with the existence of an ``establishment fixed-effect'' in advertised attributes, and it justifies the employer-level analysis we develop in the next sections.

\section{The Information Content of Job Ads}
\label{sec:learning_value}

In the previous section, we established that employers advertise many pay and non-pay attributes through job vacancy texts. How much information about the quality of employers can workers infer from these attributes? We answer this question using three empirical strategies. In Section \ref{subsec:linked_ad_to_estab}, we analyze how advertised job attributes correlate with revealed-preference measures of employer quality. In Section \ref{subsec:linked_ad_to_hire}, we link ads to hires and examine the concordance between advertised and realized job attributes, for the subset of attributes with direct counterparts in the register data. In Section \ref{subsec:linked_choice_to_ad}, we present evidence from a survey experiment in which respondents chose among experimentally designed choice sets featuring real job ads. All three strategies suggest that job ads are a reliable channel through which workers can gather information about employer quality.

\subsection{Evidence from Linked Ad-Establishment Data}
\label{subsec:linked_ad_to_estab}

We derive alternative composite measures of employer quality from the matched employer-employee data, as commonly done in revealed-preference approaches. These measures do not rely on the vacancy data and are obtained from actual labor market outcomes, such as wages and worker flows across employers. We then link these measures to the advertised job attributes retrieved in Section \ref{sec:advertised_attributes} and ask whether advertised attributes correlate with employer quality.

\subsubsection{Revealed-Preference Measures of Employer Quality}

We consider alternative revealed-preference measures previously used in the literature, each capturing workers' common valuation of alternative employers. The common theoretical underpinning of these measures is that labor market imperfections, such as search frictions or monopsony power, can sustain dispersion in employer quality in equilibrium.

We center our analysis on four measures of employer quality: (i) the employer pay premium from a standard two-sided wage regression \citep{abowd1999high}, (ii) the overall employer value in the utility-posting model of \citet[][]{sorkin2018ranking}, (iii) the poaching index of \citet[][]{bagger2019empirical}, and (iv) employer size \citep{card2018firms,burdett1998wage}.\footnote{We also report results for the flow utility value in the utility-posting model of \citet{sorkin2018ranking}.} We select these measures because they capture workers' revealed preferences within a broad class of models with heterogeneous employers. For example, in many job search models, workers have well-defined preferences over employers that are closely linked to employer size, the poaching index, and pay premiums.\footnote{Notably, the standard \cite{burdett1998wage} model with heterogeneous firm productivities yields an identical ranking of firms based on firm size, pay premiums and the metric underlying the poaching index.}

To estimate measures of employer quality, we focus on employers that are connected by worker mobility.\footnote{We compute all measures for the same subset of strongly connected employers, derived from the most stringent identification restrictions in \cite{sorkin2018ranking}.} As shown in Table \ref{tab:sample_selection} (``Employer Quality Sample''), about one-fourth of establishments satisfy this restriction.\footnote{Detailed descriptive statistics on workers and employers are in Table \ref{tab:sample_characteristics_2021_2024} of Appendix \ref{app:add_results}.} Since employment is concentrated at larger employers, however, this sample still covers 74\% of workers and 66\% of job ads. To precisely estimate employer-level parameters, we also reduce their number using a clustering algorithm as a pre-estimation step \citep{bonhomme2019distributional}, which partitions employers into groups with similar observable characteristics. The clustering provides a surjective mapping $j \rightarrow g$, where $j$ denotes the employers in the estimation sample and $g(j)$ denotes the employer-cluster to which $j$ belongs. This increases the number of movers per group, which underlies most measures, while preserving substantial employer heterogeneity. The number of clusters is selected to yield groups with 50 establishments on average, which resulted in $1,422$ unique employer clusters in our baseline.\footnote{We provide robustness using alternative clustering procedures, featuring smaller or larger clusters.}  Additional details on the clustering algorithm and the construction and estimation of the employer quality measures are in Appendix \ref{app:quality_employers}.

These four measures are imperfectly correlated, suggesting that they capture different aspects of employer quality. Most correlations fall in the 30--50\% range: for instance, the correlation between pay premiums and Sorkin values is 0.369.\footnote{For the US, the original study \citep[][Table II]{sorkin2018ranking} found this correlation to be 0.53. Our setting differs from that in \cite{sorkin2018ranking} along many dimensions: time period, institutions, labor market size (thousands vs.\ millions of employers), wage measures (yearly vs.\ hourly wages), and the definition of employer-to-employer transitions (monthly vs.\ quarterly job spells).} One exception is the strong correlation between the Sorkin value and the poaching index, which we estimate to be 0.876. Appendix Table \ref{tab:correlation_quality_measures} provides the full set of correlations across all four measures.

\subsubsection{What do Good Employers Advertise in Job Postings?}

We first investigate whether employers estimated as high- or low-quality advertise different attributes in their job postings. We estimate
\begin{align}
    A_{g(j)}^k = \alpha_{0}^{k} 
    + \alpha_{Q}^{k} \cdot \widehat{Q}_{g(j)} 
    + \varepsilon_{g(j)}^{k},
    \label{eq:reg_attribute_lhs}
\end{align}
at the employer-cluster level, where $g(j)$ denotes the mapping from an employer $j$ to a cluster $g$, $A_{g(j)}^k$ is the share of job ads from employers in cluster $g$ that advertise attribute $k$, $\widehat{Q}_{g(j)}$ is their estimated quality (one of the measures described above), and $\varepsilon_{g(j)}^{k}$ is an error term. The regression is weighted by the number of worker-years in each cluster.



The key coefficient of interest is $\alpha_{Q}^{k}$, which measures the association between employer quality $\widehat{Q}_{g(j)}$ and the propensity to advertise attribute $A_{g(j)}^k$. This descriptive approach transparently shows whether employers of different quality systematically advertise different attributes. We standardize $\widehat{Q}_{g(j)}$, so $\alpha_{Q}^{k}$ can be interpreted as the change in the propensity to advertise $A_{g(j)}^k$ associated with a one standard deviation increase in employer quality.

\begin{landscape}
\begin{figure}[htpb]
    \caption{Publicly Advertised Job Attributes By Employer Values and Pay Premiums.} \label{fig:point_estimates_main} \hspace{-2em}
    \subfloat[][Pay Premium]{\includegraphics[width=.45\textwidth]{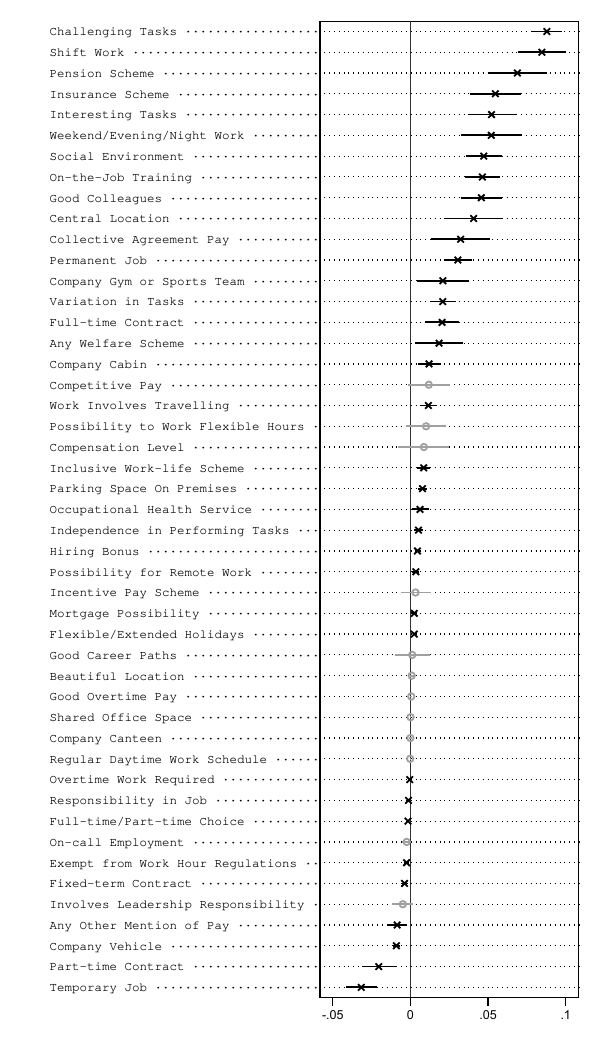}}
    \subfloat[][Overall Sorkin Value]{\includegraphics[width=.45\textwidth]{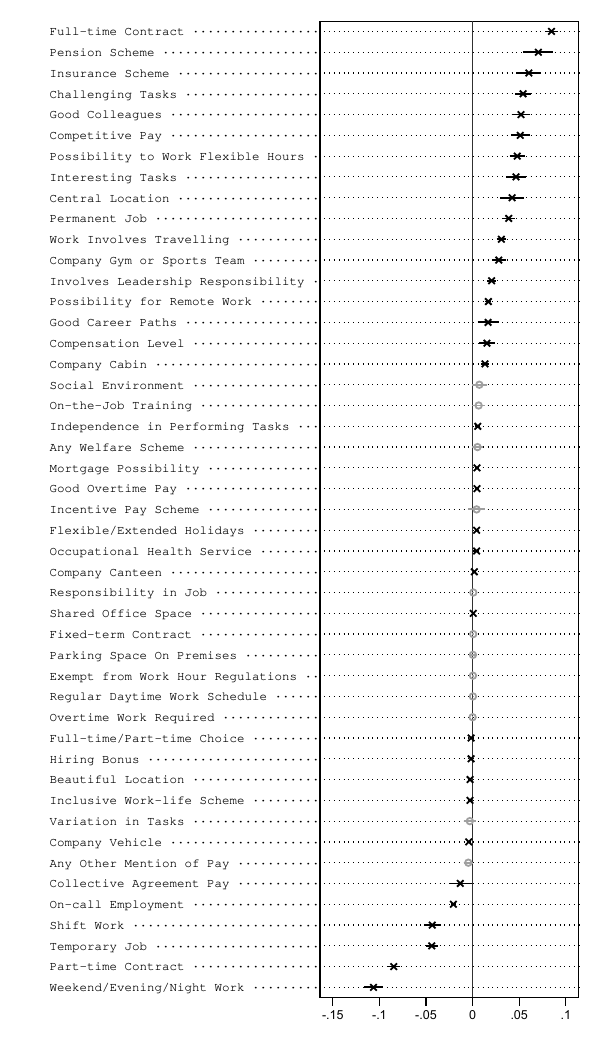}} 
    \subfloat[][Poaching Index]{\includegraphics[width=.45\textwidth]{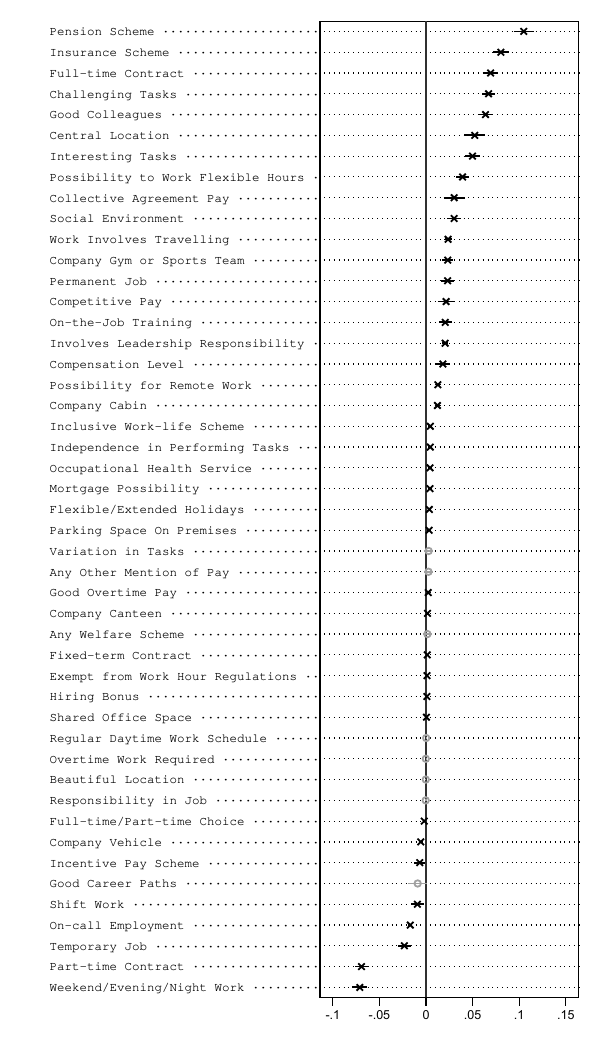}}\\
    {\footnotesize \textit{Notes}: This figure shows parameter estimates and $90\%$ confidence intervals from separate regressions of the share of each ad attribute on Employer Pay Premium, Overall Sorkin Value, and Poaching Index, capturing the change in the fraction of ads with each job attribute associated with a standard deviation increase in employer value. Estimations are done at the employer-cluster level for the period from 2021 to 2024, and are weighted by the number of worker-years. Appendix Figure \ref{fig:point_estimates_paycontrols} documents regressions on Overall Sorkin Value and Poaching Index while controlling for Pay Premium, Appendix Figure \ref{fig:point_estimates_2015_2024_main} documents the corresponding results for the period 2015-2024, while Appendix Figures \ref{fig:point_estimates_g25}-\ref{fig:point_estimates_g100} show results using smaller and larger clusters. Appendix Figures \ref{fig:point_estimates_main_flow}-\ref{fig:point_estimates_main_size} documents corresponding results by Sorkin Flow Value and Employment Size.} 
\end{figure}
\end{landscape}

Figure \ref{fig:point_estimates_main} reports $\widehat{\alpha}_{Q}^{k}$ for three revealed-preference measures of employer quality (pay premium, Sorkin value, and poaching index) and all attributes derived from the vacancy data. In each panel, attributes are ordered by $\widehat{\alpha}_{Q}^{k}$ from largest to smallest. These coefficients typically range from $-0.1$ to $0.1$, where $0.1$ implies a 10 percentage point change in the propensity to advertise a given attribute for each standard deviation increase in employer quality. As shown in Figure \ref{fig:share_attributes}, the baseline prevalence of many attributes is in the 5--30\% range. However, many estimated $\alpha_{Q}^{k}$ are not statistically different from zero, indicating no detectable association between those attributes and employer quality. 

Panel (a) shows that high-pay employers are more likely to advertise a generous compensation package (``pension scheme,'' ``insurance scheme,'' ``collective agreement pay'') and training opportunities (``on-the-job training''). As documented in Section \ref{sec:prevalence}, only 30\% of ads mention pay information (less than 10\% list an actual salary number or bracket), so pay premiums and advertised pay information are separate concepts. Several attributes related to the workplace (``good colleagues,'' ``social environment'') and tasks (``challenging tasks,'' ``interesting tasks'') are also positively associated with employer pay. In sum, high-pay employers tend to advertise a professional environment with a generous compensation package. Our estimates also show that attributes related to inconvenient hours (e.g., ``shift work'') are more common among high-pay employers, consistent with compensating differentials for inconvenient work schedules. On the other end, low-pay employers more often advertise a ``part-time contract'' or a ``temporary job''.

Panels (b) and (c) of Figure \ref{fig:point_estimates_main} show $\widehat{\alpha}_{Q}^{k}$ for the Sorkin value and the poaching index, respectively. These two measures are strongly correlated, and the overall patterns of association between $\widehat{Q}_{g(j)}$ and $A_{g(j)}^k$ are similar. Although high-quality employers advertise many of the same attributes as high-pay employers, the positive association with inconvenient work schedules disappears: ``shift work'' and ``weekend/evening/night work'' become attributes of low-quality employers. Conversely, flexible work arrangements, such as ``possibility to work flexible working hours'' and ``possibility for remote work,'' are positively associated with high-quality employers. In sum, high-quality employers tend to advertise more generous compensation and better work schedules.

We present results for a series of alternative specifications in Appendix Figures \ref{fig:point_estimates_2015_2024_main}--\ref{fig:point_estimates_main_size}. The results are robust to extending the sample period (2015--2024 instead of 2021--2024), varying cluster size (25 or 100 establishments per cluster on average instead of 50), and using alternative revealed-preference measures of employer quality (Sorkin flow value and employment size). We also show results for the flow-based measures of employer quality while controlling for pay premiums in Appendix Figure \ref{fig:point_estimates_paycontrols}. As some attributes, such as ``shift work'' and ``weekend/evening/night work,'' are positively correlated with pay (likely reflecting compensating differentials), this alternative specification reinforces their negative correlation with these measures, consistent with these attributes entering worker utility negatively.

Overall, the correlations in Figure \ref{fig:point_estimates_main} suggest that the content of job ads is informative. Attributes typically seen as positive (financial benefits, flexible schedules, job stability) are more likely to be advertised by high-quality employers, while low-quality employers more often advertise attributes seen as negative (constrained schedules, temporary positions). Advertised information therefore does not appear to be ``cheap talk.''

However, advertised attributes need not translate directly into actual job characteristics, for two reasons. First, employers may not be able to commit to all advertised attributes, particularly subjective ones such as ``social environment'' and ``good colleagues.'' These may be better interpreted as signals of workplace culture than as concrete commitments. Second, employers may omit actual job characteristics from the ad: for instance, an employer may offer a competitive salary without mentioning it explicitly upfront.

\subsubsection{Predicting the Quality of Employers from Job Ads} 

How predictive are advertised job attributes of employer quality? To quantify the predictive power of the pay and non-pay content of job ads, we estimate
\begin{align}
    \widehat{Q}_{g(j)} = \beta_0
    + \sum_k \beta_k \cdot A_{g(j)}^k
    + \beta_X X_{g(j)}
    + \varepsilon_{g(j)}^{Q},
    \label{eq:reg_attribute_rhs}
\end{align}
at the employer-cluster level, where $g(j)$ maps employer $j$ to cluster $g$, $A_{g(j)}^k$ is the share of ads from employers in cluster $g$ featuring attribute $k$, $X_{g(j)}$ is a vector of industry, location, and occupation controls capturing other commonly observed job ad characteristics, and $\varepsilon_{g(j)}^{Q}$ is an error term. The regression is weighted by the number of worker-years in each cluster.


\begin{figure}[h!]
    \caption{Predictive Power of Publicly Advertised Job Attributes.}\vspace{-1em}
    \label{fig:reg_value_on_characteristics}
    \begin{center}
    \includegraphics[width=.7\textwidth]{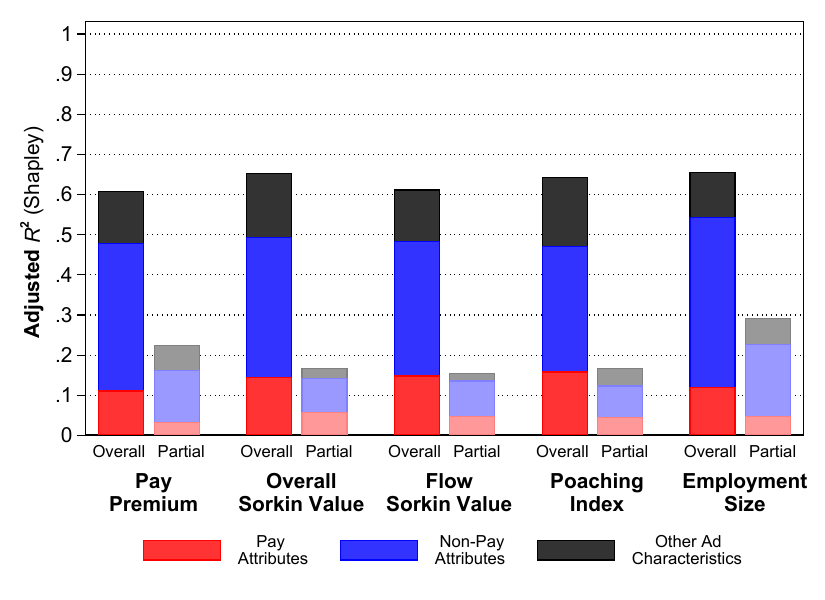}
    \end{center} \vspace{-1em}
    {\footnotesize \textit{Notes}: This figure shows adjusted $R^2$s from regressions of estimated model parameters on pay and non-pay job attributes and other ad characteristics retrieved in the text analysis, including the number of words in the ad, indicators for ad posted online, disclosure of company name, and the overall number of attributes in the ad. We decompose $R^2$s into the contributions of Pay Attributes, Non-Pay Attributes, and Other Ad Characteristics using the Shapley value decomposition. Estimations are done at the employer-cluster level for the four-year period from 2021 to 2024, and are weighted by the number of worker-years in each cluster. Overall $R^2$s are from regressions that include ad attributes and other ad characteristics, while partial $R^2$s are from regressions that also control for the composition of industry, occupation and location in each cluster. Industry and occupation controls are defined at the two-digit level. Location indicators split local municipalities into deciles, with municipalities in the same group having similar number of workers (Oslo as an own group). The sample corresponds to the ``Employer Quality Sample'' in Table \ref{tab:sample_selection}. Appendix Figure \ref{fig:r2_different_cluster_sizes} reports the corresponding results using alternative clustering procedures, while Appendix Figure \ref{fig:r2_2015-2024} provides results for the 2015-2024 period and Appendix Figure \ref{fig:r2_by_industry} provides results by main industry.}
\end{figure}

We use the adjusted $R^2$ of regression \eqref{eq:reg_attribute_rhs} as our summary measure of predictive power, reporting versions with and without the controls in $X_g$.\footnote{The formula used to calculate predictive power is $\textnormal{Adjusted }R^2=1 - (SS_{res}/df_{res})/(SS_{total}/df_{total})$ where $SS_{res}$ is the residual sum of squares, $df_{res}$ is the number of employer clusters (1,422 in our baseline) minus the number of explanatory variables, and $SS_{total}$ and $df_{total}$ are from a regression on a constant only. The partial $R^2$ is the share of variation unexplained by baseline controls alone but explained when job ad attributes are added: $\textnormal{Partial }R^2=(R^2_{full}-R^2_{baseline})/(1-R^2_{baseline})$ where $R^2_{full}$ ($R^2_{baseline}$) is the adjusted $R^2$ from a regression with (without) job attributes.\label{foot:r2_formula}} Figure \ref{fig:reg_value_on_characteristics} shows this measure for alternative estimates of employer quality $\widehat{Q}_{g(j)}$, decomposed across three broad groups of attributes: pay attributes (red bars), non-pay attributes (blue bars), and other ad characteristics (black bars).\footnote{Other ad characteristics include the number of advertised attributes, the number of words in the ad, and whether the employer name is mentioned.} Since the contribution of each group of attributes depends on the order in which they are introduced in \eqref{eq:reg_attribute_rhs}, we report a Shapley value decomposition.

Figure \ref{fig:reg_value_on_characteristics} shows that job ad content meaningfully predicts employer quality across all measures we consider. Non-pay attributes (blue bars) contribute two to three times as much as pay attributes (red bars), with other ad characteristics (black bars) contributing broadly as much as pay attributes. Based solely on job attributes, without the controls in $X_{g(j)}$, we explain about 60\% of the variation in employer quality. Conditioning on $X_{g(j)}$, job attributes still explain 15\% to 20\% of the remaining variation, as shown by the partial $R^2$ in Figure \ref{fig:reg_value_on_characteristics}. These findings are robust to using smaller or larger employer groups (Appendix Figure \ref{fig:r2_different_cluster_sizes}) and to extending the sample period to 2015--2024 (Appendix Figure \ref{fig:r2_2015-2024}).

We further explore the adjusted-$R^2$ reported in Figure \ref{fig:reg_value_on_characteristics} through the following two exercises. First, we decompose the adjusted-$R^2$ by finer attribute categories in Appendix Figure \ref{fig:decompose_r2_ag}, reporting the contribution of each of the ten categories from Table \ref{tab:broad_categories} separately. This exercise indicates that explanatory power is broadly distributed across attribute categories rather than concentrated in a few, though attributes related to compensation and inconvenient hours consistently contribute more, as shown by the Shapley value decompositions in panels (a) and (b) of Appendix Figure \ref{fig:decompose_r2_ag}.

In a similar spirit, Appendix \ref{app:restricted_models} documents that these patterns are robust to focusing on individual attributes. Specifically, we use Lasso regression in Equation \eqref{eq:reg_attribute_rhs} to select the most predictive individual job attributes for each employer quality measure. This confirms that attributes related to hours of work, contract duration, pension scheme, and collective agreement pay are highly predictive of employer quality.

\subsection{Evidence from Linked Ad-Hire Data}
\label{subsec:linked_ad_to_hire}

How well do job attributes advertised in job postings correspond to realized job characteristics observed in administrative data? For a limited set of attributes, we are able to construct counterparts of advertised attributes in the matched employer–employee data, which allow us to examine the concordance between advertised and realized job attributes. We perform this assessment using the linked ad-hire sample discussed in Section \ref{sec:posting hiring}.

We focus on a subset of attributes by constructing binary indicators from the register data that closely match those retrieved from job ad texts. The attributes with the most direct counterpart in administrative records are ``shift work'' and contract duration (``permanent job'' or ``temporary job''). From reported contractual hours, we also construct indicators for ``full-time contract'' ($\geq$30 hours per week) and ``part-time contract'' ($<$30 hours per week). Further, using detailed spell-level information on payroll and in-kind benefits, we construct three further attributes: bonus pay (``incentive pay scheme''), travel-related reimbursements (``work involves travelling''), and company car benefits (``company vehicle''). Finally, we define jobs located in one of the four largest cities in Norway as offering a ``central location.''

In most cases, the attributes we extract from administrative records are contractual in nature and unlikely to be misreported in job ads. Still, some comparisons are ambiguous: our hours threshold for full-time work may not align with the threshold employers use in postings, and our classifications of incentive pay and travel involvement depend on arbitrary cutoffs. Moreover, negotiations between employers and workers during the recruitment process may result in final contracts that deviate from what was advertised. Divergence between posted and realized attributes therefore does not necessarily imply cheap talk. Despite these limitations, we view this exercise as a novel feature of our setup.

Table \ref{ad_regdata_validation} compares attribute prevalence in job ads and in administrative records for the linked ad-hire sample. Columns (1) and (2) show that prevalence lines up closely across the two sources: attributes we rarely detect in job ads, such as ``incentive pay scheme,'' ``company vehicle,'' and ``work involves traveling,'' are also rare in register data, while common contractual attributes such as ``part-time contract,'' ``shift work,'' and ``temporary job,'' are common in both. Column (3) reports the share of linked ads and job spells where the two sources agree. We find success rates are generally above 90\%. The main exceptions are ``temporary job'' and ``part-time contract,'' where match rates are closer to 75\%. This could reflect genuine deviations from the listed attributes, but may also stem from measurement error either in the attribute definitions or in the construction of the ad-hire sample.

\begin{table}[ht!]
    \caption{Publicly Advertised Job Attributes and Observed Attributes in Register Data.}\vspace{-1em}
    \label{ad_regdata_validation}
    \begin{center}
    \scalebox{.75}{\begin{tabular}{lccccccc}
\midrule\hline 
& &&& \multicolumn{4}{c}{Cohen's $\kappa$} \\ 
& \multicolumn{2}{c}{Attribute Prevalence} && \multicolumn{4}{c}{Relative to Random Matching Within:} \\ \cmidrule(lr){2-3} \cmidrule(lr){5-8}
& Register
& Job Ads
& Success Rate
& All Jobs 
& Occuption
& Industry
& Occ. \# Ind. \\
& (1) & (2) & (3) & (4) & (5) & (6) & (7)   \\ 
\midrule
Incentive Pay Scheme       & 2.4\%    & 3.5\%    & 95.4\%   & 0.25     & 0.21     & 0.18     & 0.18     \\ 
Part-time Contract         & 23.0\%   & 22.4\%   & 78.5\%   & 0.40     & 0.25     & 0.41     & 0.24     \\ 
Shift Work                 & 13.1\%   & 12.4\%   & 87.9\%   & 0.44     & 0.19     & 0.44     & 0.17     \\ 
Temporary Job              & 30.3\%   & 27.2\%   & 75.8\%   & 0.44     & 0.34     & 0.42     & 0.34     \\ 
Work Involves Travelling   & 5.0\%    & 5.6\%    & 92.0\%   & 0.28     & 0.13     & 0.17     & 0.11     \\ 
Central Location           & 27.1\%   & 23.9\%   & 76.1\%   & 0.43     & 0.41     & 0.41     & 0.39     \\ 
Company Vehicle            & 1.0\%    & 2.0\%    & 97.7\%   & 0.18     & 0.17     & 0.16     & 0.16     \\ 
\midrule\hline 
\end{tabular}
}
    \end{center}
    {\footnotesize \textit{Notes:} This table documents attribute prevalence and the correspondence between linked attributes in register and job ads data. Columns (1) and (2) document attribute prevalence in register and job ads data. Column (3) documents how often job ad attributes correspond to the information provided in register data (success rate). Column (4) compares the success rate to the expected success rate when forming random links between ads and hires using Cohen's $\kappa$. Hired worker attributes are binary indicators extracted from matched employer-employee data, and are measured as follows: Incentive Pay Scheme indicates that bonus pay accounts for $> 5\% $ of total pay; Full-time Contract indicates contractual hours $>30$ hours a week; Shift Work and Temporary Job are directly reported by employers; Work Involves Traveling indicates that the worker had at least two recorded travel expenses; Central Location indicates that the establishment is located in one of the four largest Norwegian cities; and Company Vehicle indicates company car benefits recorded for tax purposes. For the purpose of this classification, attributes relating to Full-time Contract and Permanent Job correspond to the counterpart of Part-time Contract and Temporary Job attributes, respectively, and are thus left out from this table, although these attributes are also found in the register data. The success shares and Cohen's $\kappa$ for these attributes can be inferred from their respective counterparts. Appendix Table \ref{tab:characteristics_linked_and_survey}, Column (2), documents the sample characteristics of individuals that were uniquely linked to a job ad. }
\end{table}

Notably, the success rates in Column (3) of Table \ref{ad_regdata_validation} depend not only on the predictive value of job ads but also on the prevalence rates reported in Columns (1)-(2), i.e., differences in success rates may simply reflect differences in incidence. To address this, we use Cohen's $\kappa$-coefficient, which measures agreement between the two sources relative to what would be expected under random matching \citep{Cohen1960}.\footnote{The expected success rate with random matching is $p_{r}=s_{\textnormal{ad}}s_{\textnormal{job}} + (1-s_{\textnormal{ad}})(1-s_{\textnormal{job}})$, where $s_{\textnormal{ad}}$ is the share of ads with detected attribute, and $s_{\textnormal{job}}$ is the corresponding share of job spells with the detected attribute. The Cohen's $\kappa$-coefficient is defined as: $\textnormal{Cohen's } \kappa = (p_{o}-p_{r})/(1-p_{r})$, where $p_{o}$ and $p_{r}$ denote observed success rates and the expected success rate with random matching, respectively. Cohen's $\kappa$ is zero whenever the observed success rate $p_{o}$ equals $p_{r}$, and is one whenever $p_{o}$ is 100\%.} Columns (4)-(7) of Table \ref{ad_regdata_validation} report Cohen's $\kappa$ estimates across different levels of random matching, consistently showing that 20--40\% of the agreement between job ads and register data exceeds what could be expected purely based on random matching. We take this as evidence that posted job attributes meaningfully predict the job characteristics observed in employment contracts.

\subsection{Evidence from Survey Experiment with Job Ads}
\label{subsec:linked_choice_to_ad}

We now present evidence from a survey experiment by \citet{BhullerAEA2024}, which elicited respondents' preferences over real job ads using hypothetical choice scenarios. Each respondent $i$ faced multiple choice scenarios $s$, each presenting two alternatives $c_{is(j,j')}=\{c_{isj},c_{isj'}\}$, and stated their preferred alternative. Within each scenario, both ads were randomly drawn from the same sector and broad occupation, and therefore differences between alternatives reflect within-sector-occupation differences across employers, notably in posted job attributes. We use respondents' choices to evaluate the information content of job ads in two steps.

As a first step, we use the non-pay job attributes $A_j=(A^1_{j},\dots,A^K_{j})$ retrieved from job ad texts, as described in Section \ref{sec:advertised_attributes}, together with respondents' choices, to estimate how the presence of specific attributes affects choices and to construct willingness-to-pay (WTP) estimates for workplace amenities. Importantly, respondents could read the job ad texts associated with each alternative, so their choices may reflect preferences over the job attributes they could infer from the ads. Using the survey responses, we construct binary indicators $D_{is(j,j')}=1[c_{isj} \succ c_{isj'}]$ for respondent $i$'s preferred alternative in scenario $s(j,j')$. Using the binary indicators $D_{is(j,j')}$, non-pay job attributes $(A_j,A_{j'})$ and employer pay premiums $(w_j,w_j')$, we estimate a conditional logit model. Standard errors are clustered at the respondent level, as each respondent faced many scenarios.

Table \ref{tab:experiment_wtp} reports the results from the choice model. Column (1) presents the parameter estimates of the choice model where we used a subset of advertised non-pay attributes and employer pay premium as explanatory variables. Column (3) provides the implied WTP estimates $\gamma_k$ for these non-pay job attributes, constructed as the ratios of the attribute-specific choice parameters and the employer pay premium estimate reported in Column (1). In determining which non-pay attributes to include in the model, we use two criteria. First, we include attributes that have been emphasized in the literature as salient dimensions of employer quality, such as inconvenient or convenient hours (e.g., shift work, flexible hours) and fringe benefits (e.g., insurance scheme, mortgage possibility). Second, we focused on attributes that were estimated as significant in the preceding analysis.\footnote{As described in Section \ref{sec:advertised_attributes}, our text analysis extracts 47 distinct attributes from the job ads, while in Section \ref{subsec:linked_ad_to_estab} we show that not all of these attributes hold predictive value for the employer quality measures.} In Columns (2) and (4), we provide an additional set of estimates, where we include the remaining job attributes mentioned in Section \ref{sec:advertised_attributes} as additional controls in the choice model. We show that our WTP estimates for the salient attributes are robust to controlling for additional non-pay attributes. We use the WTPs reported in Column (4) as part of the structural analysis in Section \ref{sec:static_model}.

\begin{table}[htbp]
    \caption{Survey Experiment: Choice Parameters and Willingness-to-Pay Estimates.}\vspace{-1em}
   \label{tab:experiment_wtp}
    \begin{center}
    \scalebox{.75}{{
\def\sym#1{\ifmmode^{#1}\else\(^{#1}\)\fi}
\begin{tabular}{l*{4}{c}}
\hline\hline
                    &\multicolumn{2}{c}{\textbf{Conditional Logit}}&\multicolumn{2}{c}{\textbf{Willingness to Pay}}\\
                    &\multicolumn{2}{c}{\textbf{Choice Parameters}}&\multicolumn{2}{c}{\textbf{Estimates}}\\
                    &\multicolumn{1}{c}{(1)}&\multicolumn{1}{c}{(2)}&\multicolumn{1}{c}{(3)}&\multicolumn{1}{c}{(4)}\\
\hline
Temporary Job      &      -0.239\sym{***}&      -0.252\sym{***}&      -0.263\sym{***}&      -0.254\sym{***}\\
                    &    (0.0217)         &    (0.0226)         &    (0.0458)         &    (0.0433)         \\
Convenient Hours     &       0.151\sym{***}&       0.188\sym{***}&       0.167\sym{***}&       0.189\sym{***}\\
                    &    (0.0189)         &    (0.0203)         &    (0.0320)         &    (0.0345)         \\
Inconvenient Hours   &      -0.320\sym{***}&      -0.280\sym{***}&      -0.353\sym{***}&      -0.282\sym{***}\\
                    &    (0.0372)         &    (0.0388)         &    (0.0808)         &    (0.0667)         \\
Insurance Scheme      &      0.0390         &       0.104\sym{***}&      0.0431         &       0.105\sym{***}\\
                    &    (0.0272)         &    (0.0338)         &    (0.0304)         &    (0.0383)         \\
Mortgage Possibility     &       0.265\sym{***}&       0.116         &       0.292\sym{***}&       0.117         \\
                    &    (0.0691)         &    (0.0738)         &    (0.0962)         &    (0.0779)         \\
Career Opportunities         &      0.0600\sym{*}  &      0.0839\sym{**} &      0.0663         &      0.0845\sym{**} \\
                    &    (0.0336)         &    (0.0382)         &    (0.0406)         &    (0.0425)         \\
Possibility for Remote Work    &      0.0591\sym{***}&      0.0649\sym{***}&      0.0653\sym{***}&      0.0653\sym{***}\\
                    &    (0.0122)         &    (0.0130)         &   (0.00482)         &   (0.00482)         \\
Good Colleagues     &      0.0713\sym{***}&      0.0935\sym{***}&      0.0788\sym{**} &      0.0941\sym{***}\\
                    &    (0.0249)         &    (0.0268)         &    (0.0323)         &    (0.0333)         \\
Inclusive Work-life Scheme             &       0.204\sym{***}&       0.172\sym{**} &       0.226\sym{**} &       0.173\sym{**} \\
                    &    (0.0686)         &    (0.0743)         &    (0.0894)         &    (0.0823)         \\
Company Gym or Sports Team   &       0.237\sym{***}&       0.232\sym{***}&       0.262\sym{***}&       0.234\sym{***}\\
                    &    (0.0364)         &    (0.0402)         &    (0.0617)         &    (0.0585)         \\
Company Cabin          &       0.177\sym{***}&       0.270\sym{***}&       0.195\sym{**} &       0.272\sym{***}\\
                    &    (0.0630)         &    (0.0681)         &    (0.0828)         &    (0.0913)         \\
Employer Pay Premium          &       0.906\sym{***}&       0.993\sym{***}&        --             &        --             \\
                    &     (0.174)         &     (0.186)         &                     &                     \\
\hline
\textit{Controls:}        &              &                 &           &                  \\
Respondent Fixed Effects       &       $\checkmark$       &        $\checkmark$       &        $\checkmark$         &       $\checkmark$         \\
Survey Interview Module      &       $\checkmark$       &        $\checkmark$       &        $\checkmark$         &       $\checkmark$         \\
Travel Time to Job    &        $\checkmark$      &        $\checkmark$       &       $\checkmark$          &       $\checkmark$         \\
Additional Job Attributes    &            &        $\checkmark$       &               &       $\checkmark$         \\
\hline
Number of Respondents        &       1,040        &        1,040          &        1,040         &        1,040          \\
Number of Choice Scenarios       &       37,420         &       37,420         &       37,420         &       37,420         \\
\hline\hline
\multicolumn{5}{l}{\footnotesize Standard errors in parentheses. \sym{*} \(p<0.10\), \sym{**} \(p<0.05\), \sym{***} \(p<0.01\).}\\
\end{tabular}
}
}
     \end{center} 
    \vspace{-.5em}
    {\footnotesize \textit{Notes:} This table shows parameters estimates from conditional logit choice models of hypothetical job choices estimated using data collected in a survey experiment featuring real job ads performed by \citet{BhullerAEA2024}. Using the text analysis approach in Section \ref{sec:advertised_attributes}, we construct indicators for the presence of each job attribute listed in Figure \ref{fig:share_attributes} for every job ad. Using the retrieved job attributes, we estimate choice models allowing choices to depend on the vector of non-pay job attributes $A_j=(A^1_{j},\dots,A^K_{j})$ and employer pay premiums (as in Section \ref{subsec:linked_ad_to_estab}) associated with each job ad. Columns (1)-(2) provide the estimated choice parameters, while Columns (3)-(4) provide the implied willingness-to-pay estimates $\gamma_k$, constructed for each job attribute $k$ as the ratio of the $k$-attribute choice parameter and the pay premium estimate. All models include controls for respondent fixed effects interacted with indicators for survey modules, and travel time between job location and respondent residence. Standard errors are clustered at the respondent level.
    }
\end{table}

\begin{figure}[htbp]
\caption{Survey Experiment: Choice Probability and Employer Quality.}\vspace{-1em} \label{fig:experiment_values}
\begin{center}
    \subfloat[][Pay Premium]{\includegraphics[width=.5\columnwidth]{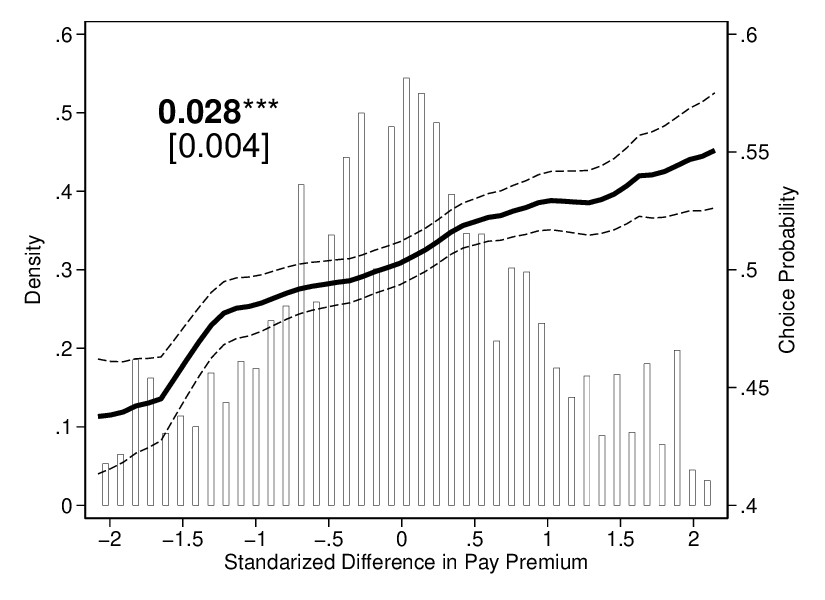}}  \\
    \subfloat[][Sorkin Overall Value]{\includegraphics[width=.5\columnwidth]{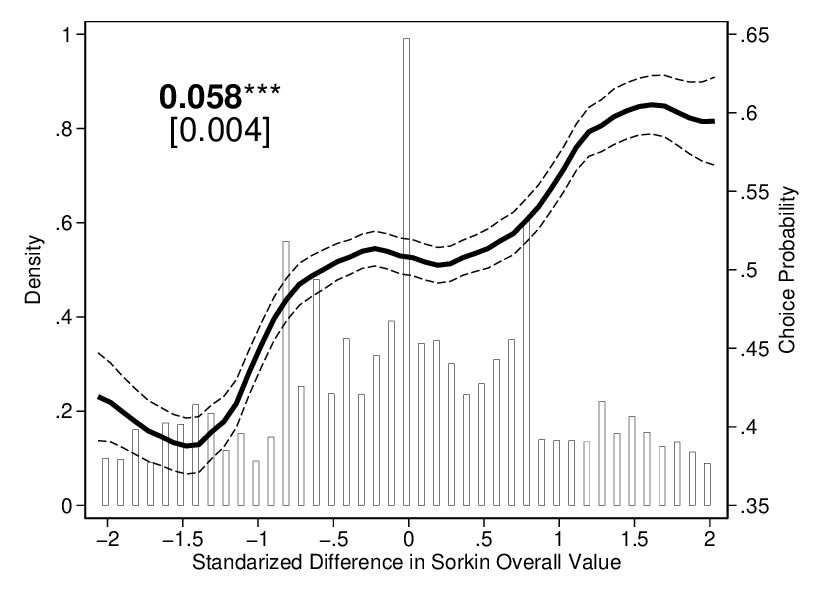}} 
    \subfloat[][Poaching Index]{\includegraphics[width=.5\columnwidth]{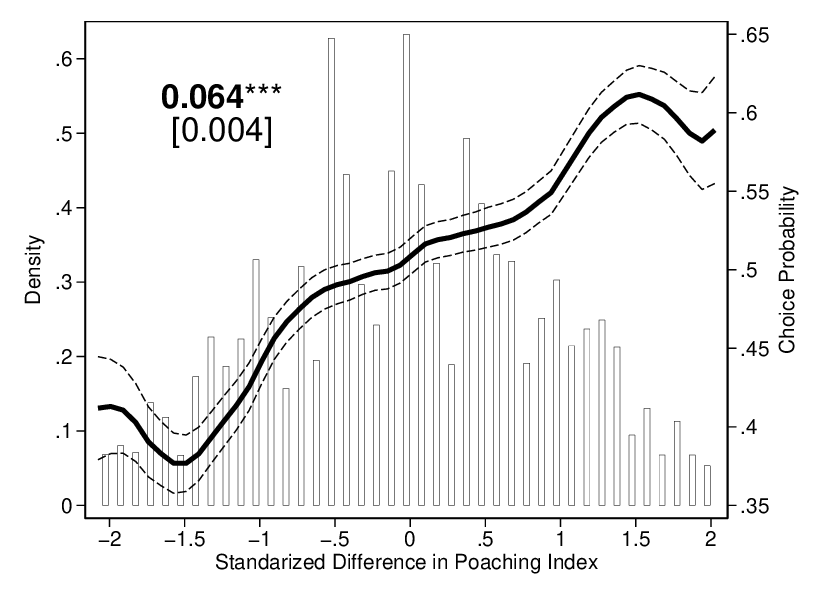}}    
\end{center}
{\footnotesize \textit{Notes}: This figure shows the relationships between the probability that survey respondents rank a job ad above the alternative option (right y-axis) and the standardized difference in employer quality across the job ad options (x-axis), as estimated using data on hypothetical choices of real job ads collected in a survey experiment performed by \citet{BhullerAEA2024}. We assign the alternative employer quality measures defined in Section \ref{subsec:linked_ad_to_estab} to each job ad, and provide the associations between Employer Pay Premium (panel a), Sorkin Overall Value (panel b) and Poaching Index (panel c), and respondents' subjective rankings of job ads based on their hypothetical choices across two randomly assigned ads in each scenario. The black lines show estimates from local polynomial regressions of choice probabilities on the standardized difference in employer quality, estimated separately for each quality measure, along with 95\% confidence intervals. Standard errors are clustered at the respondent level. In each panel, we also report the OLS regression estimates from linear probability models on each employer quality measure, providing the expected change in choice probability (reported in percentage points) associated with one standard deviation increase in the employer quality measure. The associated standard errors are reported in square brackets. We further show the distribution of standardized differences in employer quality (left y-axis), ranging between -2 and 2 in each panel.}
\end{figure}

Table \ref{tab:experiment_wtp} shows that respondents place high value on amenities such as ``convenient hours'' (e.g., ``possibility to work flexible hours''), and require substantial compensation for disamenities such as a ``temporary job'' or ``inconvenient hours'' (e.g., ``shift work''). The estimates all have the expected signs and their magnitudes are in line with prior literature estimating WTPs from similar hypothetical choice experiments. As noted in \cite{andresen2026aea}, although the survey targeted university students, the WTPs for this group closely resemble estimates found in the literature for broader populations.\footnote{\cite{andresen2026aea} provide WTP estimates for a smaller set of workplace amenities that featured in both hypothetical choices over real ads and purely hypothetical jobs that were experimentally designed, along with a literature overview. The estimates in Table \ref{tab:experiment_wtp} are similar to those in \cite{andresen2026aea}, though they use a more flexible mixed logit model with repeated choices \citep{revelttrain1998, revelttrain1999}.} For instance, we estimate a WTP for remote work at 0.065 in Column (4), which is similar in magnitude to the estimates reported by \cite{mas2017valuing} (0.078) and \cite{maestas2023value} (0.041) for the US. Similarly, the estimates in \cite{datta2019} imply a WTP of -0.242 for temporary job in the UK, while \cite{landeghem} find a WTP of 0.192 for flexible hours in Belgium, very close to the WTPs in Table \ref{tab:experiment_wtp}, Column (4). For the estimates that are specific to our setting, we find that the signs and magnitudes of our estimates are empirically plausible. For instance, jobs advertising good career opportunities have an estimated WTP of around 0.085, while working alongside qualified colleagues has a WTP of 0.094. We interpret this as evidence that respondents use advertised job attributes to guide their survey choices.

In the second step, we assess how well respondents' stated choices align with the revealed-preference measures of employer quality described in Section \ref{subsec:linked_ad_to_estab} by linking their binary choices $D_{is(j,j')}$ to these quality measures.\footnote{In recent related work, \cite{caldwell2025search} relate survey respondents' pay expectations across actual employers to pre-estimated AKM employer pay premiums, where the latter are estimated using data from the population at large \citep{abowd1999high}. We instead relate respondents' preferred choices across real-world job ads to alternative employer quality measures, including pay premiums. Besides these similarities, the survey designs depart significantly across the two studies. In particular, the survey in \cite{caldwell2025search} does not feature job ads, and is targeted at a broader population of workers employed in full-time jobs.} Specifically, for each employer quality measure, we compute the difference in employer quality between the two alternatives, $\Delta \hat{Q}_{is(j,j')}=\left\{\hat{Q}_{g(j)}-\hat{Q}_{g(j')}\right\}_{is(j,j')}$, using the notation in Section \ref{subsec:linked_ad_to_estab}. Finally, we estimate (local polynomial) regressions of $D_{is(j,j')}$ on the standardized quality differences $\Delta \hat{Q}_{is(j,j')}$.


The results are shown in Figure \ref{fig:experiment_values}. Panel (a) shows how respondents' choice probabilities relate to standardized differences in employer pay premiums. Respondents' choices are clearly positively aligned with differences in employer pay premiums: a one standard deviation difference is associated with about three percentage points higher choice probability. Notably, respondents were not provided explicit pay information in the job ad texts, so we interpret this as evidence that respondents value high-paying jobs and were able to infer signals correlated with pay premiums from the ad texts. In Panels (b)-(c), we consider two alternative measures of employer quality, the Sorkin overall value and the poaching index, respectively, and similarly show a clear alignment: a one standard deviation difference in each measure is associated with a six percentage points higher choice probability. We interpret this as evidence that respondents' stated choices are aligned with revealed-preference employer quality measures, suggesting that ad texts are informative of employer quality.\footnote{While the evidence provided above shows that survey respondents' stated choices in a controlled setting align with population-wide employer quality measures, we show in Appendix Table \ref{tab:validation_duration} that these measures are also clearly associated with the duration of posted vacancies. As described in Section \ref{sec:data}, our job ads data has information on the duration of posted vacancies, measured as the difference between the date when an ad was posted or registered and when it was either filled or removed. Recent evidence provided by \cite{mueller2024} and \cite{bassier2023}, respectively, shows that the employer pay premiums and the posted pay are negatively related to the duration of posted vacancies, consistent with the notion that high-pay employers are more attractive in the labor market. In Appendix Table \ref{tab:validation_duration}, we find an elasticity of vacancy duration with respect to the employer pay premium of about $-0.15$ (similar to \cite{mueller2024}), while vacancy durations are around 3-4\% shorter for ads with one standard deviation higher employer quality.}

\section{Job Attributes and Labor Market Inequality}
\label{sec:static_model}

Having documented that job ad attributes predict employer quality, we now quantify their contribution to employer-level inequality in the labor market. We incorporate the ad information into an otherwise standard static monopsony framework where workers value jobs at alternative employers differently \citep{card2018firms,lamadon2022imperfect}. We can then introduce a distinction between advertised amenities, derived from the ad information, and intrinsic amenities, which capture employer characteristics unobserved in job ads and valued by workers. Our goal is to separately estimate their contribution to firm-level inequality.

\subsection{A Monopsony Framework with Detailed Job Attributes}

Consider a static economy populated by a large number of workers indexed by $i$ and a large number of employers indexed by $j \in \{1, \dots, J \}$. Each employer $j$ posts a wage $w_j$ and a job attribute bundle $a_j$ that workers observe at no cost. We assume that worker $i$ with a job at employer $j$ derives indirect utility
\begin{equation}
    u_{ij} = \ln w_j + \ln a_j + \xi_{ij} = u_j + \xi_{ij},
    \label{eq:indirect_utility}    
\end{equation}
where $u_j = \ln w_j + \ln a_j$ denotes the indirect utility of working for employer $j$ common to all workers, and $\xi_{ij}$ is a worker-specific taste shock for employer $j$.

We impose standard restrictions on the idiosyncratic utility shifters $\{\xi_{ij}\}$ to obtain a closed-form expression for the labor supply curve specific to each employer $j$. The taste shocks $\{\xi_{ij}\}$ are assumed to be workers' private information, independent across ($i$,$j$) pairs, and distributed according to a Gumbel distribution with scale parameter $\sigma$. Under these assumptions, as is well-known, the probability that worker $i$ chooses employer $j$ posting wage $w_j$ and amenity bundle $a_j$ is given by
\begin{equation}
    \mathrm{Pr}\big(j = {\arg\max}_{j' \in \{1,...,J\}} \{u_{ij'}\} \big) 
    = (w_j a_j)^\sigma U^{-1},
    \label{eq:choice_probability}    
\end{equation}
where we introduce the notation $U := \sum_{k = 1}^J (w_k a_k)^\sigma$. Letting $\bar{\ell}$ denote the workforce in the economy, we arrive at the labor supply curve faced by employer $j$,
\begin{align}
    \ell_j(w_j) = 
    \bar{\ell} a_j^\sigma U^{-1}  w_j^\sigma,
    \label{eq:labor_supply_curve}
\end{align}
where, in line with prior work, we assume that employers are atomistic and ignore the effect of adjusting its own wage rate on the index $U$. As a result, the parameter $\sigma$, which scales the worker-specific utility shifter, can be interpreted as the labor supply elasticity in Equation \eqref{eq:labor_supply_curve}, and we exploit this interpretation in the model estimation.

We incorporate the non-pay job attribute information retrieved from job ads by specifying the amenity bundle offered by employer $j$ as
\begin{align}
    \ln a_j = \gamma A_j + \ln \tilde{a}_j.
    \label{eq:indirect_amenity}
\end{align}
The utility derived from the non-pay amenity bundle $\ln a_j$ is made up of two components. The first component is a vector $A_j=(A^1_{j},\dots,A^K_{j})$ of advertised non-pay attributes whose elements directly map to the ad attributes described in Section \ref{sec:advertised_attributes}. More specifically, for some job attribute $k$ at employer $j$, the element $A^k_{j}$ of $A_j$ is defined as the share of job ads posted by that employer featuring attribute $k$, similarly to the definitions used in Section \ref{subsec:linked_ad_to_estab}. With this definition, the coefficient $\gamma_k$ can be interpreted as the WTP for job attribute $k$.\footnote{If all jobs listed by employer $j$ feature attribute $k$, then $A^k_{j} = 1$ and, since the indirect utility specified in Equation \eqref{eq:indirect_utility} is in log-wage units, $\gamma_k$ is approximately the percentage wage change required to make the worker indifferent between a job with and without the attribute all else equal.} We refer to $\gamma A_j$ as the \textit{advertised component of non-pay utility}.\footnote{To the extent employers provide additional signals about workplace amenities in job ads that are not captured by our text analysis or enter into $\gamma A_j$, we may interpret this component as providing a lower bound.}

The second component of the non-pay amenity bundle, $\ln \tilde{a}_j$, is left unrestricted. It captures unobserved employer characteristics that enter workers' utility but are explicitly not listed in job ads, such as perceptions about company brand (positive) or stressful working conditions (negative). The term $\ln \tilde{a}_j$ is therefore the net effect on workers' utility of these unobserved attributes, and we refer to it as the \textit{intrinsic component of non-pay utility}.

An attractive feature of the framework described by Equations \eqref{eq:indirect_utility}-\eqref{eq:indirect_amenity} is that it does not restrict the relationship between the three components of workers' indirect utility: $\ln w_j$, $\gamma A_j$, and $\ln \tilde{a}_j$. Notably, for a given value of the labor supply elasticity $\sigma$, the advertised and intrinsic components of non-pay utility can reflect either compensating or augmenting differentials, and we test for the nature of each component in our estimation.

\subsection{Estimation Procedure and Parameter Estimates}

We combine annual matched employer-employee data and job ad data from the datasets described in Section \ref{sec:data}. We use establishments as the empirical counterpart to employers in the model. We restrict the sample to larger establishments for which we can credibly estimate the employer-specific components. Specifically, we only retain employers with at least 30 stayers and 15 movers over the period, where stayers are workers observed at the same employer in the following period, and movers are workers observed at a different employer. As shown in Table \ref{tab:sample_selection}, these restrictions leave us with a sample of 12,973 establishments. Details on the composition of this sample can be found in Appendix Table \ref{tab:sample_characteristics_2021_2024}.

The model is estimated in four broad steps. First, we reduce the number of parameters to be estimated by clustering employers paying similar wages.\footnote{The clustering provides a surjective mapping $j \rightarrow g$, where $j$ denotes the employers in the estimation sample and $g$ the employer group to which $j$ belongs. The parameters we recover are therefore cluster-level estimates. As an example, we derive estimates for $\widehat{\ln \tilde{a}}_{g(j)}$. For brevity, we use the notation $\widehat{\ln \tilde{a}}_{j}$ for $\widehat{\ln \tilde{a}}_{g(j)}$.} Second, we use well-identified pre-estimated parameters for the labor supply elasticity ($\sigma$) and the WTP vector ($\gamma$). For the labor supply elasticity, we use the pass-through estimates for Norway reported in \citet[][Table 2]{bhuller2025wage} using firm-level demand shocks. Their central pass-through estimate implies a labor supply elasticity estimate $\widehat{\sigma}$ of 3.29.\footnote{This estimate falls within the range of elasticities typically found in the literature (see \cite{azar2024monopsony} and  \cite{Kline2025} for overviews), and is close to estimates from recent studies using credible designs. For instance, \cite{Dube2025} find firm-level labor supply elasticities ranging between 1.7 and 2.6, while \cite{KroftEtAl2025} report elasticities ranging from 3.5 to 4.1 and \cite{bassier2022} find an elasticity of 4.2.} For the WTP parameters, we use the estimates derived from the survey experiment featuring real job ads described in Section \ref{subsec:linked_choice_to_ad}. In particular, we use the WTP estimates $\widehat{\gamma}$ from Column (4) of Table \ref{tab:experiment_wtp}.\footnote{Although these parameters could in principle be estimated within the model, we use the well-identified estimates from Section \ref{subsec:linked_choice_to_ad} to reduce the number of parameters that we need to estimate numerically.} Since the survey experiment is restricted to full-time job ads, we construct a WTP estimate for jobs that offer part-time contracts based on recent evidence on employer-level hours constraints in \cite{dube2022power} and \cite{lachowska2025work}.\footnote{We assign a WTP of -0.0785 for jobs that offer part-time hours, which is one minus the average WTP estimate for jobs offering full-time hours reported in \cite{dube2022power} and \cite{lachowska2025work}.} Third, we estimate a standard two-sided wage regression with worker and employer fixed effects and set the pay premiums $\ln w_j$ to the estimated employer effects. Finally, we recover the remaining parameters by maximum likelihood using Equation \eqref{eq:choice_probability} conditional on the pre-estimated parameters $\widehat{\sigma}$, $\widehat{\gamma}$, and $\widehat{\ln w}_j$. Appendix \ref{app:monopsony_model} provides details on these steps.

Table \ref{tab:dispersion_utility} reports summary statistics on the dispersion of the estimated components of workers' indirect utility. Panel A analyzes the variance of these components, using $\mathrm{Var}(\widehat{\ln w}_j)$ as a benchmark. A key statistic of interest is the dispersion of employer utility $\widehat{u}_j$ relative to employer pay premiums $\widehat{\ln w}_j$, which we estimate at 2.2. This represents a significant departure from compensating differentials, which would imply a relative dispersion close to zero. Our estimate of $\mathrm{Var}(\widehat{u}_j)/\mathrm{Var}(\widehat{\ln w}_j)$ is close to \cite{taber2020estimation}, lower than \cite{hall2018wage}, and larger than \cite{maestas2023value}. As the prior work reporting similar statistics differs along many dimensions (e.g., data source, modeling assumptions), we are cautious about making direct comparisons, but our estimate reassuringly falls within the broad range reported in the literature. Details on this comparison, including the construction of the relevant dispersion statistics in each study, are provided in Appendix \ref{app:monopsony_model}.

The remaining components of workers' indirect utility in Table \ref{tab:dispersion_utility} are specific to our framework, in which we can distinguish between the listed component of amenities $\widehat{\gamma} A_j$ and the intrinsic component $\widehat{\ln \tilde{a}}_j$. Panel A shows that the relative dispersion of the advertised component, $\mathrm{Var}(\widehat{\gamma} A_j)/\mathrm{Var}(\widehat{\ln w}_j) = 0.704$, is nearly as large as pay dispersion. The relative dispersion of the intrinsic component is even larger, with $\mathrm{Var}(\widehat{\ln \tilde{a}}_j)$ exceeding both other components by a factor of three. Panel B shows how each amenity component co-moves with the pay component. The advertised component augments pay differentials $(\mathrm{Corr}(\widehat{\gamma} A_j, \widehat{\ln w}_j) > 0)$, while the intrinsic component acts as a compensating differential $(\mathrm{Corr}(\widehat{\ln \tilde{a}}_j, \widehat{\ln w}_j) < 0)$. Taken together, the total amenity component implies compensating differentials $(\mathrm{Corr}(\widehat{\gamma} A_j + \widehat{\ln \tilde{a}}_j, \widehat{\ln w}_j) < 0)$, as the intrinsic component quantitatively dominates the advertised component. However, since $\mathrm{Corr}(\widehat{\gamma} A_j + \widehat{\ln \tilde{a}}_j, \widehat{\ln w}_j) > -1$, we still find that our model estimates imply substantial utility dispersion across employers.

\begin{table}[htbp]
    \caption{Monopsony Framework: Dispersion of Utility Components.}
    \vspace{-0.5cm}
    \label{tab:dispersion_utility}
    \begin{center}
        \begin{tabular}{l c}
\toprule \midrule
Measure & Value \\
\midrule

Panel A. Ratio of Variance to Variance of Pay Premiums ($\ln w_j$) & \\
\quad Employer Utility Value ($u_j$) & 2.200 \\
\quad Advertised Amenity ($\gamma A_j$) & 0.704 \\
\quad Intrinsic Amenity ($\ln \tilde{a}_j$) & 3.075 \\
\quad Total Amenity ($\gamma A_j + \ln \tilde{a}_j$) & 1.926 \\
\addlinespace
Panel B. Correlations of Components of Utility ($u_j$) & \\
\quad $\mathrm{Corr}(\ln w_j, \gamma A_j)$ & 0.547 \\
\quad $\mathrm{Corr}(\ln w_j, \ln \tilde{a}_j)$ & -0.469 \\
\quad $\mathrm{Corr}(\gamma A_j, \ln \tilde{a}_j)$ & -0.630 \\
\quad $\mathrm{Corr}(\ln w_j, \gamma A_j + \ln \tilde{a}_j)$ & -0.262 \\
\bottomrule \midrule
\end{tabular}
    \end{center}
    \vspace{-0.25cm}    
    {\footnotesize \textit{Notes}: The table shows summary statistics of the three employer-level components of workers' indirect utility estimated based on the model described in Equations \eqref{eq:indirect_utility}-\eqref{eq:indirect_amenity}: pay premiums, advertised non-pay amenities, and intrinsic non-pay amenities. All employer-level statistics are weighted by employment size.}
\end{table}

These results suggest that isolating advertised amenities is helpful for understanding the employer-level drivers of labor market inequality. Quantitatively, advertised amenities are a substantial source of utility dispersion and tend to reinforce pay differences. However, they only partly account for workers' preferences over employers: the intrinsic component is even more dispersed relative to pay and tends to attenuate employer pay differences. As the advertised amenities $\widehat{\gamma} A_j$ and the intrinsic amenities $\widehat{\ln \tilde{a}}_j$ are negatively correlated, the variance of total amenities ($\widehat{\gamma} A_j + \widehat{\ln \tilde{a}}_j$) is lower than the sum of variances of its components.

More broadly, our framework helps bridge the gap between the empirical labor literature analyzing specific attributes \citep[e.g.,][]{maestas2023value}, analogous to $\gamma A_j$ and its components in isolation, and the structural labor literature that collapses amenities into a scalar index \citep[e.g.,][]{morchio2024gender}, analogous to $\ln a_j = \gamma A_j + \ln \tilde{a}_j$ in our framework.

\subsection{Contribution of Amenities to Labor Market Inequality}

To further characterize the employer-side determinants of labor market inequality, we consider counterfactual scenarios in which non-pay amenities are equalized across employers. We report results for three counterfactual scenarios, switching off in turn: advertised job attributes ($A_j = 0$), intrinsic job attributes ($\ln \tilde{a}_j = 0$), and both components simultaneously. In each scenario, we recompute workers' choices and simulate a counterfactual distribution of employment size for each employer from Equations \eqref{eq:indirect_utility}--\eqref{eq:indirect_amenity}, holding all other parameters fixed at their estimated values. These counterfactuals should therefore be interpreted as partial equilibrium responses to changes in the components of non-pay utility. For instance, the ``advertised amenities turned off'' counterfactual corresponds to a policy imposing that all jobs have the baseline attributes implied by $A_j = 0$, so that workers maximize $\ln w_j + \ln \tilde{a}_j + \xi_{ij}$ over the set of employers $\{1,\dots,J\}$.

\begin{table}[ht]
    \caption{Monopsony Framework: Counterfactual Analyses.}
    \label{tab:model_counterfactuals}
    \vspace{-0.5cm}
    \begin{center}
        \begin{tabular}{l c c c}
\toprule  \midrule
 & \multicolumn{3}{c}{Counterfactual Scenarios} \\
\cmidrule(lr){2-4}
 & Advertised & Intrinsic & All \\
 & Amenities & Amenities & Amenities \\
 & Turned Off & Turned Off & Turned Off \\
 & (1) & (2) & (3) \\ 
 \midrule

Panel A. Employer Pay Premium & & & \\
\quad Average Relative to Baseline & -0.023 & 0.038 & 0.015 \\
\quad Variance Relative to Baseline & 1.132 & 0.747 & 0.772 \\
\addlinespace
Panel B. Employer Utility Value & & & \\
\quad Average Relative to Baseline & 0.048 & -0.001 & -0.047 \\
\quad Variance Relative to Baseline & 1.150 & 1.077 & 0.351 \\
\quad Variance of Utility / Variance of Pay & 2.237 & 3.170 & 1.000 \\
\addlinespace
Panel C. Employer Size & & & \\
\quad Average Relative to Baseline & -0.035 & 0.031 & 0.147 \\
\quad Variance Relative to Baseline & 1.060 & 1.021 & 0.452 \\
\bottomrule  \midrule
\end{tabular}
    \end{center}
    \vspace{-0.25cm}        
    {\footnotesize \textit{Notes}: The table reports summary statistics for the three counterfactual scenarios: Column (1) shows the outcomes when the advertised amenities are turned off ($A_j = 0$), while Column (2) considers the scenario when the intrinsic amenities are turned off ($\ln \tilde{a}_j = 0$), and Column (3) considers the scenario when all amenities are turned off (both $A_j = 0$ and $\ln \tilde{a}_j = 0$). Average relative to baseline is defined as the difference between the counterfactual mean and the baseline mean. Variance relative to baseline is defined as the ratio of the counterfactual variance to the baseline variance. Employer size is measured in logs in Panel C. Kernel density estimates of the corresponding counterfactual distributions are provided in Appendix Figure \ref{fig:counterfactual_densities}.}
\end{table}

Table \ref{tab:model_counterfactuals} reports summary measures from the baseline and counterfactual distributions for three objects of interest: employer pay premiums, employer utility differentials, and employment size.\footnote{Kernel density estimates of the corresponding distributions are in Appendix Figure \ref{fig:counterfactual_densities}.} Panel A summarizes the distribution of employer pay premiums, weighted by employment size. Switching off advertised amenities decreases average pay by 2.3 log points, while switching off intrinsic amenities increases it by 3.8 log points. These results complement the correlations in Table \ref{tab:dispersion_utility} by quantifying, in terms of pay, the extent to which advertised and intrinsic amenities are, respectively, augmenting and compensating differentials on average. Intuitively, when advertised amenities are switched off, some workers are more likely to enter lower paying jobs as higher paying jobs are now less attractive (augmenting differentials are removed). Analogously, when intrinsic amenities are switched off, some workers are more likely to enter higher paying jobs as lower paying jobs are now less attractive (compensating differentials are removed). Equalizing advertised and intrinsic amenities also has opposite effects on the dispersion of employer pay premiums. The variance of pay increases by 13\% when all employers provide the same advertised amenities, and decreases by 25\% when intrinsic amenities are equalized. This result is consistent with the estimated joint distribution of the different utility components that enter in Equation \eqref{eq:indirect_utility}.\footnote{The kernel density estimates in Appendix Figure \ref{fig:counterfactual_densities} show that many employers offer a combination of low pay, low advertised amenities, and high intrinsic amenities, which accounts for the dispersion results.}

The remaining panels of Table \ref{tab:model_counterfactuals} confirm the contribution of non-pay amenities to utility and size dispersions across employers. Panel B shows that the variance of utility goes up in both scenarios, by 15\% and 7\%, respectively.\footnote{Appendix Figure \ref{fig:counterfactual_densities}, Panel (c), shows the underlying employment distribution in each scenario. In the ``advertised amenities off'' scenario, the results are mainly driven by the bottom of the utility distribution gaining employment from the middle, while the top remains unchanged. In the ``intrinsic amenities off'' scenario, employment shifts from the bottom to the middle of the distribution, while the top shrinks markedly.} Panel C shows that non-pay utility differentials are a key factor behind the distribution of employment size across employers: in the scenario where both components are equalized, the variance of log employer size decreases by more than half. This result cannot be attributed to a specific amenity component, as separately equalizing advertised and intrinsic amenities has comparatively smaller effects.

\section{Conclusion}
\label{sec:conclusion}

In this paper, we study the pay and non-pay content of job ads. Our first set of results follows from extracting a comprehensive set of pay and non-pay attributes from vacancy texts. Nearly every job ad mentions at least one such attribute, with an average of seven attributes per ad. By considering a broad range of advertised attributes, our analysis expands on prior work focusing on specific job attributes, such as pay information \citep{batra2023online}, remote work \citep{hansen2023}, or flexible work arrangements \citep{adams2023flexible}.
 
Our second set of results follows from linking these advertised attributes to the employer posting the ad. We use a revealed-preference approach to derive measures of employer quality and find that high-quality employers tend to advertise good attributes, such as more generous pensions, flexible working hours, and permanent and full-time jobs. Ad attributes also meaningfully predict employer quality, even controlling for detailed observable characteristics such as industry and occupation. In the survey experiment featuring real-work job ads, respondents are significantly more likely to state that they prefer job ads posted by higher-quality employers. Taken together, these results indicate that job ads provide reliable signals of employer quality for workers assessing their options in the labor market.

Our last set of results follows from introducing these advertised attributes in a standard monopsony model where workers value jobs at alternative employers differently. We distinguish between advertised and intrinsic job attributes, where intrinsic attributes are those not explicitly mentioned in the ad but valued by workers, and find that the advertised component of non-pay worker utility augments pay differentials, while the intrinsic component represents a compensating differential. Our analysis therefore connects the survey-based empirical literature focusing on specific non-pay attributes \citep[e.g.,][]{mas2017valuing,maestas2023value} with the structural labor literature that collapses amenities into a scalar index \citep[e.g.,][]{hall2018wage,morchio2024gender}.

Our approach based on extracting a comprehensive set of job attributes from vacancy texts points to several directions for future work. First, the pay and non-pay content of job postings can be expected to differ across labor markets. As one example, health insurance is largely absent from job ads in Norway, where healthcare is universal, but is likely to be an important job attribute wherever employer-provided health insurance is common, such as in France or in the US. Second, there might be a cyclical element to the content of job ads, with recruiters potentially advertising different attributes when they compete for fewer job seekers. Third, while prior experimental work has focused on the salary component of vacancies \citep{belot2022wage}, our results show that there is scope to vary a much broader range of job attributes and further unpack workers' search behavior.

Finally, our results have implications for the design of policies regulating the information provided in job ads. Several jurisdictions have passed laws mandating employers to disclose a salary range in their job openings. Our analysis suggests that there is scope to extend such transparency laws beyond pay to other job characteristics, such as working arrangements. Similarly, our results can be useful for the design of online job platforms by providing guidance on the attributes employers should be encouraged to include in their postings.

\bibliographystyle{abbrvnat}
\bibliography{paper}

\clearpage
\appendix

\begin{center}
    \section*{Appendix}
\end{center}

\setcounter{page}{1}
\global\long\def\thepage{[Appendix-\arabic{page}]}

\global\long\def\thetable{A.\arabic{table}}%
\setcounter{table}{0}
\global\long\def\thefigure{A.\arabic{figure}}%
\setcounter{figure}{0}
\global\long\def\theequation{A.\arabic{equation}}%
\setcounter{equation}{0}

\section{Additional Tables and Figures} 
\label{app:add_results}

\begin{table}[h!]
    \caption{Overview of Sample Composition.} \vspace{-1em}
    \label{tab:sample_characteristics_2021_2024}
    \begin{center}
    \scalebox{.6}{\def\sym#1{\ifmmode^{#1}\else\(^{#1}\)\fi}
\begin{tabular}{lcccccc} \hline\midrule
 & & \multicolumn{3}{c}{\textbf{Main Analysis}} & \multicolumn{2}{c}{\textbf{Additional Analysis}} \\ \cmidrule(lr){3-5} \cmidrule(lr){6-7}
&\shortstack{All\\Observations}&\shortstack{Text\\Analysis\\Sample}&\shortstack{Employer\\Quality\\Sample}&\shortstack{Static\\Monopsony\\Sample}&\shortstack{Linked\\Ad-Hire\\Sample}&\shortstack{Survey\\Experiment\\Sample}  \\ 
& \multicolumn{1}{c}{(1)} & \multicolumn{1}{c}{(2)} & \multicolumn{1}{c}{(3)} & \multicolumn{1}{c}{(4)} & \multicolumn{1}{c}{(5)} & \multicolumn{1}{c}{(6)}  \\ 
\midrule 
\textbf{Worker Characteristics:} & & & & & & \\
\quad Age &42.0 &41.8 &39.9 &42.3 &42.2 &42.2 \\
\quad Immigrant &16.4\% &16.3\% &17.4\% &15.8\% &15.4\% &18.9\% \\
\quad Women &46.3\% &47.9\% &50.5\% &50.5\% &52.3\% &43.8\% \\
\quad College &49.7\% &51.2\% &56.0\% &60.8\% &55.4\% &66.8\% \\
\quad Hourly Wage &\$27.57 &\$27.77 &\$28.55 &\$30.46 &\$28.00 &\$31.32 \\
\quad Number of Workers &3,000,154 &2,792,164 &2,210,151 &1,688,064 &2,036,676 &96,529 \\
\quad Number of Worker-Years &12,220,469 &10,985,594 &8,253,567 &5,902,421 &6,999,923 &214,123 \\
\textbf{Establishment Characteristics:} & & & & & & \\
\quad Size &14.9 &26.4 &36.8 &119.4 &39.1 &137,9 \\
\quad Public Sector &13.4\% &22.5\% &27.1\% &54.5\% &29.2\% &15.5\% \\
\quad Industry Composition & & & & & & \\
\quad\quad Agriculture, Forestry and Fishing &2.9\% &1.6\% &1.2\% &0.8\% &0.8\% &0.0\% \\
\quad\quad Mining and Quarrying &0.4\% &0.4\% &0.5\% &1.5\% &0.3\% &0.0\% \\
\quad\quad Manufacturing &4.6\% &5.2\% &4.9\% &5.9\% &5.2\% &6.7\% \\
\quad\quad Energy Supply &0.4\% &0.5\% &0.4\% &0.6\% &0.5\% &1.2\% \\
\quad\quad Water Supply &0.5\% &0.7\% &0.7\% &0.6\% &0.6\% &0.4\% \\
\quad\quad Construction &12.9\% &8.6\% &7.8\% &4.7\% &6.5\% &5.0\% \\
\quad\quad Wholesale and Retail Trade &19.4\% &23.5\% &22.1\% &5.8\% &26.3\% &13.2\% \\
\quad\quad Transportation and Storage &4.6\% &3.9\% &4.1\% &3.6\% &3.0\% &4.3\% \\
\quad\quad Accommodation and Food Service &5.2\% &6.5\% &6.5\% &3.4\% &4.9\% &1.3\% \\
\quad\quad Information and Communication &4.3\% &3.4\% &3.8\% &5.5\% &2.5\% &7.0\% \\
\quad\quad Financial and Insurance &1.1\% &1.1\% &1.2\% &1.9\% &1.2\% &3.5\% \\
\quad\quad Real Estate &4.1\% &1.7\% &1.3\% &0.3\% &1.2\% &1.6\% \\
\quad\quad Professional, Scientific and Technical &10.2\% &6.6\% &6.3\% &5.7\% &5.1\% &14.1\% \\
\quad\quad Administrative and Support Service &4.5\% &4.6\% &3.4\% &3.0\% &3.0\% &14.2\% \\
\quad\quad Public Administration and Defense &2.0\% &3.4\% &3.3\% &7.6\% &3.9\% &6.3\% \\
\quad\quad Education &3.3\% &4.7\% &6.2\% &13.4\% &7.0\% &3.3\% \\
\quad\quad Health and Social Work &12.3\% &17.3\% &21.6\% &33.4\% &22.7\% &14.5\% \\
\quad\quad Arts, Entertainment and Recreation &2.7\% &2.5\% &2.3\% &1.3\% &1.9\% &0.7\% \\
\quad\quad Other Service Activities &4.5\% &3.7\% &2.7\% &1.1\% &3.3\% &2.6\% \\
\quad\quad Households as Employers &0.1\% &0.0\% &0.0\% &0.0\% &0.0\% &0.0\% \\
\quad Number of Establishments &264,084 &117,896 &70,046 &12,973 &47,877 &416 \\
\midrule\hline
\end{tabular}
} \vspace{-.5em}
     \end{center}
      {\footnotesize \textit{Notes:} This table documents the compositions of workers and employers between 2021 and 2024 across the different samples we use, as listed in Table \ref{tab:sample_selection}. ``Worker Characteristics'' (``Establishment Characteristics'') shows averages over worker-years (establishment-years), meaning that workers (employers) that are present in all years are implicitly weighted more than workers (employers) that are present in fewer years. Column (1) provides sample averages for the initial sample prior to the various sample selection steps, while the remaining columns provides sample averages for the other analysis samples we use, cf. Table \ref{tab:sample_selection}. Industry composition is based on the Classification of Economic Activities in the European Community (NACE).}
\end{table}
\FloatBarrier

\begin{table}[h!]
    \caption{Characteristics of Workers Linked to Ads and Survey Respondents.}\vspace{-1em}
    \label{tab:characteristics_linked_and_survey}
    \begin{center}
    \scalebox{.75}{\def\sym#1{\ifmmode^{#1}\else\(^{#1}\)\fi}
\begin{tabular}{lccc} \hline\midrule
\multicolumn{1}{c}{} & All Observations & Linked to Ads & Survey Respondents  \\ 
\multicolumn{1}{c}{} & \multicolumn{1}{c}{(1)} & \multicolumn{1}{c}{(2)} & \multicolumn{1}{c}{(3)}  \\ 
\midrule 
Age&42.1&34.3&21.8 \\
Immigrant&16.4\%&17.7\%&10.2\% \\
Women&46.3\%&59.3\%&69.6\% \\
College&49.7\%&55.1\%&46.7\% \\
Hourly Wage&\$27.57&\$23.36&\$18.54 \\
\midrule
Observations&3,000,154&110,932&1,040 \\
\midrule\hline
\end{tabular}
}
    \end{center}
    {\footnotesize \textit{Notes:} This table shows the average characteristics of (i) all workers in our initial sample, (ii) hired workers that were linked to a job ad in the linked ad-hire sample used in Section \ref{subsec:linked_ad_to_hire}, and (iii) survey respondents in the survey experiment sample used in Section \ref{subsec:linked_choice_to_ad}, respectively. In Column (1), we provide average worker characteristics for the initial sample prior to the various sample selection steps described in Table \ref{tab:sample_selection}, and the numbers reported here are identical to the worker-year averages reported in the first panel of Appendix Table \ref{tab:sample_characteristics_2021_2024}, Column (1). In Column (2), we report average worker characteristics recorded in administrative data records for the hired workers linked to job ads in the linked ad-hire sample, corresponding to Appendix Table \ref{tab:sample_characteristics_2021_2024}, Column (4). In Column (3), we report average self-reported background characteristics reported by respondents participating in the survey experiment, corresponding to Appendix Table \ref{tab:sample_characteristics_2021_2024}, Column (5).}
\end{table}
\FloatBarrier

\begin{table}[h!]
    \caption{Overview of Sample Selection: 2015-2024.}\vspace{-1em}
   \label{tab:sample_selection_2015_2024}
    \begin{center}
    \scalebox{.7}{\def\sym#1{\ifmmode^{#1}\else\(^{#1}\)\fi}
\begin{tabular}{lS[table-format=9.0, group-separator={,}, group-minimum-digits = 4]lS[table-format=9.0, group-separator={,}, group-minimum-digits = 4]lS[table-format=9.0, group-separator={,}, group-minimum-digits = 4]lS[table-format=9.0, group-separator={,}, group-minimum-digits = 4]l} \hline\midrule
&&&&& \multicolumn{4}{c}{Workers in Included Establishments} \\ \cmidrule(lr){6-9} 
\multicolumn{1}{c}{} & \multicolumn{2}{c}{Job Ads} & \multicolumn{2}{c}{Establishments} & \multicolumn{2}{c}{Workers} & \multicolumn{2}{c}{Worker-Years} \\\multicolumn{1}{c}{} & \multicolumn{2}{c}{(1)} & \multicolumn{2}{c}{(2)} & \multicolumn{2}{c}{(3)} & \multicolumn{2}{c}{(4)}  \\ 
\midrule 
\textbf{All Observations} & 2151994 & (100\%) & 359789 & (100\%) & 3479076 & (100\%) & 29789720 & (100\%) \\
\multicolumn{9}{l}{\textbf{Main Analysis}} \\
\quad Text Analysis Sample & 1998745 & (92.9\%) & 196231 & (54.5\%) & 3383123 & (97.2\%) & 28328064 & (95.1\%) \\
\quad Employer Quality Sample & 1660028 & (77.1\%) & 162346 & (45.1\%) & 2956702 & (85.0\%) & 23470708 & (78.8\%) \\
\midrule\hline
\end{tabular}
}
    \end{center} 
    \vspace{-.5em}
    {\footnotesize \textit{Notes:} 
    This table documents the sample sizes for different sample restrictions imposed for the 2015-2024 period. The first row shows the total number of observations in our data on job ads (Column 1) and matched employer-employee data (Columns 2-4), prior to the various sample selection steps. The ``Text Analysis Sample'' refers to the set of job ads with non-missing vacancy text and at least one section written in Norwegian. The ``Employer Quality Sample'' refers to the sample we retain by restricting the sample to workers aged 20 to 60 (inclusive) and the strongly connected set of employers used in the estimation of employer values, while removing establishments registered as recruitment or temporary employment agencies.} 
\end{table}

\begin{table}[h!] 
    \caption{Validation of Detected Pay and Non-Pay Job Attributes.}\vspace{-1em}
    \label{tab:validation_nonpay}
    \begin{center}
        \scalebox{.58}{\def\sym#1{\ifmmode^{#1}\else\(^{#1}\)\fi}

\begin{tabular}{lcccccc} \hline\midrule 
\multicolumn{1}{c}{} & \multicolumn{1}{c}{All Ads} & \multicolumn{5}{c}{Job Ads in the Validation Sample} \\ \cmidrule(lr){2-2} \cmidrule(lr){3-7} 
\shortstack{} & \shortstack{Text\\Analysis} & \shortstack{Text\\Analysis} & \shortstack{Manual\\Recognition} & \shortstack{Success\\Rate} & \shortstack{Precision\\Rate} & \shortstack{Sensitivity\\Rate} \\ 
 & (1) & (2) & (3) & (4) & (5) & (6) \\ 
\midrule 
Compensation Scheme&&&&&& \\  
\quad –– Competitive Pay&15.8&17.5&19.2&94.2&88.6&80.5 \\  
\quad –– Collective Agreement Pay&24.3&30.8&33.8&87.0&83.7&76.3 \\  
\quad –– Incentive Pay Scheme&2.7&2.0&1.5&98.5&50.0&66.7 \\  
\quad –– Hiring Bonus&0.2&0.0&0.0&100.0&–&– \\  
\quad –– Good Overtime Pay&0.4&0.2&0.8&99.5&100.0&33.3 \\  
\quad –– Any Other Mention of Pay&14.5&14.5&8.8&86.2&32.8&54.3 \\  
Financial Attributes&&&&&& \\  
\quad –– Pension Scheme&30.4&34.8&32.8&98.0&94.2&100.0 \\  
\quad –– Insurance Scheme&19.1&23.8&21.5&96.2&87.4&96.5 \\  
\quad –– Mortgage Possibility&1.4&2.0&3.8&98.2&100.0&53.3 \\  
Career Opportunities&&&&&& \\  
\quad –– Good Career Paths&12.6&16.8&6.2&85.0&23.9&64.0 \\  
\quad –– On-the-Job Training&41.6&46.8&39.2&78.0&68.4&81.5 \\  
Hours of Work&&&&&& \\  
\quad –– Full-time Contract&59.3&55.8&53.2&95.0&93.3&97.7 \\  
\quad –– Part-time Contract&30.9&34.0&28.2&91.2&78.7&94.7 \\  
\quad –– Full-time/Part-time Choice&4.1&3.0&3.0&94.0&0.0&0.0 \\  
Convenient Hours&&&&&& \\  
\quad –– Possibility to Work Flexible Hours&6.5&6.8&7.5&98.2&92.6&83.3 \\  
\quad –– Regular Daytime Work Schedule&14.0&8.8&8.8&97.0&82.9&82.9 \\  
\quad –– Exempt from Work Hour Regulations&0.5&1.2&0.2&98.5&0.0&0.0 \\  
Inconvenient Hours&&&&&& \\  
\quad –– Shift Work&15.4&15.0&11.8&95.8&75.0&95.7 \\  
\quad –– Weekend/Evening/Night Work&18.4&21.8&17.0&94.8&77.0&98.5 \\  
\quad –– On-call Employment&3.4&3.8&2.2&98.0&53.3&88.9 \\  
\quad –– Overtime Work Required&0.2&0.0&0.2&99.8&–&0.0 \\  
Contract Duration&&&&&& \\  
\quad –– Permanent Job&40.6&38.8&40.8&93.0&93.5&89.0 \\  
\quad –– Temporary Job&29.6&32.2&29.2&91.0&81.4&89.7 \\  
\quad –– Fixed-term Contract&0.4&0.8&0.8&100.0&100.0&100.0 \\  
Workplace Attributes&&&&&& \\  
\quad –– Social Environment&41.4&44.0&40.8&80.8&74.4&80.4 \\  
\quad –– Good Colleagues&21.7&26.0&28.2&77.2&60.6&55.8 \\  
\quad –– Possibility for Remote Work&0.4&0.5&0.5&100.0&100.0&100.0 \\  
\quad –– Shared Office Space&0.2&0.0&0.0&100.0&–&– \\  
\quad –– Inclusive Work-life Scheme&7.1&7.2&6.0&98.2&79.3&95.8 \\  
Task-Related Attributes&&&&&& \\  
\quad –– Interesting Tasks&34.5&38.2&22.2&82.5&56.2&96.6 \\  
\quad –– Challenging Tasks&28.6&32.0&19.0&85.0&56.2&94.7 \\  
\quad –– Variation in Tasks&10.9&10.2&15.8&87.5&65.9&42.9 \\  
\quad –– Responsibility in Job&1.6&1.8&10.2&90.5&71.4&12.2 \\  
\quad –– Independence in Performing Tasks&2.5&2.8&5.0&95.8&63.6&35.0 \\  
\quad –– Involves Leadership Responsibility&11.8&13.5&9.0&87.0&35.2&52.8 \\  
\quad –– Work Involves Travelling&8.2&10.2&3.5&93.2&34.1&100.0 \\  
Minor Perks&&&&&& \\  
\quad –– Beautiful Location&0.8&2.2&4.8&96.5&77.8&36.8 \\  
\quad –– Central Location&33.0&37.5&5.2&65.8&11.3&81.0 \\  
\quad –– Company Gym or Sports Team&3.7&5.0&6.5&95.5&70.0&53.8 \\  
\quad –– Parking Space On Premises&0.5&0.5&0.2&99.8&50.0&100.0 \\  
\quad –– Company Vehicle&1.2&1.5&1.0&99.5&66.7&100.0 \\  
\quad –– Any Welfare Scheme&4.2&4.8&2.5&97.2&47.4&90.0 \\  
\quad –– Company Cabin&1.9&2.0&3.2&98.8&100.0&61.5 \\  
\quad –– Occupational Health Service&1.4&1.2&1.2&100.0&100.0&100.0 \\  
\quad –– Company Canteen&0.3&0.8&0.8&99.0&33.3&33.3 \\  
\quad –– Flexible/Extended Holidays&0.5&0.0&0.0&100.0&–&– \\  
\midrule 
Number of Job Ads&860,467&400&400&400&–&– \\  
\midrule\hline 
\end{tabular}}
    \end{center}
      {\footnotesize \textit{Notes:} See details in the notes to Table \ref{tab:attribute_distribution}.}    
\end{table}
\FloatBarrier

\begin{table}[h!] 
    \caption{Association Between Employer Quality and Vacancy Duration.}\vspace{-1em}
    \label{tab:validation_duration}
    \begin{center}
        \scalebox{.9}{{
\def\sym#1{\ifmmode^{#1}\else\(^{#1}\)\fi}
\begin{tabular}{l*{2}{c}}
\hline\hline
                    &\multicolumn{2}{c}{Outcome: Log Vacancy Duration}                                                           \\
                    &\multicolumn{1}{c}{(1)}&\multicolumn{1}{c}{(2)}\\
\hline

Pay Premium         & -0.150\sym{***}    &  -0.164\sym{**}   \\
                    &     (0.036)        &  (0.055)          \\
[1em]
Overall Sorkin Value& -0.028\sym{***}    & -0.040\sym{***} \\
                    &  (0.004)           & (0.008) \\
[1em]
Poaching Index      &  -0.029\sym{***}   & -0.033\sym{***}  \\
                    &   (0.004)          & (0.007)     \\
[1em]
\hline
Occupation FE       &         &  \checkmark   \\
[1em]
Industry FE         &          &  \checkmark   \\
[1em]
Location FE        &           &  \checkmark  \\
\hline
Number of Job Ads                 &    1,587,634     &  1,587,634    \\
\hline\hline
\multicolumn{3}{l}{\footnotesize Standard errors in parentheses}\\
\multicolumn{3}{l}{\footnotesize \sym{*} \(p<0.10\), \sym{**} \(p<0.05\), \sym{***} \(p<0.01\)}\\
\end{tabular}
}
}
    \end{center}
      {\footnotesize \textit{Notes:} This table documents associations between employer quality measures and vacancy duration. Each column reports point estimates from three separate regressions of alternative revealed-preference employer quality measures defined in Section \ref{subsec:linked_ad_to_estab}, namely Employer Pay Premium, the Overall Sorkin Value, and the Poaching Index, on Log Vacancy Duration. Vacancy Duration is the number of days between the dates when the ad is posted and unlisted, right censored at 90 days to avoid outliers. For this analysis, we drop a small share of ads with zero or negative durations. Pay Premium is measured in logs, so the estimated coefficient on this measure can be interpreted as an elasticity, while Overall Sorkin Value and Poaching Index are scaled by their standard deviation, so the estimated coefficient for each of these represents a standard deviation increase in employer quality. Estimations are done at the ad level for the period between 2015 and 2024. In Column (2), we control for occupation and industry using two-digit occupation and two-digit industry fixed effects and location based on deciles of the number of workers in the municipality of the firm (Oslo as an own group), using separate categories for missing occupation, industry, and locations.}
\end{table}
\FloatBarrier

\begin{figure}[ht!]
    \caption{Event Study: Hires and Job Postings.}\vspace{-1em}
    \label{fig:hiring_pattern}
    \begin{center}
     \subfloat[][Hires All Occupations]{\includegraphics[width=.35\textwidth]{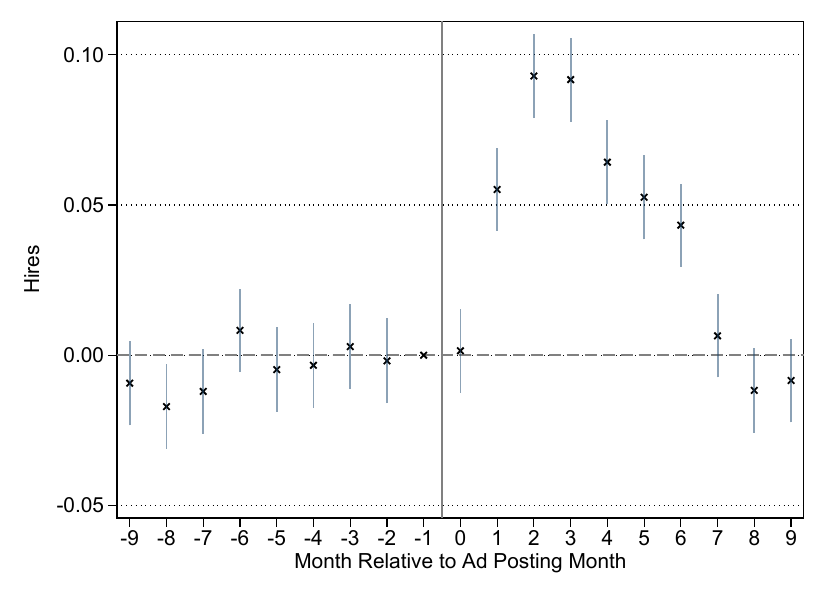}} \\
     \subfloat[][Within 1-digit Occupation]{\includegraphics[width=.35\textwidth]{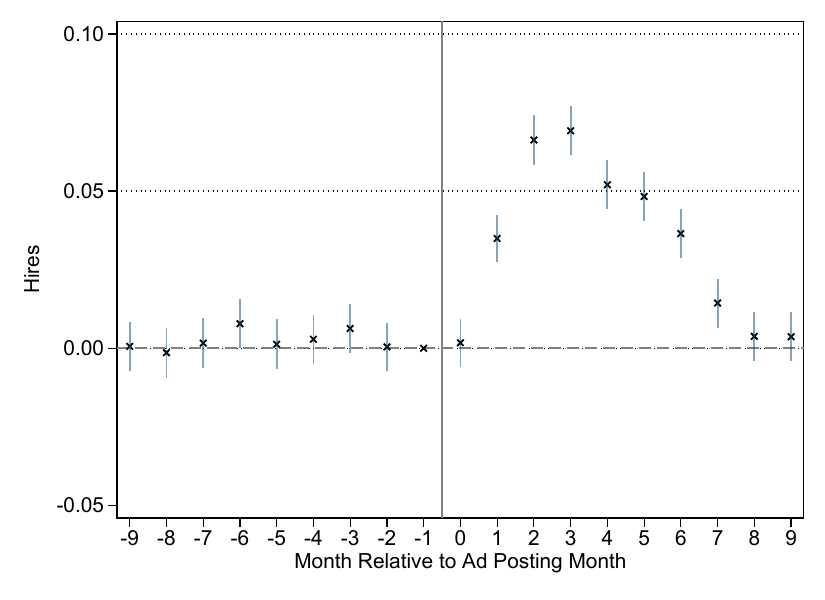}} 
     \subfloat[][Within 2-digit Occupation]{\includegraphics[width=.35\textwidth]{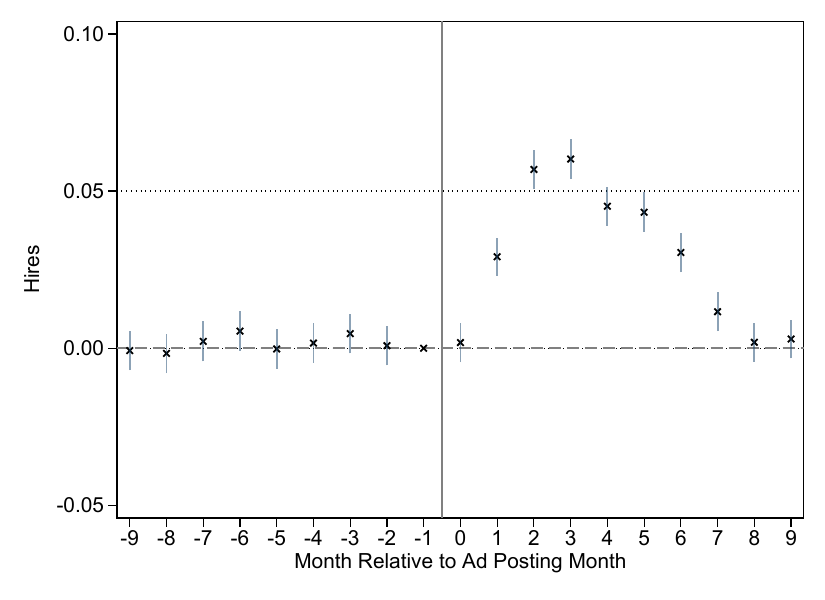}} \\
     \subfloat[][Within 3-digit Occupation]{\includegraphics[width=.35\textwidth]{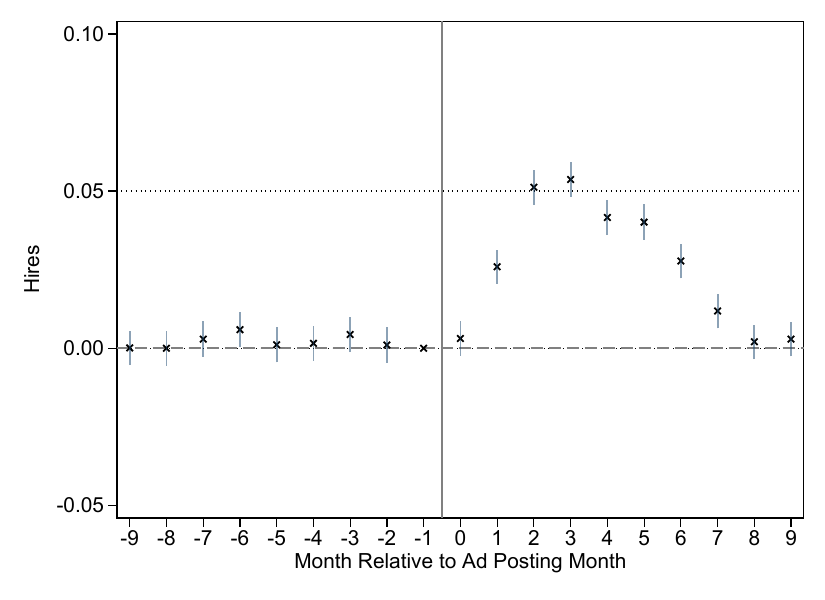}} 
     \subfloat[][Within 4-digit Occupation]{\includegraphics[width=.35\textwidth]{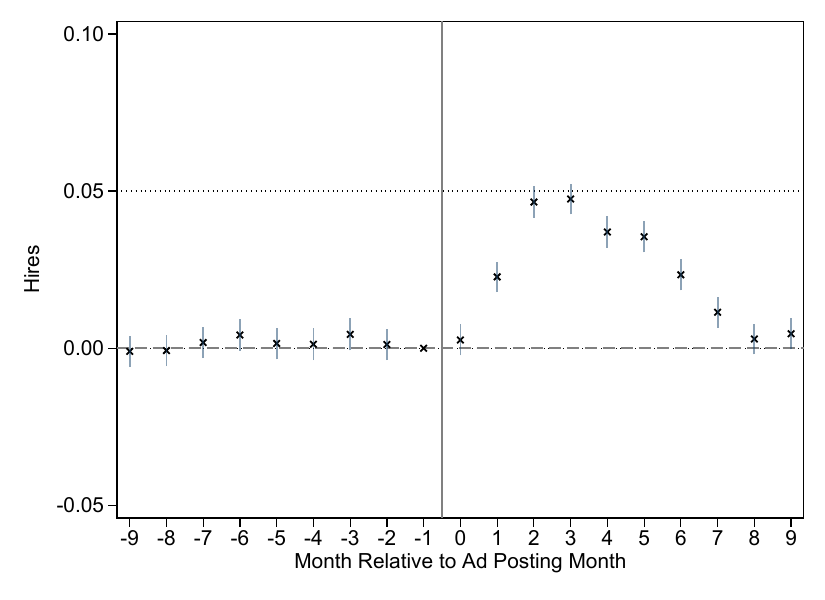}} \\
    \end{center} 
    {\footnotesize \textit{Notes}: This figure shows the number of hires around the month of job ad posting within establishments and occupations, constructed using subsets of the text analysis sample as listed in Table \ref{tab:sample_selection} from 2021 to 2024. We start by identifying establishment-occupation cells with posted ads and merge in the number of hired workers in surrounding months. The number of hired workers is constructed by counting workers with a reported starting date in each month at the establishment and occupation. We estimate an event study regression controlling for year–month fixed effects to account for seasonality in ad posting and hiring: $$\textnormal{hires}_{jom}=\sum_{k\ne-1} \beta_{k} 1(m-m_e=k)+\gamma_m + \varepsilon_{jom},$$where hires$_{jom}$ is the number of hires in month $m$ in establishment $j$ and occupation $o$, $k$ indicates month relative to the month of ad posting ($m_e$), and $\gamma_m$ are year-month fixed effects. The sum of point estimates to the right of the vertical lines relative to the left of the vertical lines implies an increase of about 0.43 hires at the establishment level (Panel (a)), and 0.22 hires at the establishment-occupation level (Panel (e)). }
\end{figure}


\begin{landscape}
\begin{figure}[h!]
\caption{Survey Experiment: Job Choice Scenario with Real Ads.}\vspace{-1em} \label{fig:experiment_scenario} \vspace{1em} 
    \centering
    \includegraphics[width=\linewidth]{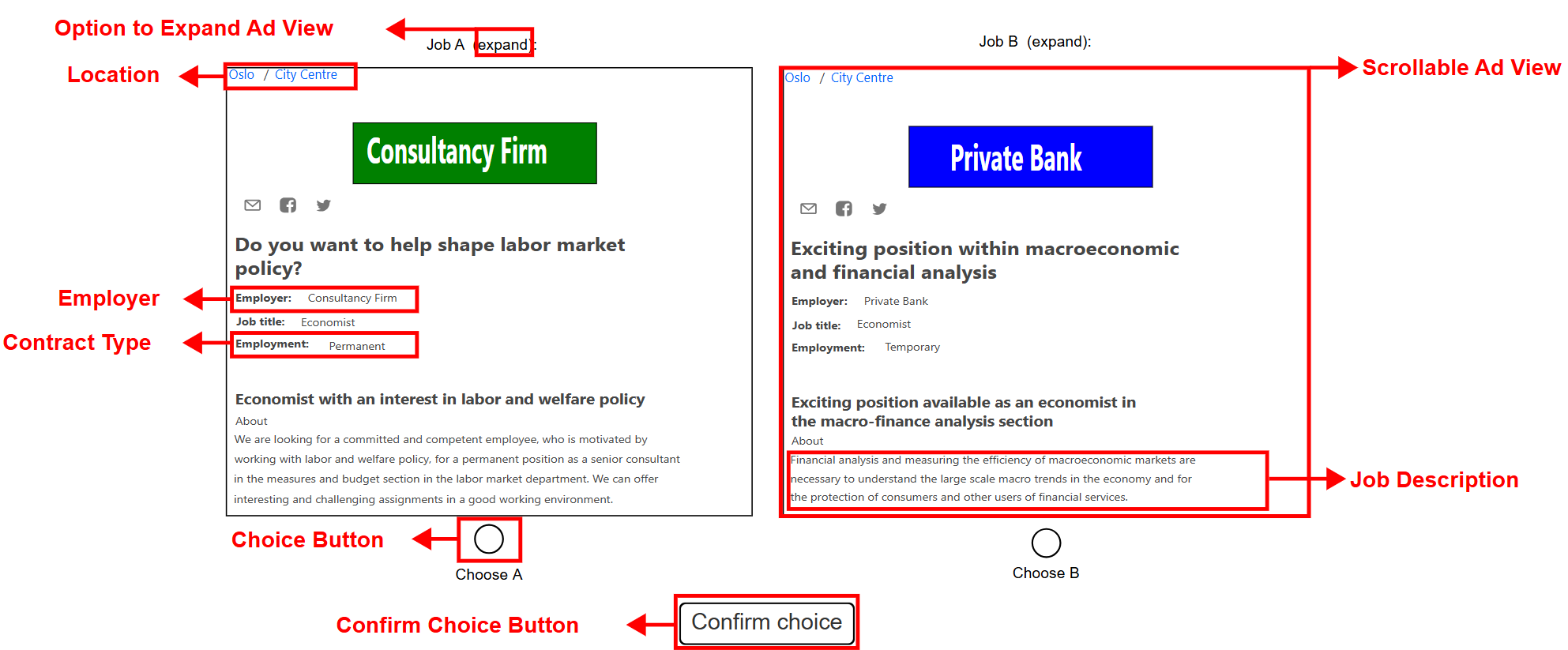}
{\footnotesize \textit{Notes}: This figure shows an example of a choice scenario with job ads that respondents faced in the survey experiment performed by \cite{BhullerAEA2024}. This is simplified version of the choice scenarios shown in \cite{andresen2026aea}, who introduce additional experimental features.}
\end{figure}
\end{landscape}


\begin{figure}[h!]
    \caption{Sample Vacancy Text: Teacher Substitute.} \vspace{-3em}
    \label{fig:sample_ads_teacher}
    \begin{center}
        \include{figures/sample_ads/teacher_substitute.tex}  \vspace{-2em}
    {\footnotesize \textit{Notes:} Translation from Norwegian by the authors.}
    \end{center}
    \vspace{-1em}
\end{figure}
\FloatBarrier

\begin{figure}[h!]
    \centering
    \caption{Sample Vacancy Text: Civil Engineer.} \vspace{-1em} 
    \label{fig:sample_ads_engineer}
    \include{figures/sample_ads/civil_engineer}  \vspace{-2em}
    {\footnotesize \textit{Notes:} Translation from Norwegian by the authors.}
\end{figure} 
\FloatBarrier

\begin{figure}[h!]
    \caption{The Prevalence of Job Attributes Advertised in Vacancy Texts Across Samples.}\vspace{-1em}
    \label{fig:balance_attributes}
    \begin{center}
        \includegraphics[width=.7\textwidth]{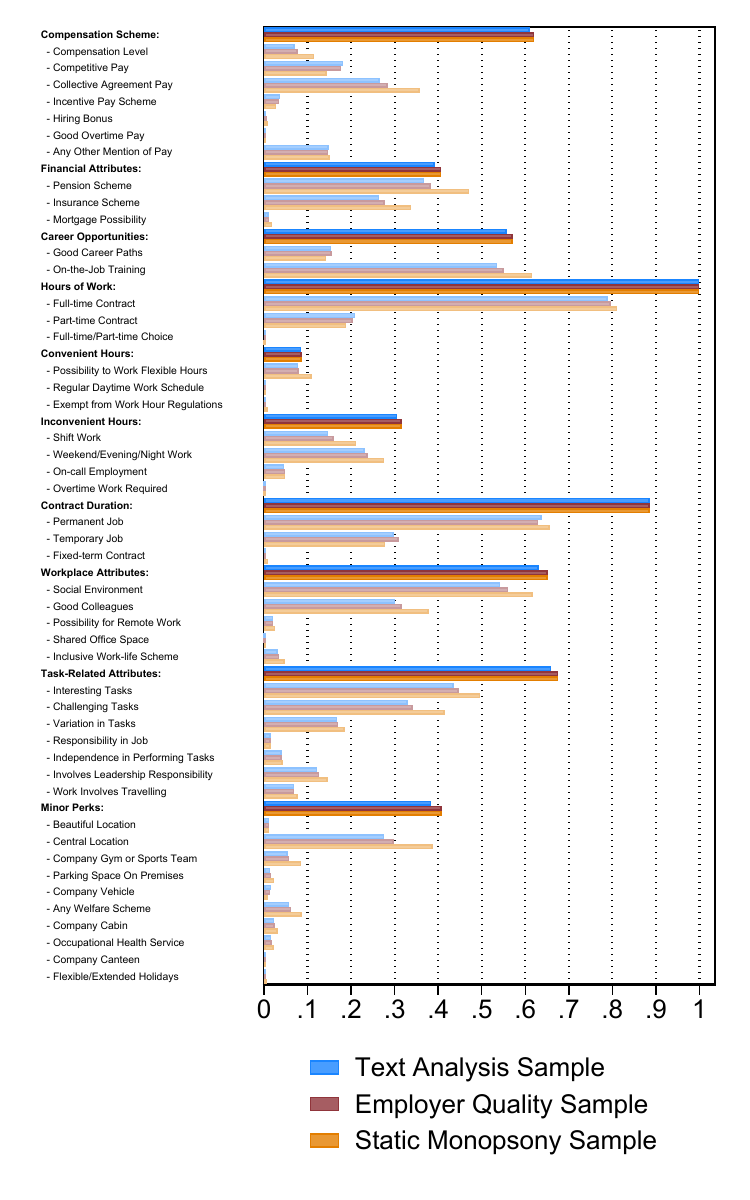}
    \end{center}
    \vspace{-1em}
    {\footnotesize \textit{Notes:} This figure documents the prevalence of job attributes detected in job ads posted in Norway between 2021 and 2024, separately for the three main analysis samples. The light bars show the share of ads detected with each distinct job attribute, while dark bars show the share of ads detected with at least one attribute within ten broad categories. Sample sizes are reported in Table \ref{tab:sample_selection}.} 
\end{figure}

\begin{figure}[h!]
    \caption{The Prevalence of Job Attributes Advertised in Vacancy Texts: 2015-2024.} \vspace{-1em} 
    \label{fig:share_attributes_2015_2024}
    \begin{center}
    \includegraphics[width=.7\textwidth]{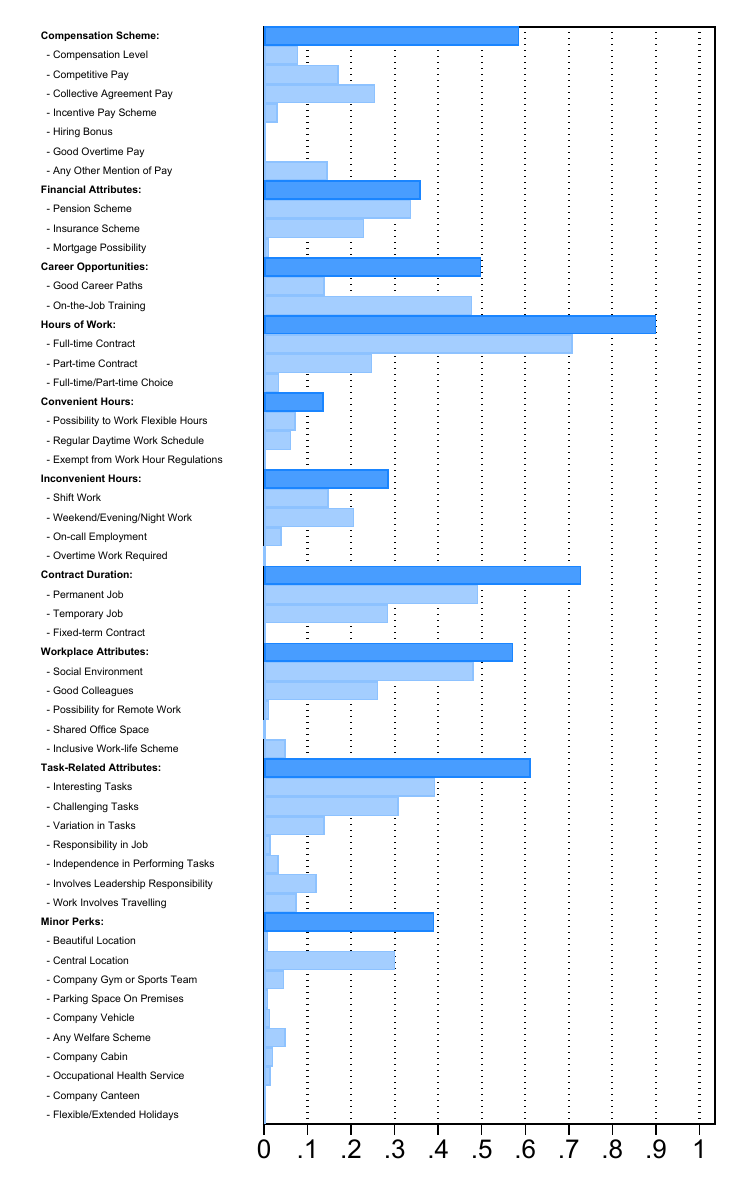}
    \end{center}
    \vspace{-1em}
    {\footnotesize \textit{Notes:} This figure documents the prevalence of job attributes detected in job ads posted in Norway between 2015 and 2024 [N=1,998,745]. The light blue bars show the share of ads detected having each of the distinct job attributes. The dark blue bars show the share of ads detected with at least one attribute within ten broad categories. Figure \ref{fig:share_attributes} shows the corresponding results for the 2021-2024 period.}
\end{figure}
\FloatBarrier

\begin{figure}
    \begin{center}
    \caption{Explained Variation in Publicly Advertised Job Attributes: Shapley Values.}
    \label{fig:explained_variation_shapley}
    \vspace{-0.5em}\includegraphics[width=.85\textwidth]{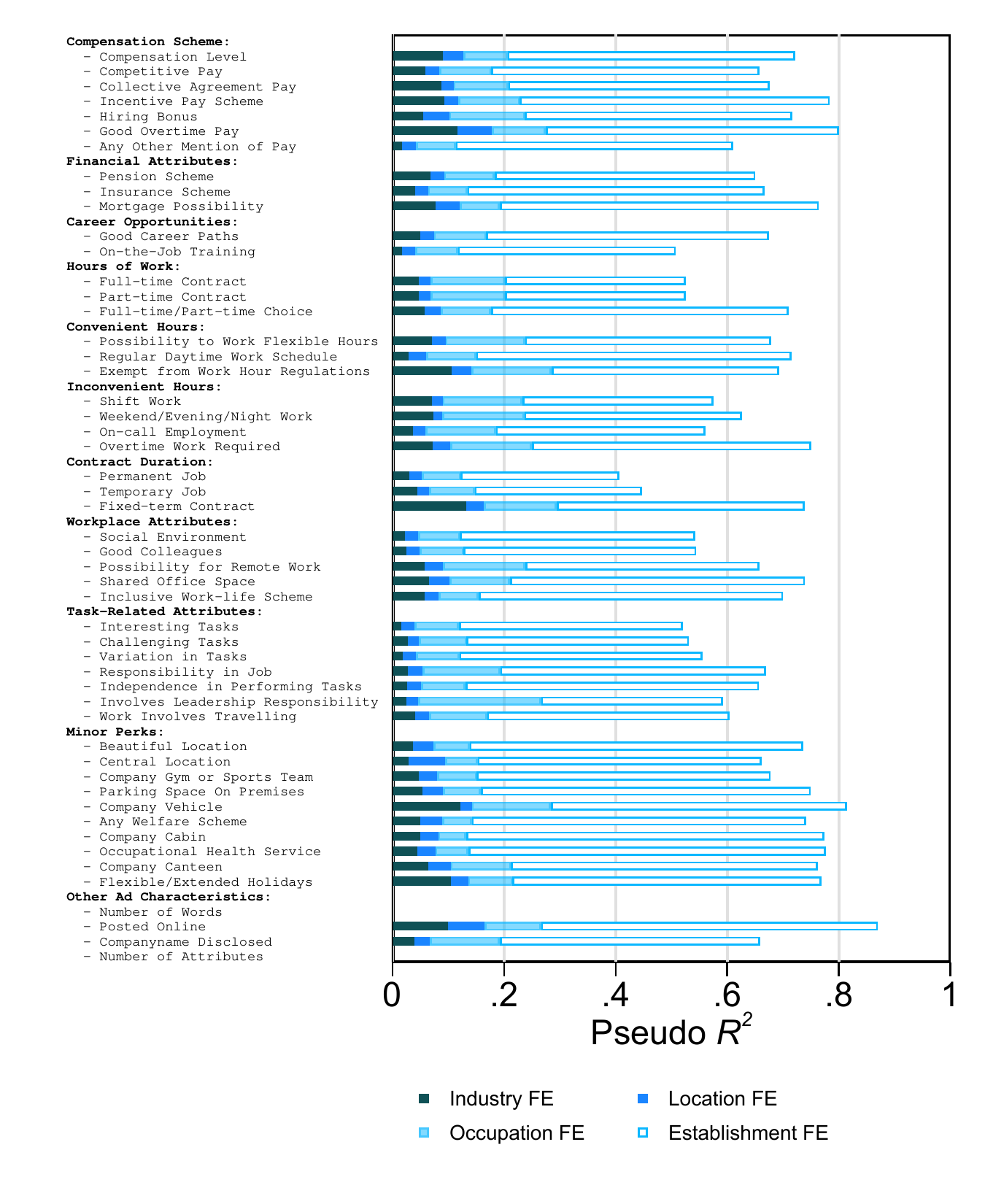}
    \end{center} 
    \vspace{-3em}
    {\footnotesize \textit{Notes:} This figure shows the Shapley Value Decomposition of the pseudo $R^2$ \citep{mcfadden1974} from logistic regressions of binary job attributes detected in job ads posted in Norway between 2021 and 2024 [N=748,870] on fixed effects denoting unique combinations of 2-digit industries (88 groups), location indicators (10 groups), 2-digit occupations (44 groups), and 57,656 establishments. Every regression considered by the Shapley Value Decomposition consists of a single categorical variable indicating unique combinations of included industries, location categories, occupations, and establishments. The fully interacted regression with all four categories controls for 133,964 unique combinations. The regression excludes 194,851 ads with a unique combination of establishment, occupation, location, and industry indicators, since these observations are perfectly predicted by the last set of regressions by construction. Location indicators group 422 municipalities into 10 groups based on the number of workers, so municipalities assigned to the same group have a similar number of workers, with a specific indicator for job postings in Oslo. Figure \ref{fig:explained_variation} shows the corresponding results for sequentially adding industry, location, occupation, and establishment FE.}
\end{figure}

\FloatBarrier
\begin{figure}[h!]
    \begin{center}
    \caption{Explained Variation in Publicly Advertised Job Attributes: 2015-2024.}
    \label{fig:explained_variation_pseudo_all}
    \includegraphics[width=.85\textwidth]{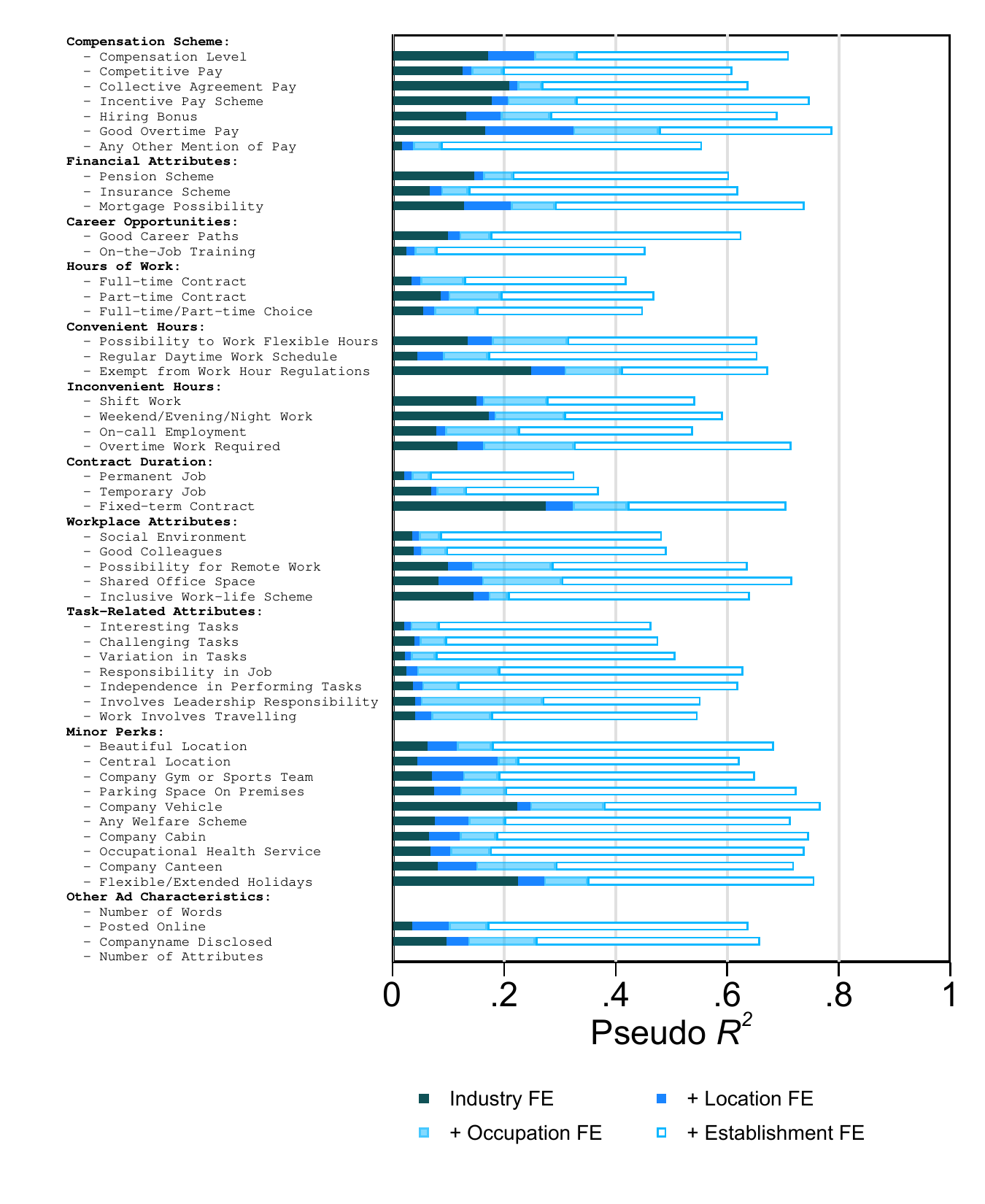}
    \end{center} \vspace{-1em}
    {\footnotesize \textit{Notes:} 
    This figure shows pseudo $R^2$ \citep{mcfadden1974} from separate logistic regressions of binary job attributes detected in job ads posted in Norway between 2015 and 2024 [N=1,677,211], on fixed effects denoting unique combinations of 2-digit industries (88 groups), location indicators (10 groups), 2-digit occupations (44 groups), and 88,887 establishments. We start by including industry fixed effects, continue by including industry$\times$location fixed effects, and so on. The regression excludes 321,534 ads with a unique combination of establishment, occupation, location, and industry indicators, as these observations are precisely predicted by the last set of regressions by construction. The last set of regressions controls for 260,241 unique combinations. Location indicators group 422 municipalities into ten groups based on the number of workers, such that municipalities assigned to the same group have a similar number of workers, with a specific indicator for job postings in Oslo. Figure \ref{fig:explained_variation} shows the corresponding results for the 2021-2024 period.}
\end{figure}
\FloatBarrier

\begin{figure}[htpb]
    \begin{center}
    \caption{Job Attributes and Employer Quality: Controlling for Pay Premiums.} \label{fig:point_estimates_paycontrols} \hspace{-2em}
    \subfloat[][Overall Sorkin Value]{\includegraphics[width=.5\textwidth]{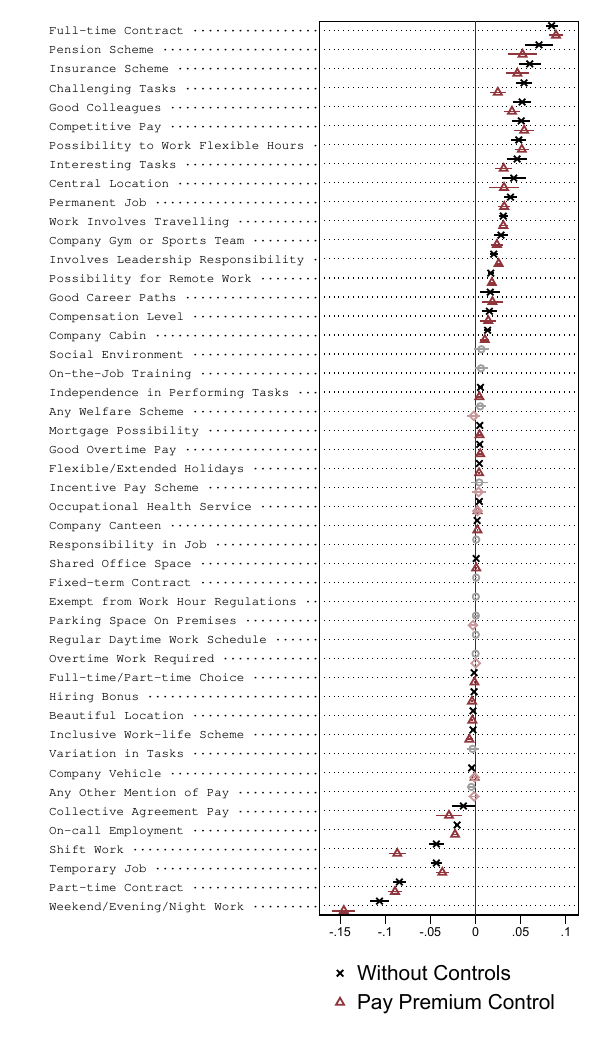}} 
    \subfloat[][Poaching Index]{\includegraphics[width=.5\textwidth]{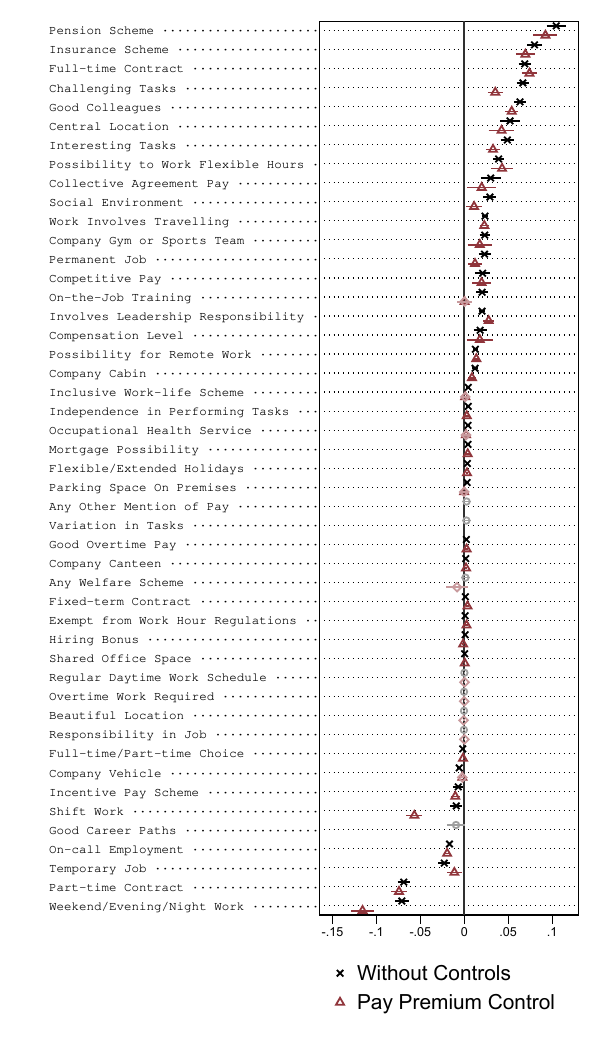}}\\
    \end{center}
    {\footnotesize \textit{Notes}: This figure shows parameter estimates and $90\%$ confidence intervals from separate regressions of the share of each ad attribute on Overall Sorkin Value and Poaching Index while controlling for Pay Premium, capturing the change in the fraction of ads with each job attribute associated with a residualized standard deviation increase in employer value. Estimations are done at the employer-cluster level for the period from 2021 to 2024, and are weighted by the number of worker-years.} 
\end{figure}

\begin{landscape}
\begin{figure}[ht!]
    \caption{Publicly Advertised Job Attributes By Employer Quality Measures: 2015-2024.} \label{fig:point_estimates_2015_2024_main} \hspace{-2em}
    \subfloat[][Pay Premium]{\includegraphics[width=.45\textwidth]{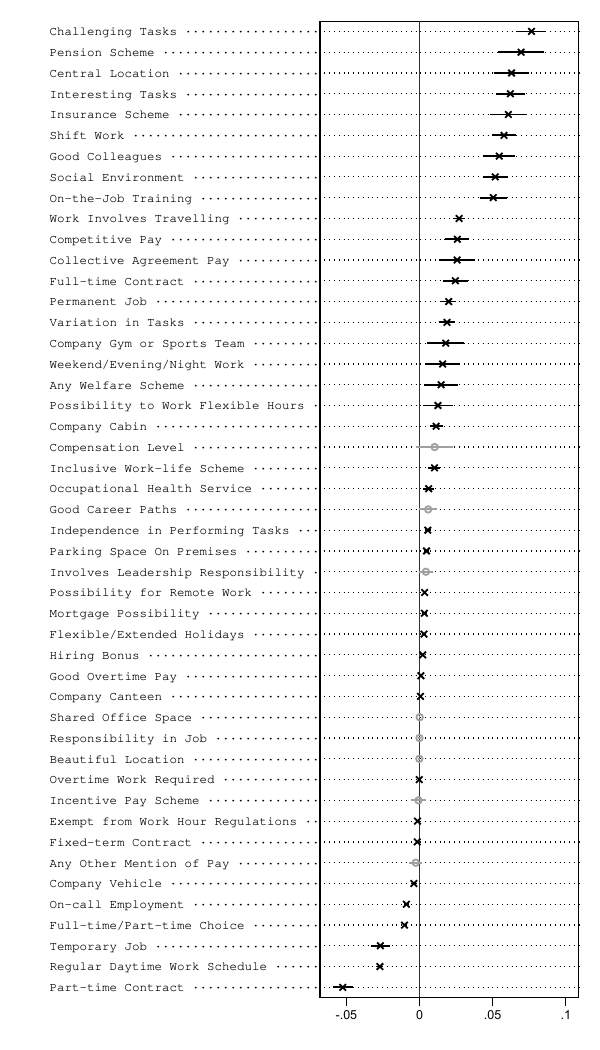}}
    \hfill
    \subfloat[][Overall Sorkin Value]{\includegraphics[width=.45\textwidth]{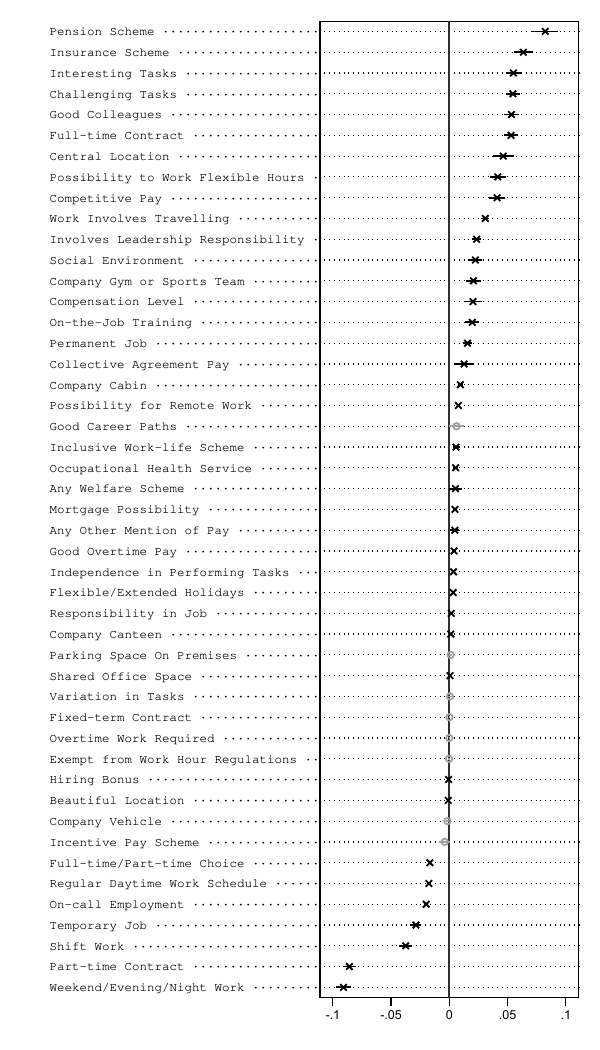}}
    \hfill
    \subfloat[][Poaching Index]{\includegraphics[width=.45\textwidth]{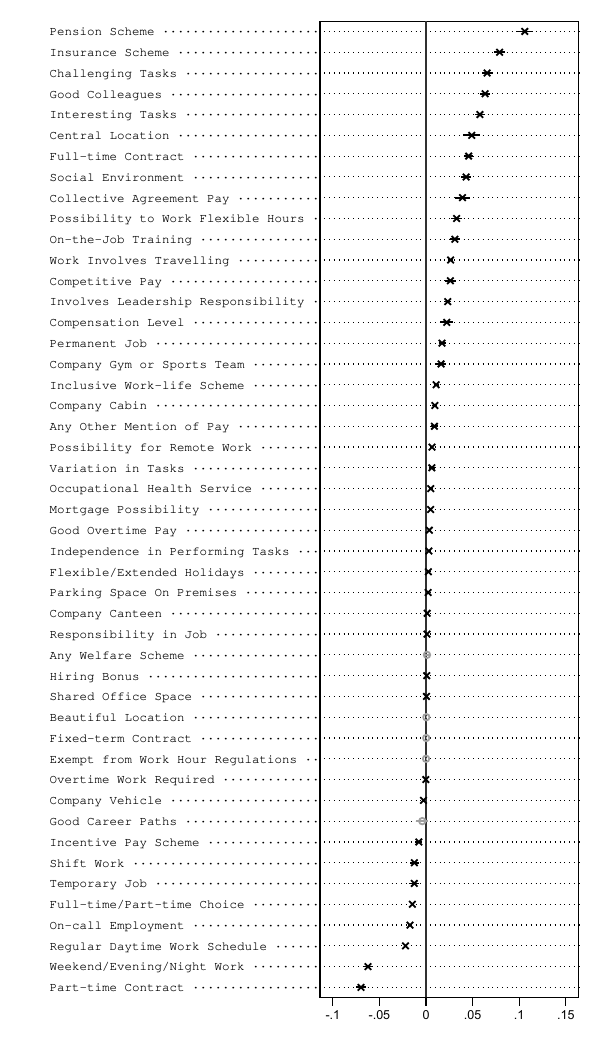}}\\
    {\footnotesize \textit{Notes}: This figure shows parameter estimates and $90\%$ confidence intervals from separate regressions of the share of each ad attribute on Employer Pay Premium, Employer Overall Sorkin Value, and Employer Poaching Index, capturing the change in the fraction of ads with each job attribute associated with a standard deviation increase in employer value. Estimations are done at the employer-cluster level for the period from 2015 to 2024, and are weighted by the number of worker-years.} 
\end{figure}
\end{landscape}


\begin{landscape}
\begin{figure}[ht!]
    \caption{Publicly Advertised Job Attributes By Employer Quality Measures: Smaller Clusters.} \label{fig:point_estimates_g25} \hspace{-2em}
    \hfill
    \subfloat[][Pay Premium]{\includegraphics[width=.45\textwidth]{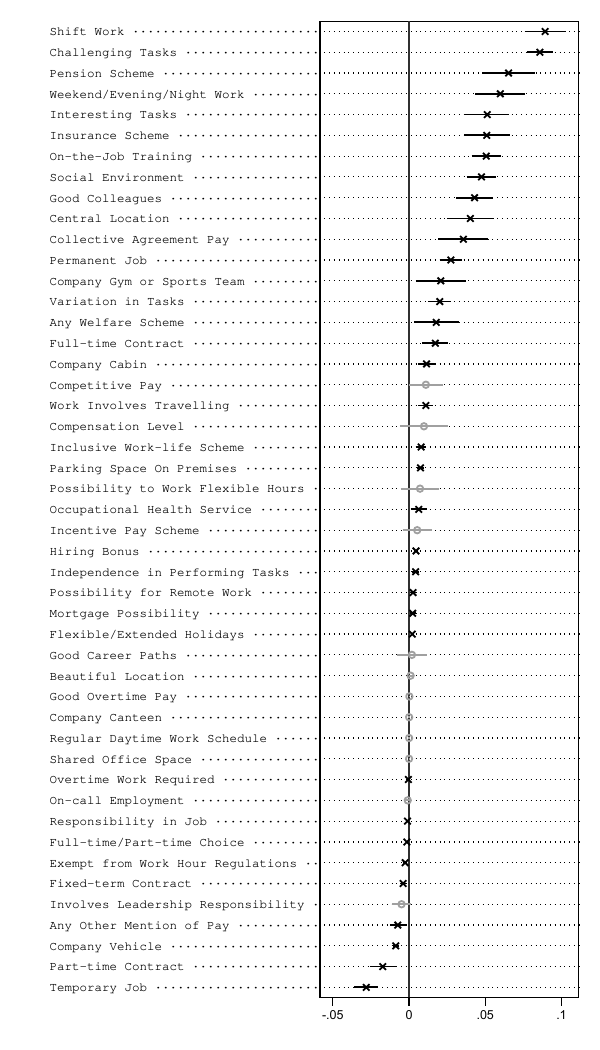}}
    \hfill
    \subfloat[][Overall Sorkin Value]{\includegraphics[width=.45\textwidth]{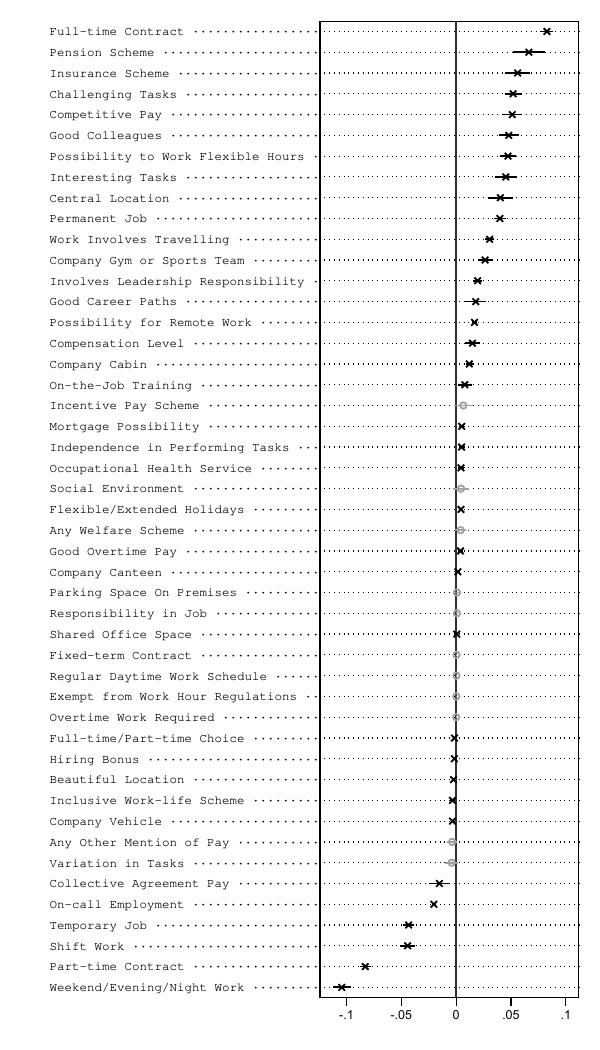}} 
    \hfill
    \subfloat[][Poaching Index]{\includegraphics[width=.45\textwidth]{figures/reg_attr_on_v/reg_attr_2021_2024_poaching_index_glevel_factor50_controlsreg_controls0.pdf}}\\
    {\footnotesize \textit{Notes}: This figure shows parameter estimates and $90\%$ confidence intervals from separate regressions of the share of each ad attribute on Employer Pay Premium, Employer Overall Sorkin Value, and Employer Poaching Index, capturing the change in the fraction of ads with each job attribute associated with a standard deviation increase in employer value. Estimations are done at the employer-cluster level for the period from 2021 to 2024, and are weighted by the number of worker-years. On average, each cluster has 25 establishments, i.e., $G=$round$(J/25)=2,843$ unique employer clusters, while Figure \ref{fig:point_estimates_main} shows the corresponding results in our baseline with $G=$round$(J/50)=1,422$ unique employer clusters.} 
\end{figure}
\end{landscape}


\begin{landscape}
\begin{figure}[ht!]
    \caption{Publicly Advertised Job Attributes By Employer Quality Measures: Larger Clusters.} \label{fig:point_estimates_g100} \hspace{-2em}
    \hfill
    \subfloat[][Pay Premium]{\includegraphics[width=.45\textwidth]{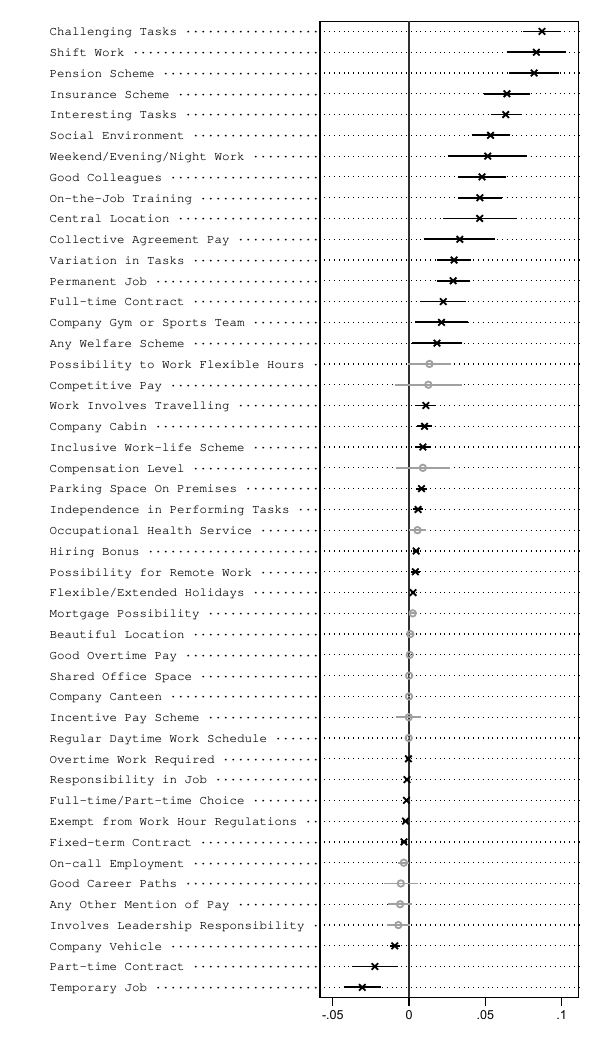}}
    \hfill
    \subfloat[][Overall Sorkin Value]{\includegraphics[width=.45\textwidth]{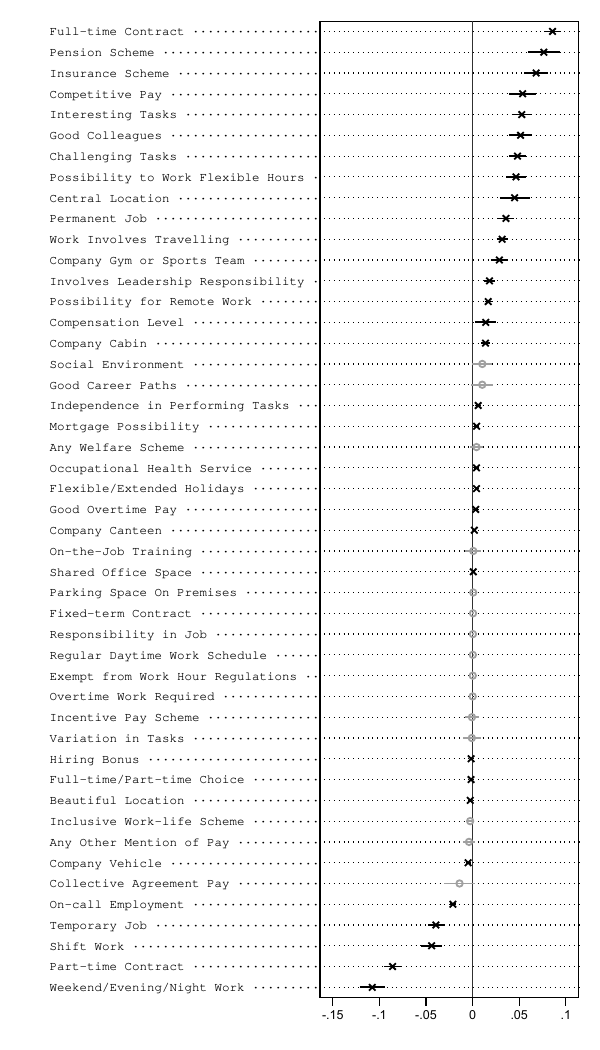}} 
    \hfill
    \subfloat[][Poaching Index]{\includegraphics[width=.45\textwidth]{figures/reg_attr_on_v/reg_attr_2021_2024_poaching_index_glevel_factor50_controlsreg_controls0.pdf}}\\
    {\footnotesize \textit{Notes}: This figure shows parameter estimates and $90\%$ confidence intervals from separate regressions of the share of each ad attribute on Employer Pay Premium, Employer Overall Sorkin Value, and Employer Poaching Index, capturing the change in the fraction of ads with each job attribute associated with a standard deviation increase in employer value. Estimations are done at the employer-cluster level for the period from 2021 to 2024, and are weighted by the number of worker-years. On average, each cluster has 100 establishments, i.e., $G=$round$(J/100)=711$ unique employer clusters, while Figure \ref{fig:point_estimates_main} shows the corresponding results in our baseline with $G=$round$(J/50)=1,422$ unique employer clusters.} 
\end{figure}
\end{landscape}


\begin{landscape}
\begin{figure}[ht!]
    \caption{Publicly Advertised Job Attributes By Sorkin Flow Value.} \label{fig:point_estimates_main_flow} \hspace{-2em}
    \subfloat[][Baseline]{\includegraphics[width=.45\textwidth]{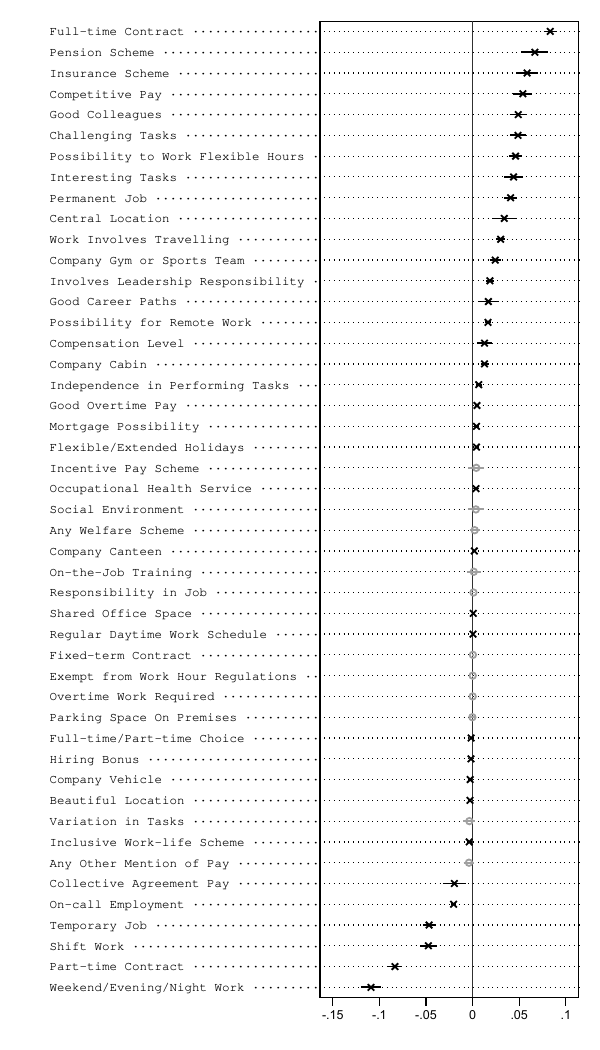}}
    \hfill
    \subfloat[][Smaller Clusters]{\includegraphics[width=.45\textwidth]{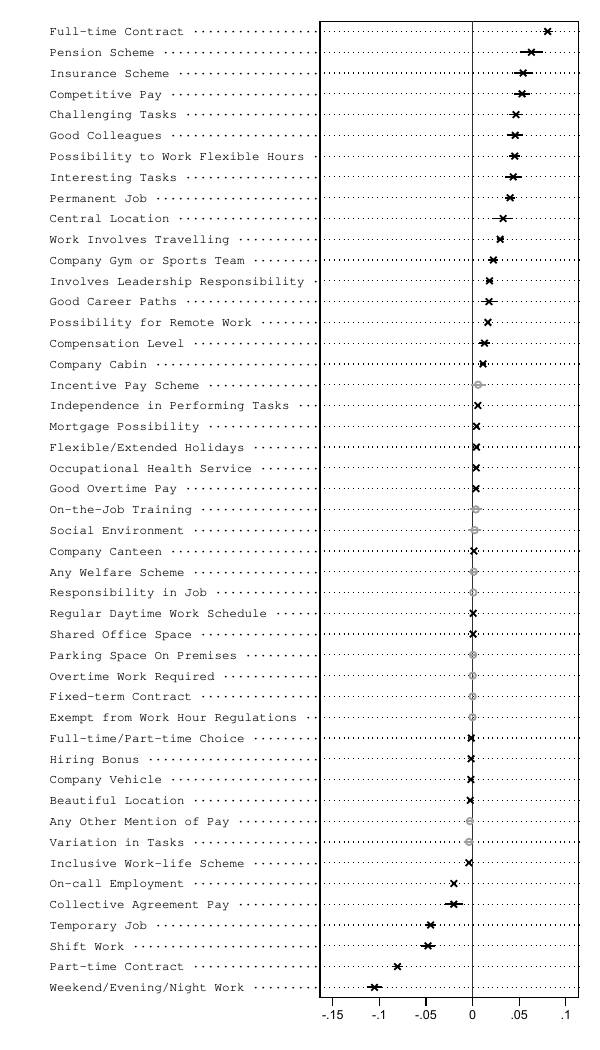}}    
    \hfill
    \subfloat[][Larger Clusters]{\includegraphics[width=.45\textwidth]{figures/reg_attr_on_v/reg_attr_2021_2024_u_glevel_factor25_controlsreg_controls0.pdf}}\\
    {\footnotesize \textit{Notes}: This figure shows parameter estimates and $90\%$ confidence intervals from separate regressions of the share of each ad attribute on Employer Sorkin Flow Value, capturing the change in the fraction of ads with each job attribute associated with a standard deviation increase in employer value. Estimations are done at the employer-cluster level for the period from 2021 to 2024, and are weighted by the number of worker-years. The baseline model in panel (a) features $G = \text{round}(J/50)=1,422$ unique employer clusters, while panel (b) shows results using smaller clusters ($G = \text{round}(J/25)=2,843$ unique employer clusters) and panel (c) shows results using larger clusters ($G = \text{round}(J/100)=711$ unique employer clusters).} 
\end{figure}
\end{landscape}

\begin{landscape}
\begin{figure}[ht!]
    \caption{Publicly Advertised Job Attributes By Employment Size.} \label{fig:point_estimates_main_size} \hspace{-2em}
    \subfloat[][Baseline]{\includegraphics[width=.45\textwidth]{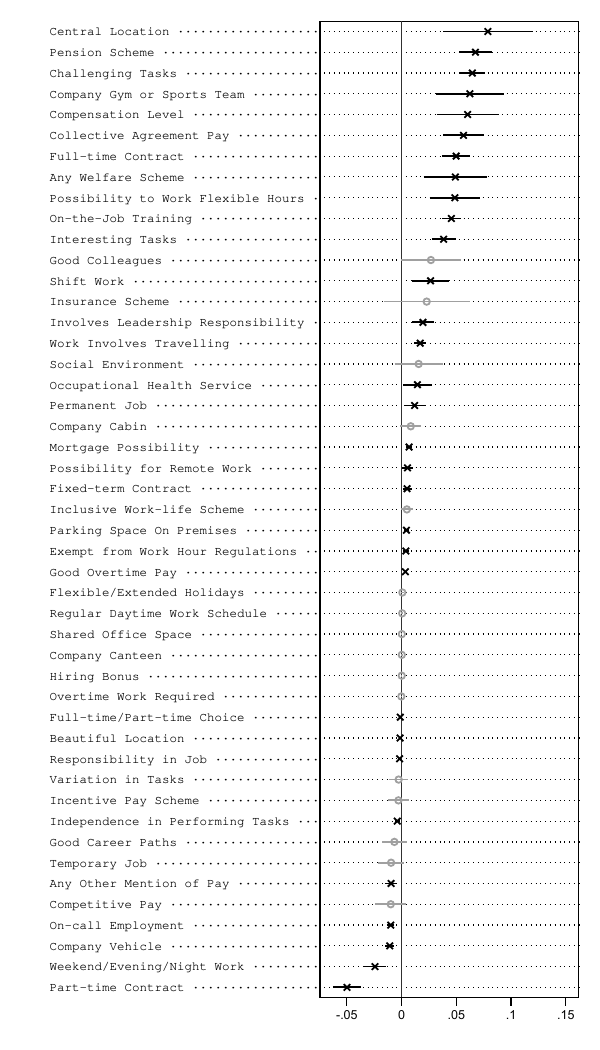}}
    \hfill
    \subfloat[][Smaller Clusters]{\includegraphics[width=.45\textwidth]{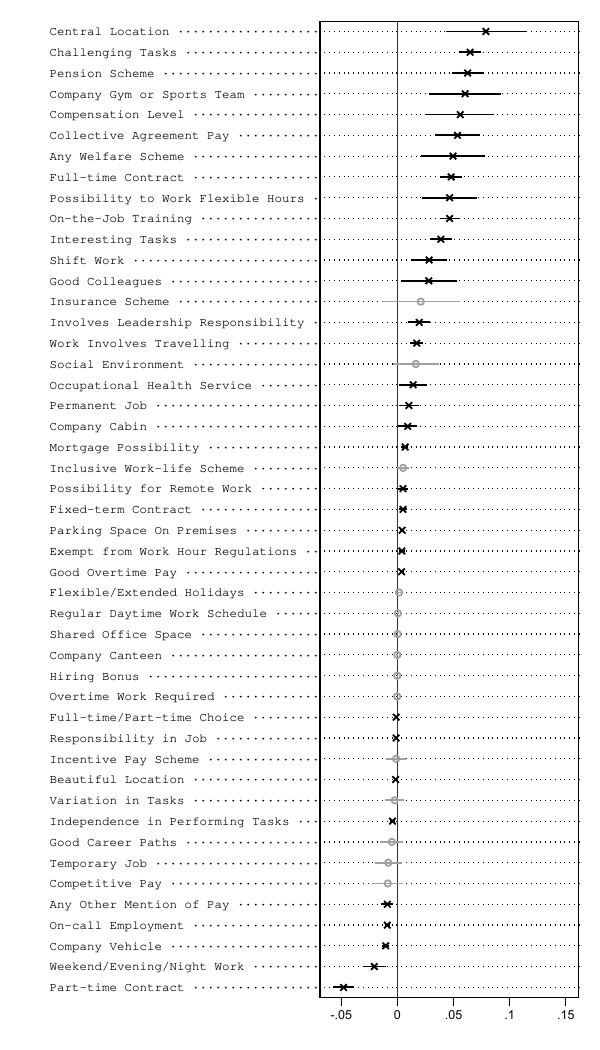}}
    \hfill    
    \subfloat[][Larger Clusters]{\includegraphics[width=.45\textwidth]{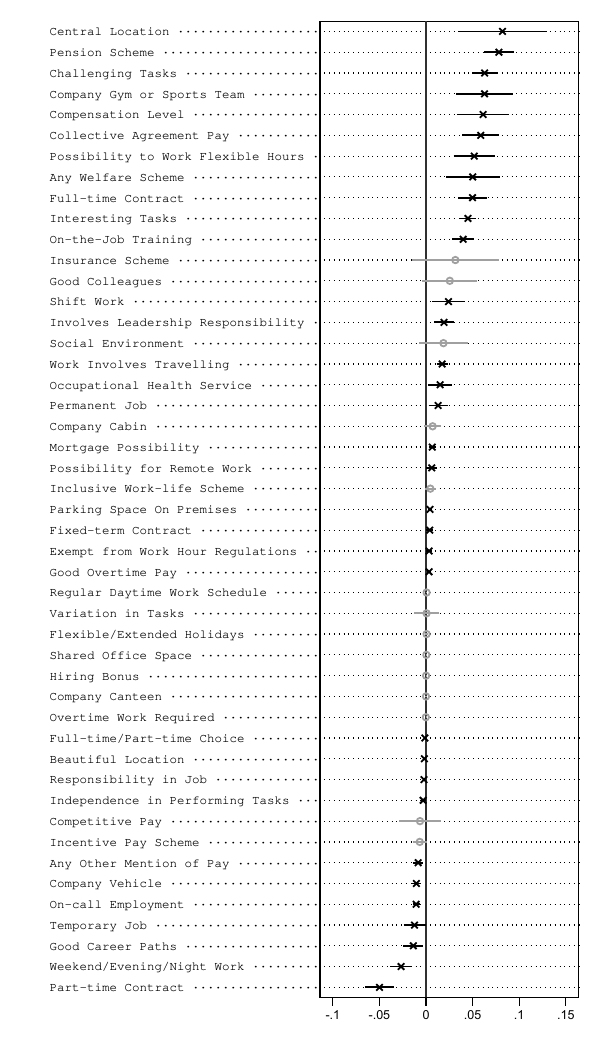}}\\
    {\footnotesize \textit{Notes}: This figure shows parameter estimates and $90\%$ confidence intervals from separate regressions of the share of each ad attribute on Employer Size (in logs), capturing the change in the fraction of ads with each job attribute associated with a standard deviation increase in size. Estimations are done at the employer-cluster level for the period from 2021 to 2024, and are weighted by the number of worker-years. The baseline model in panel (a) features $G = \text{round}(J/50)=1,422$ unique employer clusters, while panel (b) shows results using smaller clusters ($G = \text{round}(J/25)=2,843$ unique employer clusters) and panel (c) shows results using larger clusters ($G = \text{round}(J/100)=711$ unique employer clusters).} 
\end{figure}
\end{landscape}

\begin{figure}[p!]
    \caption{Predictive Power of Publicly Advertised Job Attributes: By Cluster Size.} \vspace{-1em}\label{fig:r2_different_cluster_sizes}
        \begin{center}
        \subfloat[][Baseline Model: $G=\text{round}(J/50)$]{\includegraphics[trim={0 1.2cm 0 0},clip, width=.5\columnwidth]{figures/shapley/shapley_2021_2024_glevel_factor50_groupall_controlsreg_main.pdf}} \\ \vspace{1em}
        \subfloat[][Smaller Clusters: $G=\text{round}(J/25)$]{\hspace{-2em}\includegraphics[trim={0 1.2cm 0 0},clip, width=.5\columnwidth]{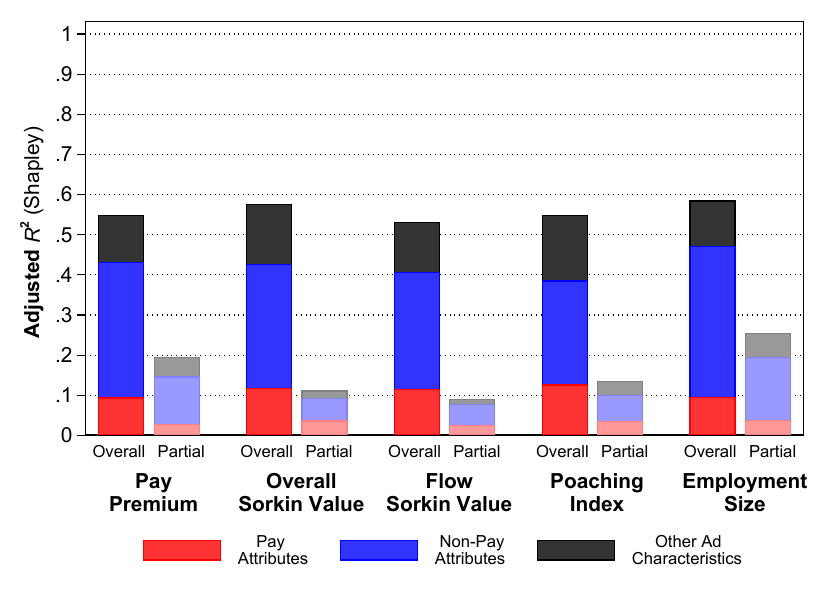}}
        \subfloat[][Larger Clusters: $G=\text{round}(J/100)$]{\includegraphics[trim={0 1.2cm 0 0},clip, width=.5\columnwidth]{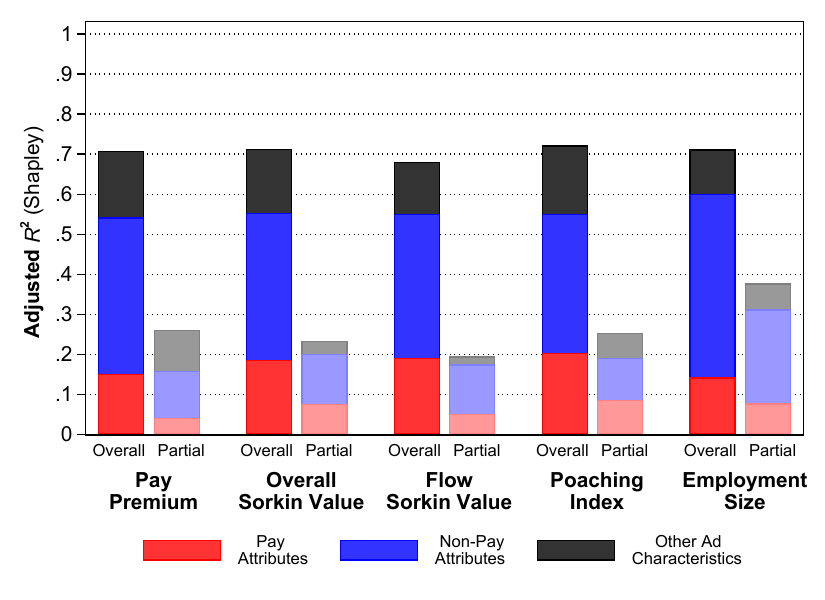}}  \\
        \vspace{-1em}
        \captionsetup[subfigure]{labelformat=empty} 
        \subfloat[][]{\includegraphics[trim={0 0 0 8.5cm},clip, width=.7\columnwidth]{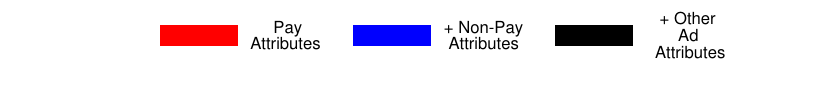}}  
        \end{center} 
     \vspace{-1em}         
    {\footnotesize \textit{Notes}: This figure documents adjusted $R^2$ from regressions of cluster-size specific estimated model parameters on ad attributes and characteristics in the text analysis as in Figure \ref{fig:reg_value_on_characteristics}. Estimations are done at the employer-cluster level for the 2021-2024 sample and are weighted by the total number of worker-years in each cluster and period. Overall $R^2$s are from regressions that include ad attributes and other ad characteristics, while partial $R^2$s are from regressions that also control for the composition of industry, occupation and location in each cluster. Industry and occupation controls are defined at the two-digit level. Location indicators split local municipalities into deciles, with municipalities in the same group having similar number of workers (Oslo as an own group). The baseline model in panel (a) features $G = \text{round}(J/50)=1,422$ unique employer clusters, while panel (b) shows results using smaller clusters ($G = \text{round}(J/25)=2,843$) and panel (c) shows results using larger clusters ($G = \text{round}(J/100)=711$).} 
\end{figure}

\begin{figure}[h!]
    \caption{Predictive Power of Publicly Advertised Job Attributes, 2015-2024.} \label{fig:r2_2015-2024}
    \begin{center}
        \subfloat[][Baseline Sample: 2021-2024]{\includegraphics[trim={0 1.2cm 0 0},clip, width=.5\columnwidth]{figures/shapley/shapley_2021_2024_glevel_factor50_groupall_controlsreg_main.pdf}} \vspace{1em}
        \subfloat[][Extended Sample: 2015-2024]{\includegraphics[trim={0 1.2cm 0 0},clip, width=.5\columnwidth]{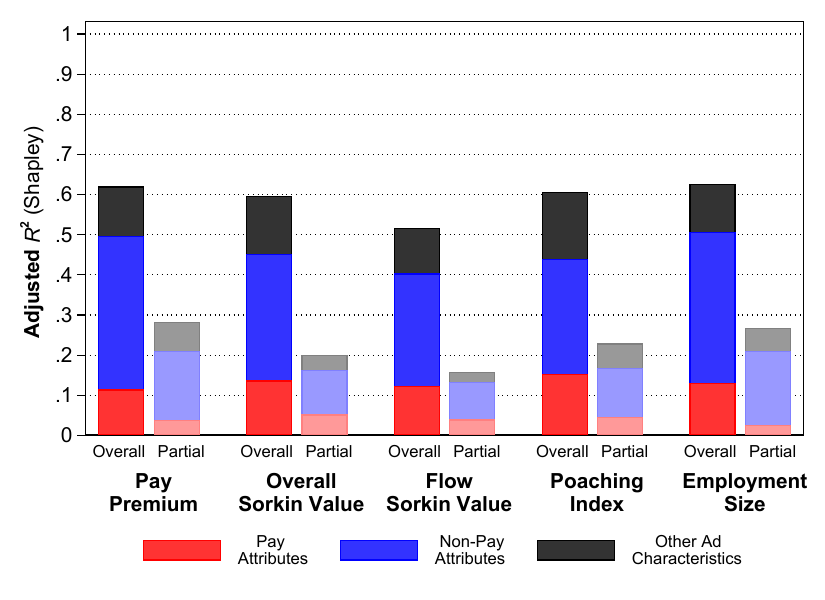}} \vspace{1em}
        \captionsetup[subfigure]{labelformat=empty} 
        \subfloat[][]{\includegraphics[trim={0 0 0 8.5cm},clip, width=.7\columnwidth]{figures/heterogeneity/r2_heterogeneity_educ_0_weights_by_group.pdf}}    
    \end{center} \vspace{-1em}
    {\footnotesize \textit{Notes}: This figure documents adjusted $R^2$ from regressions of estimated population group-specific model parameters on ad attributes and characteristics in the text analysis as in Figure \ref{fig:reg_value_on_characteristics}. Estimations are done at the employer-cluster level for the baseline sample period from 2021 to 2024 (Panel a) and from 2015 to 2024 (Panel b), and are weighted by the total number of worker-years in each cluster and period. Overall $R^2$s are from regressions that include ad attributes and other ad characteristics, while partial $R^2$s are from regressions that also control for the composition of industry, occupation and location in each cluster. Industry and occupation controls are defined at the two digit level. Location indicators split local municipalities into deciles, with municipalities in the same group having similar number of workers (Oslo as an own group).}
\end{figure}
\FloatBarrier

\begin{figure}[hp]
    \caption{Decomposing the Predictive Power of Publicly Advertised Job Attributes.} \vspace{-1em}\label{fig:decompose_r2_ag}
        \begin{center} 
        \subfloat[][Pay Premium]{\includegraphics[width=.35\textwidth]{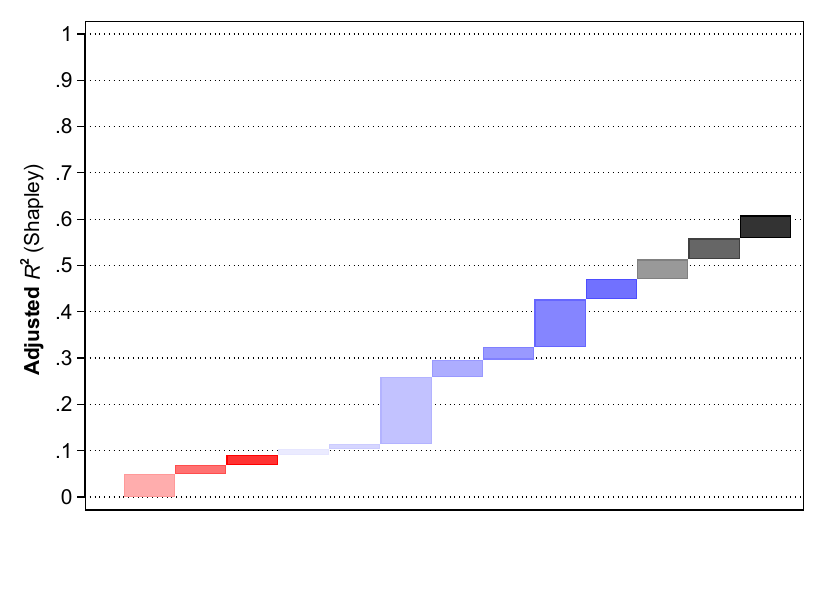}} \\
        \subfloat[][Overall Sorkin Value]{\includegraphics[width=.35\textwidth]{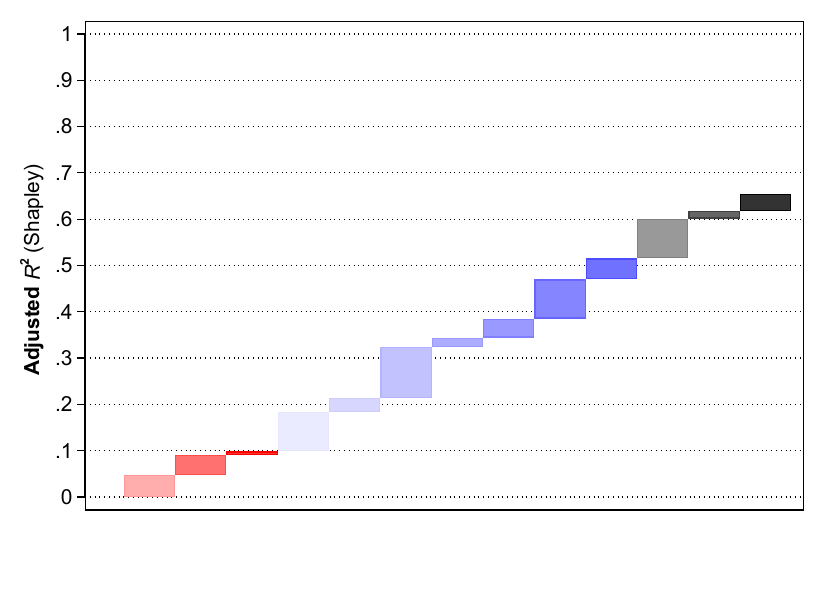}} 
        \subfloat[][Poaching Index]{\includegraphics[width=.35\textwidth]{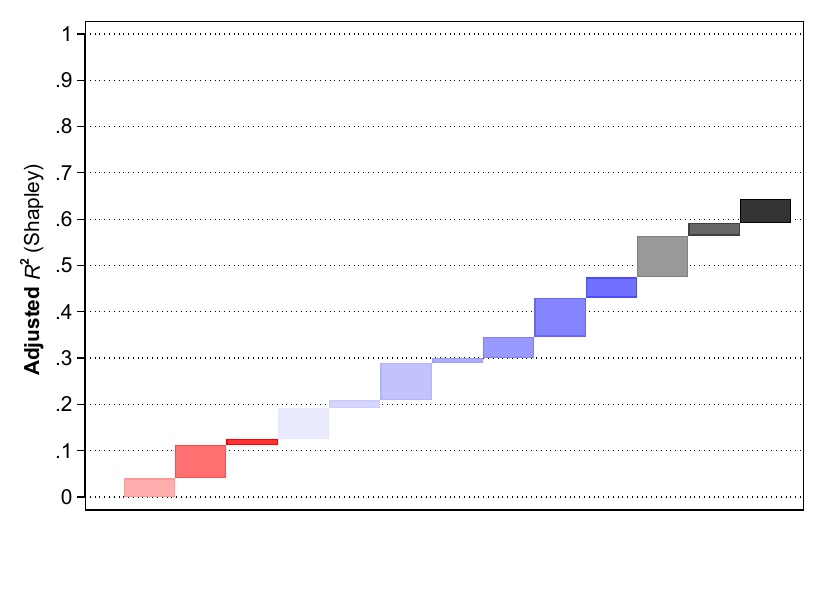}} \\
        \subfloat[][Flow Sorkin Value]{\includegraphics[width=.35\textwidth]{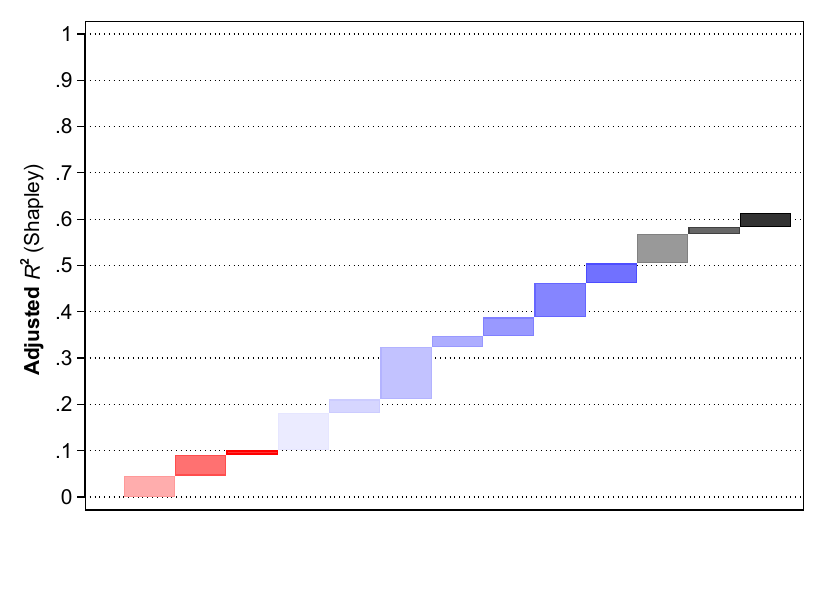}} 
        \subfloat[][Employment Size]{\includegraphics[width=.35\textwidth]{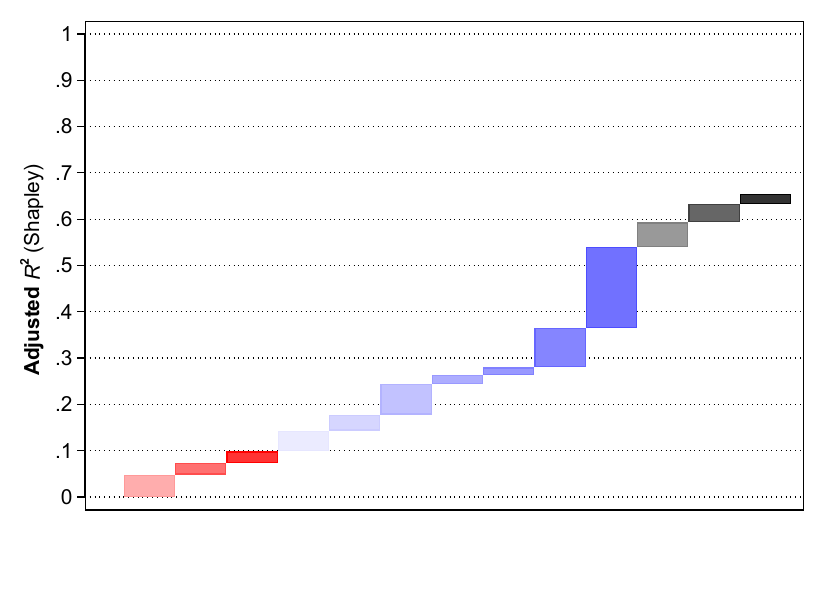}} \\
        \captionsetup[subfigure]{labelformat=empty} \vspace{-1.5em} 
        \subfloat[][]{\hspace{-2em}\includegraphics[trim={0bp 0bp 0bp 15em}, clip, width=.8\columnwidth]{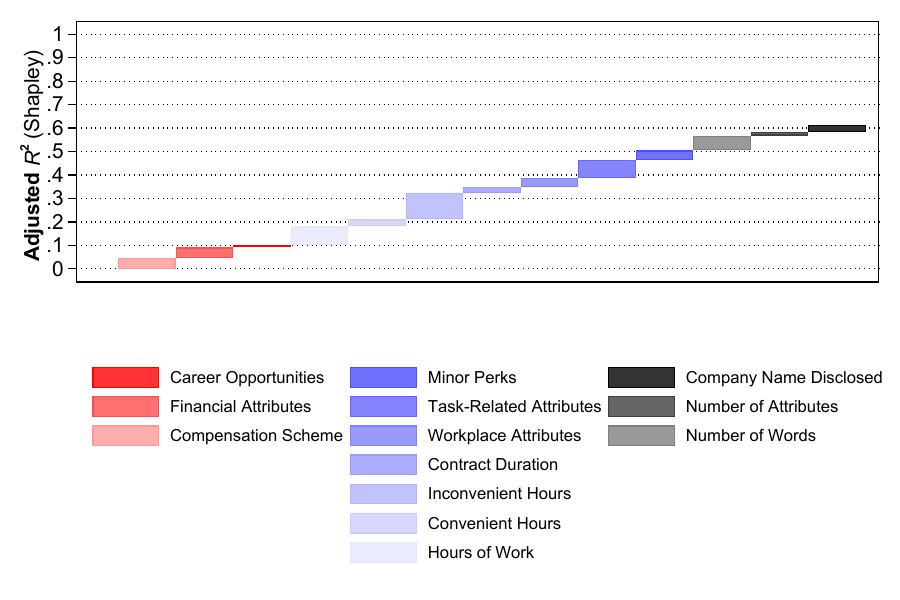}}
        \end{center}
    \vspace{-1em}         
    {\footnotesize \textit{Notes}: This figure documents adjusted $R^2$s from regressions of estimated employer values on pay and non-pay job attributes and other ad characteristics detected in the text analysis as in Figure \ref{fig:reg_value_on_characteristics}, decomposing the overall adjusted $R^2$s into ten broad categories of pay and non-pay attributes using the Shapley Value Decomposition. Estimations are done at the employer-cluster level for the period from 2021 to 2024, and are weighted by the number of worker-years in each cluster.}
\end{figure}
\FloatBarrier
\begin{figure}[hp!]
     \caption{Predictive Power of Publicly Advertised Job Attributes: By Industry.} \vspace{-1em}\label{fig:r2_by_industry}
         \begin{center}
         \subfloat[][Pay Premium]{\includegraphics[width=.4\linewidth]{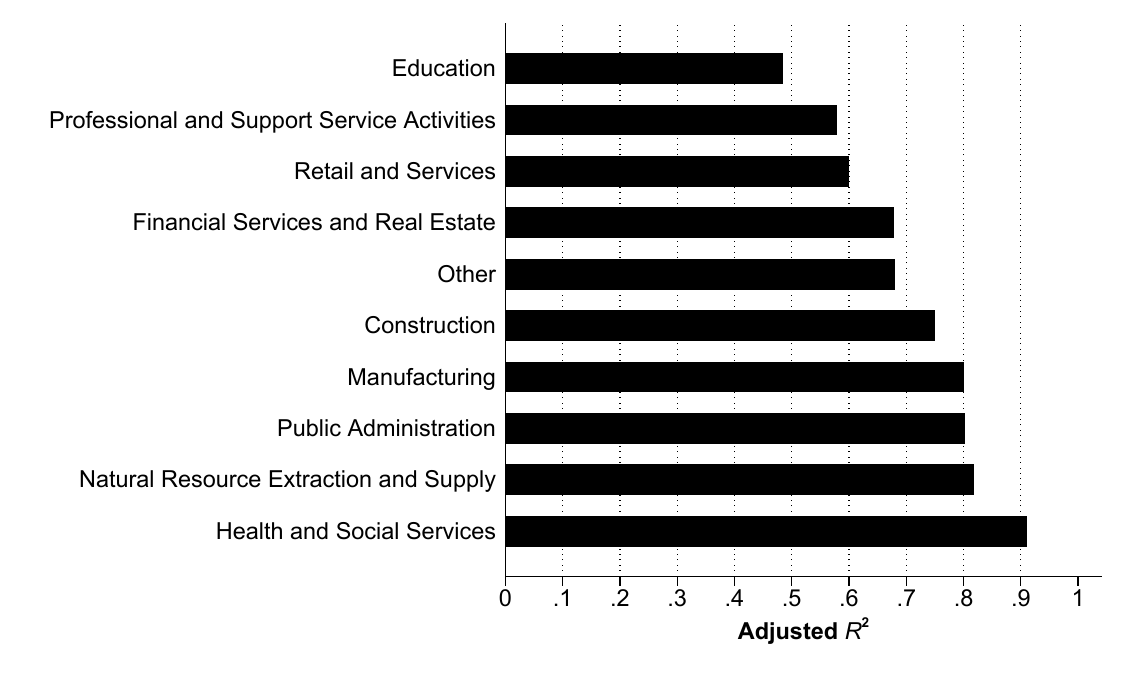}} \\
         \subfloat[][Overall Sorkin Value]{\includegraphics[width=.4\linewidth]{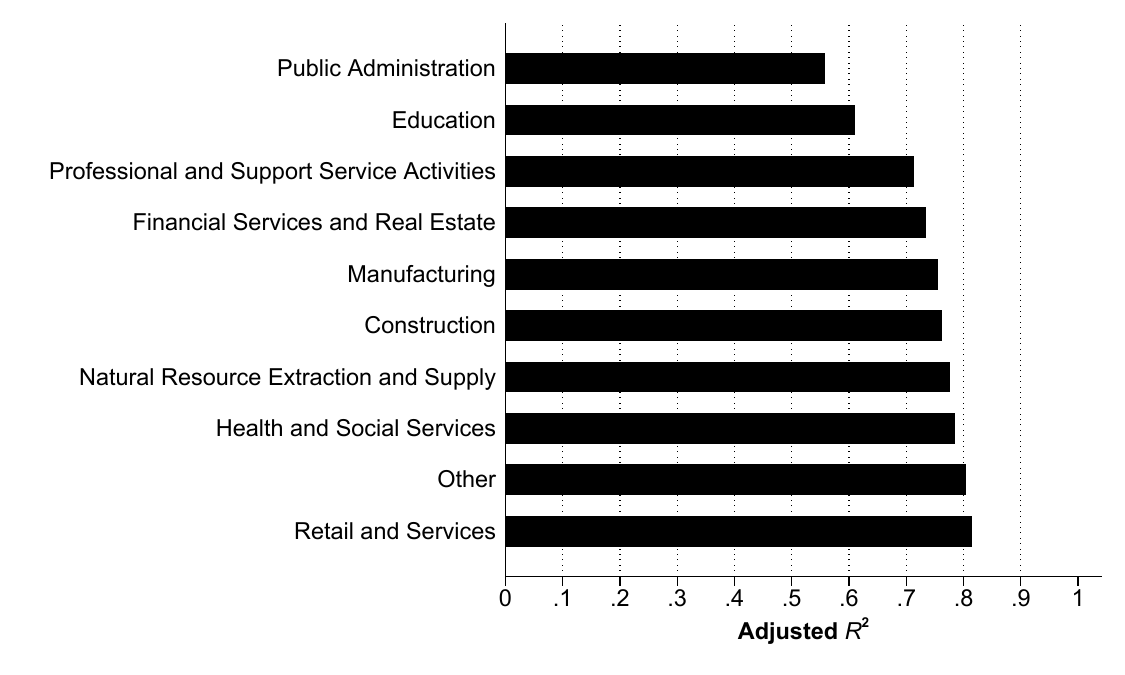}} 
         \subfloat[][Poaching Index]{\includegraphics[width=.4\linewidth]{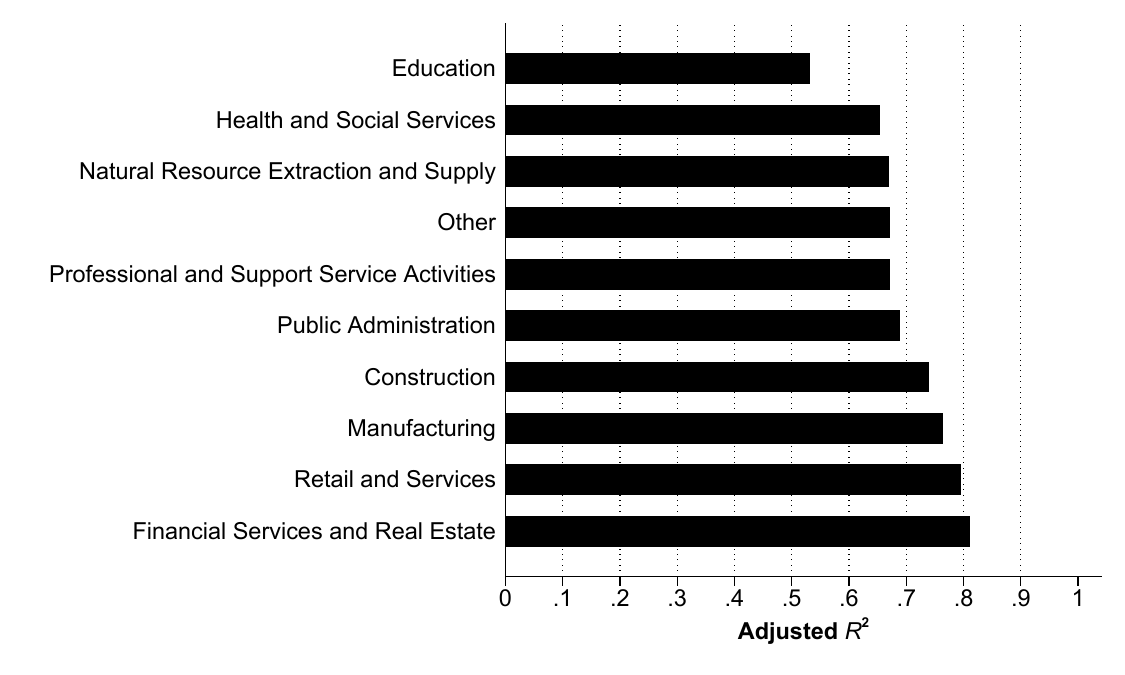}} \\
         \subfloat[][Flow Sorkin Value]{\includegraphics[width=.4\linewidth]{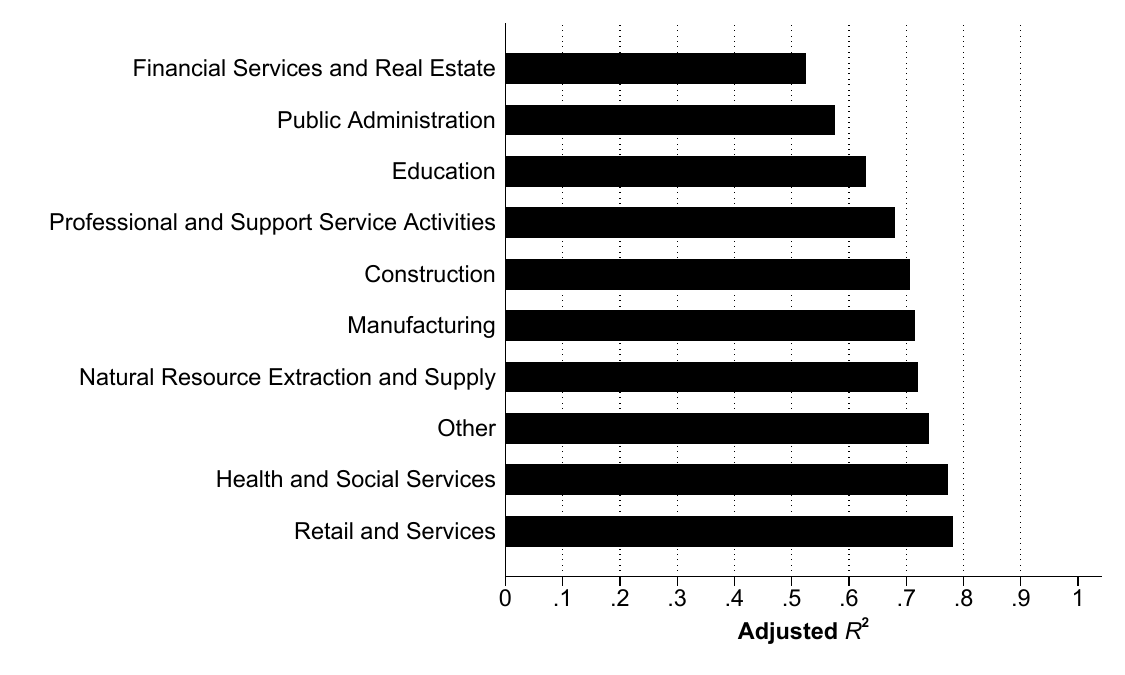}} 
         \subfloat[][Employment Size]{\includegraphics[width=.4\linewidth]{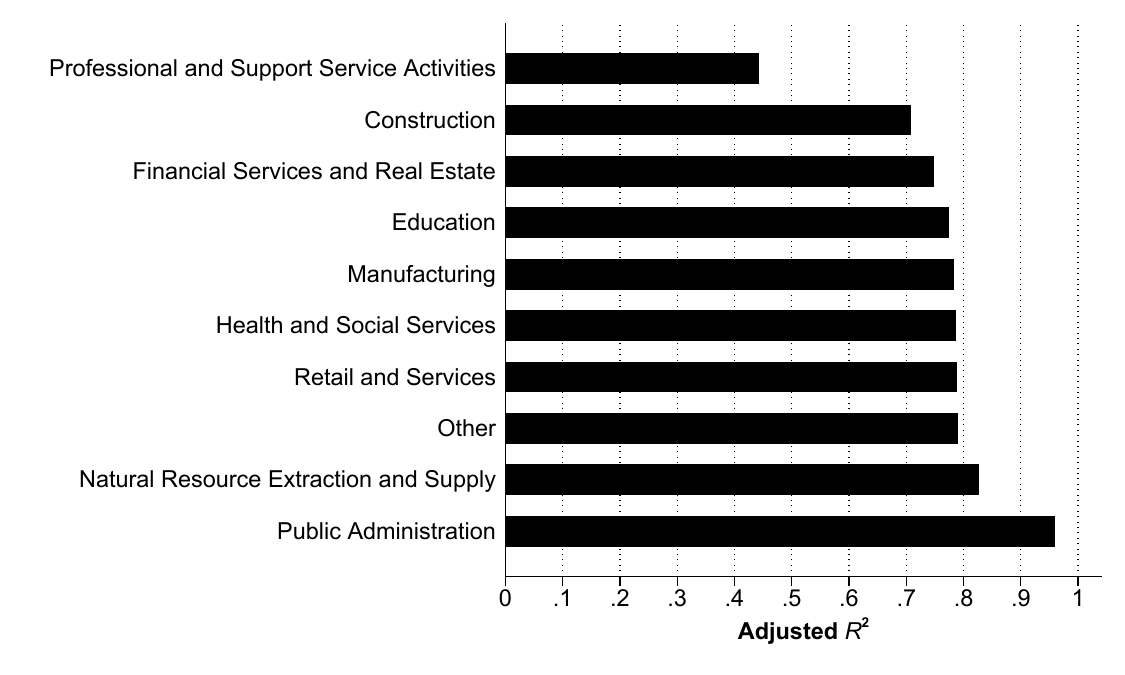}}
         \end{center}
      \vspace{-1em}         
     {\footnotesize \textit{Notes}: This figure documents adjusted $R^2$s from regressions of estimated employer values on pay and non-pay job attributes and other ad characteristics detected in the text analysis as in Figure \ref{fig:reg_value_on_characteristics}. Estimations are done separately for the ten main industry groups at the employer-cluster level for the period from 2021 to 2024, and are weighted by the number of worker-years in each cluster. We assign an industry to each cluster $g$ based on the modal industry among the establishments in the cluster.} 
\end{figure}
\FloatBarrier

\global\long\def\thetable{B.\arabic{table}}%
\setcounter{table}{0}
\global\long\def\thefigure{B.\arabic{figure}}%
\setcounter{figure}{0}
\global\long\def\theequation{B.\arabic{equation}}%
\setcounter{equation}{0}

\newpage
\section{Details on the Text Analysis} 
\label{app:text_analysis}

Section \ref{app:text_cleaning} documents how we clean the text of job ads before detecting job attributes. Section \ref{app:extract_lists} describes one of the processes we use to generate the complete lists of attributes we search for in vacancy texts. Section \ref{app:desc_attributes} provides descriptions of detailed attributes and displays some of the phrases used to detect these attributes. 

\subsection{Text Cleaning of Job Ads} 
\label{app:text_cleaning}

Before extracting attributes from job ads, we apply some simple text cleaning. We first search for and remove HTML tags, which always take the form ``$<$...$>$'', to avoid treating HTML code as potential job attributes and to reduce the number of words processed by the phrase-generation model. We keep a record of which ads contain such tags.

We replace numbers in ads with flags indicating their range and whether they are followed by a percentage sign. This allows us to group similar phrases that differ only in their specific numbers. For example, ``50\,\%'' and ``40\,\%'' both commonly indicate part-time jobs, so we group all phrases of the form \{number below 100\}\,\% under a single flag rather than searching for each value individually. Specifically, we flag numbers below 100 (and 100 itself) followed by a percentage sign or the word ``percentage''. We then assign separate flags to each integer from one to four, remaining numbers in $(0, 100)$, the number 100, and numbers greater than 100.

Before storing the ads, we split each ad into sentences and each sentence into words. We do this because the language model we use to expand the dictionary of target phrases expects input in this form. Splitting into sentences also prevents words at the end of one sentence from being conflated with words at the beginning of the next. We use the Python package \verb|nltk| for this, with a tokenization model trained on a Norwegian corpus. The model correctly handles periods that do not mark sentence boundaries, such as those appearing in common abbreviations.

\subsection{Extracting Common Attributes from ``We Offer'' Lists}
\label{app:extract_lists}

Our choice of pay and non-pay attributes to include in the text analysis is partly informed by attributes commonly advertised in ``we offer'' sections of job ads. We document our approach here, and show that social environment, on-the-job training, tasks, pension schemes, and insurance schemes are among the most commonly advertised attributes in such sections. Our approach has four steps: (i) we extract lists from the corpus of job ads; (ii) we sort these lists into four topics using a latent Dirichlet allocation (LDA) topic model; (iii) we identify the topic corresponding to ``we offer'' sections by inspecting the words and lists most closely associated with it; and (iv) we extract common phrases from the lists assigned to this topic. We then assign phrases to attribute categories and add these categories to the set of attributes we search for in the ads.

\paragraph{Extracting lists from vacancies}
To identify ``we offer'' sections, we start by extracting lists from the full corpus of job ads. Lists are commonly used in job ads to group related information, such as skill requirements, tasks, and pay and non-pay attributes. We extract lists by identifying sequences of the HTML tag \texttt{\textless li\textgreater}, which denotes list elements in web pages, or plain-text lines beginning with a centered dot~(·), hyphen~(-), or star~(*). We combine each such sequence with the preceding sentence, which typically serves as the list header, into a single observation. This process allows us to extract sections of ads that are likely to focus on a single topic. We extract close to 1.5 million lists from around 50\% of the ads, though the approach may omit some lists that do not follow these formatting conventions.

\paragraph{Sorting lists into different topics}
To distinguish ``we offer'' lists from lists covering other topics, we apply Latent Dirichlet Allocation (LDA) to the collection of lists \citep{blei2003latent}. LDA is commonly used in natural language processing to partition a collection of texts into topics. It assigns texts to $K$ categories such that words frequently occurring in topic $k$ are unlikely to occur in other topics. The model performs well when topics are associated with distinctive vocabulary, but it ignores word order, which can make individual words less informative. For example, ``good'' commonly appears in descriptions of workplace amenities or pay, but carries little information about the specific attribute without its context. Additionally, some lists combine ``we offer'' content with other topics, such as skill requirements, so we expect to lose some ``we offer'' content that the model assigns to other topics.

LDA models each word as drawn from a mixture of $K$ topic-specific word distributions. Each text is associated with its own mixture over the $K$ topics, so the probability that a word is drawn from topic $k$ varies across texts. That is, the probability that word $i$ in text $j$ ($w_{ij}$) equals ``offer'' is
\begin{align*}
    \Pr(w_{ij}=\textnormal{``offer''})=
    \sum_{k=1}^K
    \Pr(w_{ij}=\textnormal{``offer''}|\textnormal{topic}_{ij}=k)
    \Pr(\textnormal{topic}_{ij}=k)
\end{align*}
We set the number of topics to $K=4$ and estimate the model by maximum likelihood using the Python package \verb|gensim|.

\paragraph{Labelling topics}
After estimating the model, we label the four topics by inspecting the word distributions and the list-specific topic mixtures. Table~\ref{tab:topic words} reports the ten words most strongly associated with each topic, i.e., those with the highest probability of being drawn from that topic. Words in topics 1--3 are associated with (1) pay and non-pay job attributes, (2) skill requirements, and (3) task descriptions, while topic 4 contains a mixture of requirements (e.g., education and police certificate), additional information (e.g., contact details), and ambiguous words (numbers and ``work''). We label topics 1--3 accordingly, and topic 4 as a ``Residual Category''.

The pay and non-pay attributes topic contains words commonly used to indicate a good social environment, information about pay, and task-related attributes that may be attractive to workers (e.g., challenging tasks). We label this topic ``We Offer''. We note that some words in this topic are ambiguous outside their original context; for example, ``development'' can refer to personal development or to project-based work tasks.

We validate these labels by inspecting the lists most strongly associated with each topic, shown in Figures~\ref{fig:sample list we offer}--\ref{fig:sample list residual}. Unlike the word distributions, these lists preserve the original phrasing and context, addressing the bag-of-words limitation of LDA. The lists largely confirm the assigned labels. The list most strongly associated with the ``We Offer'' topic begins with the header ``we offer:'' and contains information about various non-pay amenities. The list most strongly associated with the ``Residual Category'' contains miscellaneous job information, such as details about the hiring process and pay. Overall, lists in the ``We Offer'' category appear to contain dense information about amenities, though we are likely to lose some pay and non-pay content that the model assigns to other categories.

\paragraph{Extracting common phrases from ``We Offer" sections}
To identify specific job attributes commonly advertised in vacancies, we focus on common phrases in lists assigned to the ``We Offer'' topic. We assign each list to the topic with the highest probability in its topic distribution, and extract approximately 260,000 lists under the ``We Offer'' label, representing 17\% of all lists.

We then extract and inspect the 200 most common unigrams, bigrams, and trigrams, excluding phrases that include stopwords (e.g., ``and''). We add phrases indicating pay or non-pay attributes to our dictionary, and create new attribute categories for any attribute not currently included.

Table~\ref{tab:common phrases in lists} reports the 10 most common unigrams, bigrams, and trigrams in ``We Offer'' lists, excluding phrases without a clear reference to a specific job attribute. The table shows that these lists commonly refer to the social environment, pay, pension and insurance schemes, and on-the-job training. The procedure uncovers relatively few attributes related to hours of work, which are likely described in other sections of the ads, such as the practical information found in lists assigned to the ``Residual Category.''

\FloatBarrier
\begin{table}[h!]
    \caption{Words Used to Separate Lists Into Topics}
    \label{tab:topic words}
    \hspace{-1em}\scalebox{.65}{\begin{tabular}{rlrlrlrl} \hline\midrule 
 \multicolumn{2}{c}{Topic 1: We-Offer} & \multicolumn{2}{c}{Topic 2: Skill Requirements} & \multicolumn{2}{c}{Topic 3: Tasks} & \multicolumn{2}{c}{Topic 4: Residual Category} \\ \cmidrule(lr){1-2} \cmidrule(lr){3-4} \cmidrule(lr){5-6} \cmidrule(lr){7-8} 
 \multicolumn{1}{c}{\shortstack{Word}} & \multicolumn{1}{c}{\shortstack{Probability}} & \multicolumn{1}{c}{\shortstack{Word}} & \multicolumn{1}{c}{\shortstack{Probability}} & \multicolumn{1}{c}{\shortstack{Word}} & \multicolumn{1}{c}{\shortstack{Probability}} & \multicolumn{1}{c}{\shortstack{Word}} & \multicolumn{1}{c}{\shortstack{Probability}}\\ 
 \multicolumn{1}{c}{(1)} & \multicolumn{1}{c}{(2)} & \multicolumn{1}{c}{(3)} & \multicolumn{1}{c}{(4)} & \multicolumn{1}{c}{(5)} & \multicolumn{1}{c}{(6)} & \multicolumn{1}{c}{(7)} & \multicolumn{1}{c}{(8)}\\ 
 \midrule 
``offer''                   & 0.037 & ``experience''     & 0.038 & ``tasks''            & 0.025 & Number $>99$ & 0.044 \\ 
``good (plural)''           & 0.032 & ``good (plural)''  & 0.027 & ``responsibility''   & 0.011 & Number in [5-99] & 0.024 \\ 
``workplace environment''   & 0.029 & ``good''           & 0.024 & ``to follow up''     & 0.010 & ``have to'' & 0.016 \\ 
``good (alternative form)'' & 0.026 & ``qualifications'' & 0.021 & ``contribute''       & 0.009 & ``Field of education'' & 0.009 \\ 
``pay''                     & 0.019 & ``ability''        & 0.016 & ``cooperation''      & 0.009 & ``police certificate'' & 0.008 \\ 
``professional''            & 0.017 & ``Norwegian''      & 0.014 & ``other''            & 0.008 & ``contact information'' & 0.007 \\ 
``exciting''                & 0.015 & ``relevant''       & 0.014 & ``tasks''            & 0.007 & ``level of education'' & 0.007 \\ 
``tasks''                   & 0.014 & ``written''        & 0.014 & ``participate''      & 0.007 & ``The position'' & 0.007 \\ 
``developement''            & 0.014 & ``oral''           & 0.013 & ``customers''        & 0.006 & ``work'' & 0.006 \\ 
``challenging''             & 0.010 & ``traits''         & 0.013 & ``developement''     & 0.005 & ``two'' & 0.006 \\ 
\hline\hline 
\end{tabular}
} 
    \vspace{.7em} \\
    {\footnotesize \textit{Notes:} This table documents the words that are most strongly associated with the topics identified by applying Latent Dirichlet Allocation to our collection of lists. Columns (2), (4), (6), and (8) contain estimated probabilities of drawing the corresponding word given that the word is drawn from the distribution associated with that topic. The table shows the words with the 10 largest probabilities for each topic. ``Number $>99$'' and ``Number in [5-99]'' are flags that we have inserted into the vacancy in the cleaning step to group numbers by range.  Translation from Norwegian by the authors. Each entry is a single Norwegian word, although the English translations sometimes require more than one word. Clarifications in parentheses are added whenever multiple distinct Norwegian words have identical English translations. The table is constructed using the period from 2002 to 2021.}
\end{table}

\FloatBarrier
\begin{figure}[h]
    \caption{Sample List Assigned to Topic ``We Offer''.}\vspace{-1em}
    \label{fig:sample list we offer}
    \begin{center}
    \vspace{-3em}
    \include{figures/sample_ads/example_list_0}
    \end{center}
    \vspace{-2em}
    {\footnotesize \textit{Notes:} This figure shows the list that the model most strongly associates with the ``we offer'' topic. This means that the model assigns a probability of almost  1 for words in this list to be drawn from the distribution associated with ``we offer.'' Translation from Norwegian by the authors.} 
\end{figure}

\begin{figure}[h]
    \caption{Sample List Assigned to Topic ``Skill Requirements''.}\vspace{-1em}
    \label{fig:sample list skill}
    \begin{center}
    \vspace{-3em}
    \include{figures/sample_ads/example_list_3}
    \end{center}
    \vspace{-2em}
    {\footnotesize \textit{Notes:} This figure shows the list that the model most strongly associates with the ``skill requirement'' topic. This means that the model assigns a probability of almost  1 for words in this list to be drawn from the distribution associated with ``skill requirement.'' Translation from Norwegian by the authors.} 
\end{figure}

\begin{figure}[h]
    \caption{Sample List Assigned to Topic ``Tasks''.}\vspace{-1em}
    \label{fig:sample list tasks}
    \begin{center}
    \vspace{-3em}
    \include{figures/sample_ads/example_list_2}
    \end{center}
    \vspace{-2em}
    {\footnotesize \textit{Notes:} This figure shows the list that the model most strongly associates with the topic ``tasks''. This means that the model assigns a probability of almost  1 for words in this list to be drawn from the distribution associated with the topic ``tasks.'' Translation from Norwegian by the authors.} 
\end{figure}

\begin{figure}[h]
    \caption{Sample List Assigned to Topic ``Residual Category''.}\vspace{-1em}
    \begin{center}
    \vspace{-3em}
    \label{fig:sample list residual}
    \include{figures/sample_ads/example_list_1}
    \end{center}
    \vspace{-2em}
    {\footnotesize \textit{Notes:} This figure shows the list that the model most strongly associates with the residual category. This means that the model assigns a probability of almost  1 for words in this list to be drawn from the distribution associated with the residual category. Translation from Norwegian by the authors.} 
\end{figure}
\FloatBarrier

\FloatBarrier
\begin{table}[h!]
    \caption{Most Frequent Phrases in ``We-Offer'' Lists.}
    \label{tab:common phrases in lists}
    \begin{center}
        \scalebox{.6}{\def\sym#1{\ifmmode^{#1}\else\(^{#1}\)\fi}

\begin{tabular}{cllcl} \hline\midrule 
\multicolumn{1}{c}{\shortstack{Rank\\ \phantom{a}}} & \multicolumn{1}{c}{\shortstack{Phrases\\ \phantom{a}}} & \multicolumn{1}{c}{\shortstack{English\\Translation}} & \multicolumn{1}{c}{\shortstack{Occurrences\\ \phantom{a}}} & \multicolumn{1}{c}{\shortstack{Assigned\\Attribute Category}} \\ 
\multicolumn{1}{c}{(1)} & \multicolumn{1}{c}{(2)} & \multicolumn{1}{c}{(3)} & \multicolumn{1}{c}{(4)} & \multicolumn{1}{c}{(5)} \\ 
\midrule 
\multicolumn{5}{l}{\textbf{Panel A: Unigrams}} \\ 
3&``arbeidsmiljø''                      &``workplace environment''        &138,459&Social Environment \\  
5&``lønn''                               &``pay''                          &91,477&Any Other Mention of Pay \\  
6&``spennende''                          &``exciting''                     &88,695&Interesting Tasks \\  
12&``utfordrende''                       &``challenging''                  &59,382&Challenging Tasks \\  
21&``opplæring''                         &``training''                     &37,351&On-the-Job Training \\  
24&``pensjons''                          &``pension''                      &32,520&Pension Scheme \\  
25&``pensjonsordning''                   &``pension scheme''               &32,434&Pension Scheme \\  
26&``utviklingsmuligheter''              &``opportunities for development''&31,724&Good Career Paths \\  
28&``forsikringsordninger''              &``insurance schemes''            &30,366&Insurance Scheme \\  
36&``oslo''                              &``oslo''                         &23,336&Central Location \\  
\midrule 
\multicolumn{5}{l}{\textbf{Panel B: Bigrams}} \\ 
1&``godt arbeidsmiljø''                  &``good workplace environment''   &65,143&Social Environment \\  
2&``konkurransedyktige betingelser''     &``competitive conditions''       &38,016&Competitive Pay \\  
3&``gode pensjons''                      &``good pension''                 &22,344&Pension Scheme \\  
4&``faglig utvikling''                   &``professional development''     &18,415&On-the-Job Traning \\  
5&``varierte arbeidsoppgaver''           &``varying tasks''                &18,085&Variation in Tasks \\  
7&``utfordrende arbeidsoppgaver''        &``challenging tasks''            &17,777&Challenging Tasks \\  
8&``fleksibel arbeidstid''               &``flexible work hours''          &16,700&Possibility to Work Flexible Hours \\  
10&``personlig utvikling''               &``personal development''         &15,617&On-the-Job Traning \\  
11&``trivelig arbeidsmiljø''             &``pleasant working environment'' &14,591&Social Environment \\  
12&``god pensjonsordning''               &``good pension scheme''          &12,854&Pension Scheme \\  
\midrule 
\multicolumn{5}{l}{\textbf{Panel C: Trigrams}} \\ 
1&``høyt faglig nivå''                   &``high professional level''      &6,021&Good Colleagues \\  
2&``god pensjonsordning gjennom''        &``good pension scheme through '' &3,716&Pension Scheme \\  
7&``godt faglig miljø''                  &``good professional environment''&2,134&Good Colleagues \\  
8&``pensjonsordning gjennom klp''        &``pension scheme through klp''   &2,009&Pension Scheme \\  
9&``godt sosialt miljø''                 &``good social environment''      &1,905&Social Environment \\  
10&``gjennom statens pensjonskasse''     &``public pension fund''          &1,805&Pension Scheme \\  
17&``faglig dyktige kollegaer''          &``professionaly skilles colleagues ''&1,195&Good Colleagues \\  
18&``pensjonsordning gjennom statens''   &``pension scheme through public''&1,183&Pension Scheme \\  
19&``ulykkesforsikring samt fritidsulykkeforsikring''&``sport and general accident insurance''&1,159&Insurance Scheme \\  
20&``svært godt arbeidsmiljø''           &``very good social environment''&1,129&Social Environment \\  
\midrule\hline 
\end{tabular}} 
        \vspace{.5em} \\
    \end{center}
    {\footnotesize \textit{Notes:} This table documents the words occurring most frequently in the ``we offer'' lists extracted from the vacancies. Column 3 shows which attribute category we have assigned the word to. These categories are added to the list of attributes we search for in the vacancies. Not every word is assigned to an attribute category. The translations are provided by the authors. Every phrase is originally one word in Norwegian. The table is constructed using the period from 2002 to 2021.}
\end{table}
\FloatBarrier

\subsection{Description of Detailed Pay and Non-Pay Attributes} 
\label{app:desc_attributes}

Tables~\ref{tab:attributes_pay_explained} to~\ref{tab:attributes_nonpay2_explained} describe each attribute we search for in job ads and provide examples of the phrases used to detect them. In our final dataset, we record each attribute as a binary indicator equal to one whenever the text of the associated job ad contains any of the phrases associated with that attribute.\footnote{The exception is the attribute ``Any Other Mention of Pay'', which we construct by searching for generic pay-related words and flagging the attribute as detected, but then setting the indicator to zero if any other attribute listed under ``Compensation Scheme'' was already detected by the same set of words.}

\FloatBarrier
\begin{table}[h!]
    \caption{Description of Pay Related Attributes.}
    \label{tab:attributes_pay_explained}
    \hspace{-2em}\scalebox{.58}{\begin{tabular}{p{7cm}>{\raggedright}p{9cm}>{\centering\arraybackslash}p{7cm}>{\centering\arraybackslash}p{7cm}} \toprule \multicolumn{1}{c}{Attribute} & \multicolumn{1}{c}{Description} & \multicolumn{1}{c}{Example Phrases (translated)} & \multicolumn{1}{c}{Example Phrases (Norwegian)} \\ \midrule \textbf{Compensation Scheme:} & & & \\ -- Compensation Level & Indication of pay, pay level, or pay range (e.g., steps in collective bargaining agreements). & wage step, wage range, wage rate & lønnstrinn, lønnsspenn, lønnssats \\[1.5em] -- Competitive Pay & Indication of good or competative salary. & good pay conditions, competitive wage, favorable pay conditions & gode lønnsbetingelser, konkurransedyktig lønn, gunstige lønnsbetingelser \\[1.5em] -- Collective Agreement Pay & Explicit reference to a collective bargaining agreement between trade unions and employers' associations. & wage step, tariff agreement, the main tariff agreement [public sector] & lønnstrinn, tariffavtale, hovedtariffavtalen \\[1.5em] -- Incentive Pay Scheme & Pay scheme dependents on an individual worker's or company's performance. & commission, bonus scheme, sales bonus & provisjon, bonusordning, salgsbonus \\[1.5em] -- Hiring Bonus & One time bonus at time of hiring, usually to encourage applicants. & recruitment supplement, establishment supplement & rekrutteringstillegg, rekrutteringstilskudd \\[1.5em] -- Good Overtime Pay & Reference to the level of overtime pay. & overtime pay, paid overtime & overtidsbetalt, betalt overtid \\[1.5em] -- Any Other Mention of Pay & Any other reference to pay, excluding any of the attributes listed above. & pay, income, fixed salary & lønn, inntekt, fastlønn \\[1.5em] \textbf{Financial Attributes:} & & & \\ -- Pension Scheme & Position covered by a pension scheme. Includes reference to mandatory and additional pension schemes. & pension scheme, occupational pension scheme, contribusion based pension savings & pensionsordning, tjenestepensjonsordning, inskuddspensjon \\[1.5em] -- Insurance Scheme & Employer-sponsored insurance. Includes mandatory and additional insurance schemes, and coverage of proparty damage, various health insurances and similar. & insurance scheme, group life insurance, life insurance scheme, disability insurance, health insurance & forsikriningsordning, gruppelivsforsikring, livsforsikring, uføreforsikring, helseforsikring \\[1.5em] -- Mortgage Possibility & Employer provides access to mortgage schemes with favorable conditions. & mortgage scheme, favorable mortgage & boliglånsordning, gunstig lånetilbud \\[1.5em] \textbf{Career Opportunities:} & & & \\ -- Good Career Paths & Suggests that the position is a good step in advancing a career, or that there are good opportunities to advance within the employer. & career opportunities, promotions, internal career & karrieremuligheter, forfremmelse, intern karriere \\[1.5em] -- On-the-Job Training & The position includes a training program or internship or opportunities to learn new skills. & training program, professional developement, internship & etterutdaningsmuligheter, faglig utvikling, praksisplass \\[1.5em] \bottomrule \end{tabular} } \vspace{.25em} \\
    {\footnotesize \textit{Notes:} This table describes each of the pay-related attributes detected in vacancies and documents some of the phrases used to detect these attributes. The phrases in Norwegian are used in the analysis, and the translations to English are for illustration. Comments to individual phrases are displayed in brackets.}
\end{table}
\FloatBarrier

\begin{table}[h!]
    \caption{Description of Non-Pay Related Attributes (Part 1).}
    \label{tab:attributes_nonpay1_explained}
    \hspace{-2em}\scalebox{.58}{\begin{tabular}{p{7cm}>{\raggedright}p{9cm}>{\centering\arraybackslash}p{7cm}>{\centering\arraybackslash}p{7cm}} \toprule \multicolumn{1}{c}{Attribute} & \multicolumn{1}{c}{Description} & \multicolumn{1}{c}{Example Phrases (translated)} & \multicolumn{1}{c}{Example Phrases (Norwegian)} \\ \midrule \textbf{Hours of Work:} & & & \\ -- Full-time Contract & Full time position, usually 37.5 hours per week. & full-time position, full-time employee, 100\% & fulltidsstilling, fulltidsansatt, 100\% \\[1.5em] -- Part-time Contract & Part-time position. & part time, part-time position, 50\% [includes all numbers less than 100] & deltidsstilling, deltidsjobb, 50\% \\[1.5em] -- Full-time/Part-time Choice & Employer open for both full-time and part-time hires. & full-time or part-time, full-time/part-time & heltid eller deltid, fulltid/deltid \\[1.5em] \textbf{Convenient Hours:} & & & \\ -- Possibility to Work Flexible Hours & Flexible work hours, commonly reflecting workers' discretion to choose when to start or stop work during a shift. & flexible working hours, flexible hours scheme & fleksibel arbeidstid, fleksibel arbeidstidsordning \\[1.5em] -- Regular Daytime Work Schedule & Regular, day-time work hours, i.e. working on evenings, nights or weekends is not required. & regular working hours & normal arbeidsdag \\[1.5em] -- Exempt from Work Hour Regulations & The job is exempt from the work hours regulation covering most workers, usually leaders, academic staff or similar. & exempt from the working environment act, leader position [legal term], doctoral fellow [always exempted] & untatt arbeidsmiljøloven, ledende stilling, doktorgradsstilling \\[1.5em] \textbf{Inconvenient Hours:} & & & \\ -- Shift Work & Position requires shift work where employees usually switch between working day, evening, and night shifts. & shift work, rotation scheme, night shift & skiftarbeid, turnusordning, natturnus \\[1.5em] -- Weekend/Evening/Night Work & Position requires shift work involving either working on weekends, evening, and/or night. & night shift, evening work,late night shift & nattursnus, kveldsarbeid, senvakt \\[1.5em] -- On-call Employment & Positions where the employer needs additional on-call employees, often hired temporarily. & on-call help, on-call substitute, calling help [norwegian expression], & ekstrahjelp, tilkallingsvikar, ringevikar \\[1.5em] -- Overtime Work Required & Indication that some overtime work is required in the position. & overtime work, some overtime expected, overtime in periods & overtidsarbeid, påregnes noe overtid, overtid i perioder \\[1.5em] \textbf{Contract Duration:} & & & \\ -- Permanent Job & Permanent job. & permanent position, permanent employment & fast stilling, fast ansettelse \\[1.5em] -- Temporary Job & Temporary job, including substitute positions. & temporary job, substitute position, one year substitute & midlertidig stilling, vikariat, ettårsvikariat \\[1.5em] -- Fixed-term Contract & This position is restricted to a fixed period, and is exempted from Norwegian restriction of the length of temporary spells. & fixed-term contract [legel term] & åremål \\[1.5em] \textbf{Workplace Attributes:} & & & \\ -- Social Environment & Indication of good social environment. & good social environment, good working community, great colleagues & godt arbeidsmiljø, godt arbeidsfellessakap, flotte kolleger \\[1.5em] -- Good Colleagues & Professional, knowledgable, helpful, or in other ways good colleagues. & skilled colleagues, good colleagues, knowledgeable colleagues, helpful colleagues & dyktige kolleger, gode kolleger, kunnskapsrike kolleger, hjelpsomme kolleger \\[1.5em] -- Possibility for Remote Work & Possibility to work remotely. & home office, home office scheme, work from home & hjemmekontor, hjemmekontorløsning, arbeide hjemmefra \\[1.5em] -- Shared Office Space & Shared office space. & shared office space, open office space & kontorlandskap, åpent landskap \\[1.5em] -- Inclusive Work-life Scheme & Indication that the employer provides an inclusive work-life scheme, which is a formal work-life scheme that provides more generous sick-leave arrangements. & inclusive work-life scheme, IA [abbrevation] & inkluderende arbeidsliv, IA \\[1.5em] \bottomrule \end{tabular} } \vspace{.25em}\\
    {\footnotesize \textit{Notes:} This table and the next describe each of the non-pay-related attributes detected in vacancies and document some of the phrases used to detect these attributes. The phrases in Norwegian are used in the analysis, and the translations to English are for illustration. Comments to individual phrases are displayed in brackets.}
\end{table}
\FloatBarrier

\begin{table}[h!]
    \caption{Description of Non-Pay Related Attributes (Part 2).}
    \label{tab:attributes_nonpay2_explained}
    \hspace{-2em}\scalebox{.58}{\begin{tabular}{p{7cm}>{\raggedright}p{9cm}>{\centering\arraybackslash}p{7cm}>{\centering\arraybackslash}p{7cm}} \toprule \multicolumn{1}{c}{Attribute} & \multicolumn{1}{c}{Description} & \multicolumn{1}{c}{Example Phrases (translated)} & \multicolumn{1}{c}{Example Phrases (Norwegian)} \\ \midrule \textbf{Task-Related Attributes:} & & & \\ -- Interesting Tasks & Indicates interesting, exciting, or meaningful tasks. Includes description of company, sector, or industry. & interesting, exciting, meaningful & interessante, spennende, meninsfylt \\[1.5em] -- Challenging Tasks & Indicates challenging tasks. & challenging, demanding & utfordrende, krevende \\[1.5em] -- Variation in Tasks & Varying, nonmonotonous tasks. & varied tasks, versatile tasks & varierte arbeidsoppgaver, allsidige oppgaver \\[1.5em] -- Responsibility in Job & Involves responsibilities. Includes responsibility as leader or project manager. & full responsibility, a lot of responsibilities, leadership responsibilities & totalansvar, mye ansvar, lederansvar \\[1.5em] -- Independence in Performing Tasks & Freedom to choose how to approach and solve tasks. & manage the working day, influence your own, shape your own & styre arbeidsdagen, påvirke egen, utforme egen \\[1.5em] -- Involves Leadership Responsibility & The job involves leadership responsibility. & looking for CEO, leadership role, communications director & søker daglig leder, lederrolle, kommunikasjonsdirektør \\[1.5em] -- Work Involves Travelling & The job requires traveling. & business trips, travel days, travel activity & arbeidsreiser, reisedøgn, reiseaktivitet \\[1.5em] \textbf{Minor Perks:} & & & \\ -- Beautiful Location & Describes the job location's environment as beautiful. & beautiful nature, magnificent nature, great tracking environment & flott natur, praktfull natur, flott turterreng \\[1.5em] -- Central Location & Central location, including mentions of major cities. & oslo [capial], centrally located, close to the city center & oslo, sentral beliggenhet, sentrumsnært \\[1.5em] -- Company Gym or Sports Team & The company offers the opportunity for paid exercise during work hours, access to fitness center/equipment, and/or has a sports team or the like. & company sports team, exercise during work-hours, fitness room, fitness center membership & bedriftsidrettslag, trening i arbeidstiden, treningsrom, treningsavtale \\[1.5em] -- Parking Space On Premises & Available/free parking spaces on premises. & parking facilities, free parking & parkeringsmuligheter, gratis parkering \\[1.5em] -- Company Vehicle & Access to company vehicle during work hours, sometimes involves company vehicle available full-time. & company vehicle, company vechicle scheme, leasing vechicle & firmabil, bilordning, leasingbil \\[1.5em] -- Any Welfare Scheme & Generic description of company welfare scheme. & welfare schemes, personnel schemes, employee benefits & velferdsordninger, personalordninger, personalgoder \\[1.5em] -- Company Cabin & The company has access to a cabin usable for employees. These cabins can be accessable for company arrangements or for employees' access during holidays or vacations. & company cabins, personnell cabin, vacation apartments & firmahytte, personalhytte, ferieleiligheter \\[1.5em] -- Occupational Health Service & Access to health professionals and doctors. Involves preventions and treatment of injuries/sickness. & occupational health service, company doctor & bedriftshelsetjeneste, bedriftslege \\[1.5em] -- Company Canteen & Access to canteen. & canteen scheme, personnell canteen & kantineordning, personalkantine \\[1.5em] -- Flexible/Extended Holidays & Description of the total length of vacation, or of extended length compared to mandatory vacation. & x week vacation [where x is a number], extra vacation & x ukers ferie, ekstra feie \\[1.5em] \bottomrule \end{tabular} } \vspace{.25em}\\
    {\footnotesize \textit{Notes:} This table and the previous describe each of the non-pay-related attributes detected in vacancies and document some of the phrases used to detect these attributes. The phrases in Norwegian are used in the analysis, and the translations to English are for illustration. Comments to individual phrases are displayed in brackets.}
\end{table}
\FloatBarrier

\global\long\def\thetable{C.\arabic{table}}%
\setcounter{table}{0}
\global\long\def\thefigure{C.\arabic{figure}}%
\setcounter{figure}{0}
\global\long\def\theequation{C.\arabic{equation}}%
\setcounter{equation}{0}

\newpage
\section{Details on Employer Quality Measures}
\label{app:quality_employers}

\subsection{Alternative Measures of Employer Quality}

\paragraph{Employer Pay Premium}
We recover the pay premium from a standard two-sided unobserved heterogeneity wage regression with worker and firm effects \citep{abowd1999high}.
We estimate the regression
\begin{equation}
    \label{eq:twoway_fe}
    \ln w_{it} = X_{it}'\beta + \alpha_i + \psi_{j(i,t)} + e_{it},    
\end{equation}
where $w_{it}$ are hourly wages, $X_{it}$ are exogenous worker characteristics, $\alpha_i$ is the unobserved worker effect, $\psi_{j(i,t)}$ is the unobserved firm effect of worker $i$'s employer $j$ in period $t$, and $e_{it}$ is an idiosyncratic shock. As is standard in the literature, the vector $X_{it}$ includes a set of year-by-education fixed effects and a separate cubic age polynomial by education level restricted to be flat at age 40 \citep{card2013workplace}. 

\paragraph{Employer Value from the Utility-Posting Model}
We use the employer value $V_j$ and flow value $u_j$ from the utility-posting model of \citet[][]{sorkin2018ranking}, estimated from the full model with employer-specific destruction and reallocation shocks, as described in Section V of the original paper.\footnote{The literature is often ambiguous about which version of the model is reported. The model is sometimes described as PageRank, which corresponds to a different normalization of the worker mobility transition matrix. Many contributions also do not clarify whether they use all worker flows, only job-to-job flows, or the full utility-posting model.} For completeness, we summarize this model in Appendix Section \ref{app:sorkin_model}.

\paragraph{Employer Poaching Index}
We use the poaching index defined in \cite{bagger2019empirical}. The index is constructed as
\begin{equation}
    \label{eq:poaching_index}
    \text{Poaching Index}_j = \frac{
        \sum_{k,t} \mathrm{EE}_{kj,t}
    }{
        \sum_{k,t} \mathrm{EE}_{kj,t} + \mathrm{NE}_{j,t}
    }.
\end{equation}
In Equation \eqref{eq:poaching_index}, $\mathrm{EE}_{kj,t}$ denotes job-to-job flows from employer $k$ to employer $j$, and $\mathrm{NE}_{j,t}$ denotes flows from non-employment to employer $j$.

\paragraph{Employment Size}
We define employment size as the number of worker-year observations at each employer over the sample period.

\subsection{Summary of the Utility-Posting Model}
\label{app:sorkin_model}

\paragraph{Environment}
Consider an economy in discrete time populated by a fixed measure of infinitely-lived workers and a fixed discrete number $J$ of employers.
Workers have discount factor $\beta \in (0,1)$ and can be either employed at one of the $J$ employers or non-employed.
Let $j \in \mathcal{J} = \{1,\dots,J\}$ denote employers and $V_j$ the value to a worker of being currently employed at employer $j$.
Let $N$ denote non-employment and $V_N$ the value to a worker of being currently non-employed.
Both employed and non-employed workers search for better employers.
Search is random. 
When searching, workers make contact with one of the $j \in \mathcal{J}$ employers with probability $f_j$, where $f$ captures the relative weight of employers in workers' search.  
We maintain the wage-posting (here ``utility-posting'') assumption from \cite{burdett1998wage}.
The value of working at $j$ is summarized by $V_j$ and is not renegotiated even in the event of a change in the workers' outside option.

\paragraph{Worker flows between employers}
Let $M_{jk}$ denote the measure of workers making a job-to-job move from employer $j$ to employer $k$.
$M_{jk}$ is assumed to be the sum of involuntary relocation flows $M_{jk}^{R}$ and voluntary flows $M_{jk}^{V}$.
Relocation job-to-job flows are given by
\begin{align}
    \label{eq:mobility_relocation}
    M_{jk}^{R} = L_j \cdot \rho_j \cdot f_k,   
\end{align}
where $L_j$ is employment at $j$, $\rho_j$ is the probability that workers employed at $j$ are forced to reallocate to another employer, and $f_k$ is the probability that they find a job at employer $k$.\footnote{There are no such relocation shocks in \cite{burdett1998wage}, but they are a common feature in random search models with on-the-job search (see, e.g., \cite{jolivet2006empirical} and \cite{bagger2019empirical}).}
In Equation \eqref{eq:mobility_relocation}, workers do not make a choice and are forced to relocate to employer $k$.
This is in contrast with voluntary job-to-job flows, which are given by
\begin{align}
    \label{eq:mobility_voluntary}
    M_{jk}^{V} = L_j 
    \cdot 
    (1 - \rho_j - \delta_j) 
    \cdot 
    \lambda_1 
    \cdot 
    f_k 
    \cdot 
    \text{Pr}(k \succ j). 
\end{align}
Voluntary worker flows from $j$ to $k$ are given by the measure of workers employed at $j$ ($L_j$), who are neither forced to reallocate (with probability $\rho_j$) nor forced to move to non-employment (with probability $\delta_j$), search for an alternative employer (with probability $\lambda_1$), and get an offer from employer $k$ with probability $f_k$.\footnote{
    The probabilities $\rho_j$ and $\delta_j$ are assumed to be mutually exclusive.
    }
In Equation \eqref{eq:mobility_voluntary}, workers choose whether to remain with employer $j$ or move to $k$, which occurs with probability $\text{Pr}(k \succ j)$.
This is the sense in which these flows are voluntary and reveal workers' preferences over the set of employers $\mathcal{J}$.
We follow \cite{sorkin2018ranking} and assume there are i.i.d. taste shocks $\{\varepsilon_i\}_{i \in \{\mathcal{J},N\}}$ that make the value of working at employer $j$ and of non-employment worker-specific in each period. 
We further assume that these shocks are drawn from a Gumbel distribution with location parameter normalized to zero and scale parameter $\sigma^{-1}$.\footnote{Normalizing the location parameter to zero is without loss of generality because it shifts the value of working at each employer $j$ and of being in non-employment by the same amount.}   
With this assumption, the acceptance probability for a worker with a job offer from employer $k$ currently working at $j$ is given by 
\begin{align*}
    \text{Pr}(k \succ j) 
    = \text{Pr}(V_k + \varepsilon_k \geq V_j + \varepsilon_j)
    = \frac{\exp(\sigma V_j)}{\exp(\sigma V_j) + \exp(\sigma V_k)}.
\end{align*}

\paragraph{Worker flows to non-employment} 
The flow measure of workers from employer $j$ to non-employment $N$ is the sum of a relocation part $M_{jN}^R$ and a voluntary part $M_{jN}^V$.
The relocation flow to non-employment is simply the fraction of employment at firm $j$ hit by an exogenous separation shock $\delta_j$
\begin{align}
    \label{eq:mobility_jN_relocation}
    M_{jN}^R = L_j \cdot \delta_j.
\end{align}
The voluntary flow to non-employment is given by the fraction of workers who decide to leave employer $j$ 
\begin{align}
    \label{eq:mobility_jN_voluntary}
    M_{jN}^V 
    = L_j 
    \cdot (1 - \delta_j - \rho_j) 
    \cdot (1 - \lambda_1)
    \cdot \text{Pr}(N \succ j),        
\end{align}
where $1 - \delta_j - \rho_j$ is the fraction of workers who are not forced to reallocate and $1 - \lambda_1$ is the fraction of workers who do not have an alternative offer in the current period.
Similarly to the case of voluntary flows between employers \eqref{eq:mobility_voluntary}, the probability that a worker employed at $j$ moves to non-employment is 
\begin{equation}
    \text{Pr}(N \succ j)
    = \text{Pr}(V_N + \varepsilon_N \geq V_j + \varepsilon_j)
    = \frac{\exp(V_N)}{\exp(V_j) + \exp(V_N)},   
\end{equation}
given the maintained assumption of i.i.d. idiosyncratic taste shocks $(\varepsilon_j,\varepsilon_N)$ from an Extreme Value Type I distribution. 

\paragraph{Worker flows from non-employment} 
All transitions from non-employment $N$ to employer $j$ are voluntary.
The decision to accept a job opportunity is again subject to i.i.d. taste shocks $(\varepsilon_N,\varepsilon_j)$.
The flow measure of workers moving from non-employment to employer $j$ is given by
\begin{align}
\label{eq:mobility_Nj}
    M_{Nj} = L_N 
    \cdot \lambda_0
    \cdot f_j
    \cdot \frac{\exp(V_j)}{\exp(V_N) + \exp(V_j)}.
\end{align}
In Equation \eqref{eq:mobility_Nj}, $L_N$ is the measure of workers in non-employment, $\lambda_0$ is the probability that a non-employed worker gets an offer, and $f_j$ is the probability that the offer is from employer $j$.
For the identification result, it is required that the offer distribution $f$ is the same for employed and non-employed workers, which we assume.

\paragraph{Value of employment}
The sequence of events described by the relocation and voluntary worker flows in Equations \eqref{eq:mobility_relocation}-\eqref{eq:mobility_voluntary} implies the following expression for the value $V_j$ of working at employer $j$.
Let $u_j$ denote the flow utility of working at $j$.
$V_j$ can be expressed as
\begin{equation}
\label{eq:value_j}
\begin{split}
    V_j  =  u_j  
    & + \beta   \delta_j  \mathbf{E} 
    \big[ 
        V_N + \varepsilon_N
    \big] \\  
    & + \beta  \rho_j  \sum_{k \in \mathcal{J}} f_k  
    \mathbf{E}
    \big[ 
        V_k + \varepsilon_k
    \big]  \\
    & + \beta  (1 - \delta_j - \rho_j)  \lambda_1  
    \sum_{k \in \mathcal{J}} f_k 
    \mathbf{E}
    \big[ 
        \max 
        \big\{
            V_j + \varepsilon_j,V_k + \varepsilon_k
        \big\}  
    \big] \\
    & + \beta   (1 - \delta_j - \rho_j)  
    (1 - \lambda_1)            
    \mathbf{E}
    \big[ 
        \max \big\{
            V_N + \varepsilon_N, V_j + \varepsilon_j
        \big\} 
    \big].
\end{split}
\end{equation}
In Equation \eqref{eq:value_j}, the continuation value is made of the four following terms.
With probability $\delta_j$, the worker transitions to non-employment, an exogenous shock.
With probability $\rho_j$, the worker is hit by a relocation shock and is forced to move to an alternative employer by drawing from the offer distribution $f$, an exogenous shock.
With probability $(1 - \delta_j - \rho_j) \lambda_1$, the worker gets an offer from a potential alternative employer by drawing from the offer distribution $f$ and decides whether to stay or move, an endogenous choice.
With probability $(1 - \delta_j - \rho_j) (1 - \lambda_1)$, the worker decides whether to stay with their current employer or to move to non-employment, also an endogenous choice.
All expectations $\mathbf{E}[.]$ are taken over the i.i.d. taste shocks $(\varepsilon_k,\varepsilon_N)$.

\paragraph{Value of non-employment}
With the notation introduced in Equation \eqref{eq:mobility_Nj}, the value function of non-employment follows directly as 
\begin{equation}
\label{eq:value_N}
    V_N = u_N + \beta 
    \left \{
        \lambda_0  \sum_k f_k \mathbf{E}
        \big[ 
            \max 
            \big\{
                V_k + \varepsilon_k, V_N + \varepsilon_N
            \big\}  
        \big]
        + (1 - \lambda_0) \mathbf{E}
        \big[ 
            V_N + \varepsilon_N
        \big]
    \right \},
\end{equation}
where $u_N$ is the flow value of non-employment and $\beta$ the discount factor.
The continuation value is made of two terms.
With probability $\lambda_0$, non-employed workers get a draw from the offer distribution $f$, which they decide to accept or reject given their draw of taste shocks.
With probability $1-\lambda_0$, they remain in non-employment.
All expectations $\mathbf{E}[.]$ are taken over the i.i.d. taste shocks $(\varepsilon_k,\varepsilon_N)$.

\paragraph{Identification}

The model has several employer-level parameters, such as workers' valuations of alternative employers $\{V_j\}_{j \in \mathcal{J}}$, and aggregate-level parameters, such as the offer arrival rate for employed workers $\lambda_1$. These parameters are identified using the matched employer-employee data described in Section \ref{sec:data}, which contain the relevant information on worker flows (transitions in and out of non-employment and between employers). 

The employer-level parameters $\{\sigma V_j,L_j,f_j,\rho_j,\delta_j\}_{j \in \mathcal{J}}$ and the transition parameters $\lambda_0$ and $\lambda_1$ are identified from (i) the realized mobility flows between employers $M_{jk}$, and (ii) which employers are shrinking or growing \citep[see][Section V.B]{sorkin2018ranking}. The notation $\sigma V_j$ emphasizes that worker flows pin down the value of working at employer $j$ up to the scale of the idiosyncratic taste shocks.

The main source of identification for $\sigma V_j$ is the aggregation of the realized mobility flows $M_{jk}$.
An estimate for $L_j$ is obtained directly from the data as the share of worker-year observations at employer $j$.
$f_j$ is identified from the share of non-employment flows to employer $j$.
The separation shocks $\delta_j$ and $\rho_j$ are identified from the separation flows at shrinking firms, which give information on involuntary worker flows.
Finally, the transition parameters $\lambda_0$ and $\lambda_1$ are pinned down by accounting for the aggregate transition flows, respectively from non-employment to employment and between employers.

Two sample restrictions are required for identification at this step.
First, the aggregation of mobility flows only identifies $V_j$ within the strongly connected sets of employers.\footnote{
    Within a strongly connected set of employers, each employer has at least one worker moving in and one worker moving out.
    This condition is required for the fixed-point associated with the appropriately scaled matrix of worker flows to exist. See, e.g., \cite{jackson2008social} for details on social network definitions.}
Second, each employer must hire at least one worker from non-employment for its sampling weight $f_j$ to be well-defined.

Given $\{\sigma V_j,L_j,f_j,\rho_j,\delta_j\}_{j \in \mathcal{J}}$ and the aggregate transition rate $\lambda_1$, $\sigma u_j$ can be recovered directly from Equation \eqref{eq:value_j}, again up to the scale of the i.i.d. taste shocks.
This calculation is straightforward since, given that $\{\sigma \varepsilon_i\}_{i \in \{\mathcal{J},N\}}$ are i.i.d. draws from a Type 1 Extreme Value distribution with scale one, all expectations in Equation \eqref{eq:value_j} admit the following closed-form solutions
\begin{equation}
    \begin{split}
        \sigma \mathbf{E}
        \big[ 
            V_j + \varepsilon_j
        \big]
        & = \sigma V_j + \gamma, 
        \\
        \sigma \mathbf{E}
        \big[ 
            \max 
            \big\{
                V_j + \varepsilon_j, V_k + \varepsilon_k
            \big\}  
        \big] 
        & = \ln \big(\exp (\sigma V_j) + \exp (\sigma V_k) \big)+ \gamma,
    \end{split}
\end{equation}
where $\gamma$ denotes Euler's constant.\footnote{\cite{gyetvai2022identification} make a related point in the context of a continuous time random search model.}
This step further requires us to assume a value for the discount factor $\beta$, which we set in line with a 5\% annual discount rate.

As in the original paper, the scale of the taste shocks $\sigma$ is not separately identified, so the model is identified up to scale. We set $\sigma = 1$.

\paragraph{Solution algorithm}
The model is solved using a fixed-point iteration algorithm. The algorithm is described in Online Appendix G of \citet{sorkin2018ranking}. 

\subsection{Sample Selection}

The model is estimated on the Norwegian matched employer-employee data described in Section \ref{sec:data}. We use establishments as the empirical counterpart to employers (``$j$'') in the model. We follow the employer-employee fixed-effect literature and assign each individual one main employer in each year by selecting the establishment with their largest annual earnings. Our data feature the exact start date and end date of each employment spell with an employer in each year, so we can make a precise distinction between employer-to-employer moves and moves with a non-employment spell in-between. Moves between two main employers that include an intervening spell with another employer also qualify as employer-to-employer moves.\footnote{We define employer-to-employer flows as consecutive employment spells with a gap of at most 31 days.}

We restrict the sample to the period 2021-2024 and retain workers aged 20 to 60 (inclusive). We compute all measures for the same subset of employers, derived from the most stringent identification restrictions in \citet{sorkin2018ranking}. Specifically, we restrict the sample to the largest strongly connected set of establishments and impose that each of these establishments hires at least one individual from non-employment.\footnote{Focusing on the largest connected set is common in the literature. See, among others, \cite{card2013workplace} and \cite{bonhomme2020much} in the context of the two-sided unobserved heterogeneity regression model \eqref{eq:twoway_fe} and \cite{sorkin2018ranking} and \cite{morchio2024gender} in the context of the utility-posting model.} These last two restrictions are interdependent, and they are imposed recursively until the set of establishments converges.

As shown in Table \ref{tab:sample_selection}, around a quarter of employers satisfy identification restrictions in \cite{sorkin2018ranking}. Since employment is concentrated at larger employers, this sample covers the majority of workers (74\% of unique workers) and job ads (68\% of posted job ads). See Appendix Table \ref{tab:sample_characteristics_2021_2024} for additional descriptive statistics on workers and firms.

\subsection{Clustering}

To precisely estimate employer parameters, we reduce their number using a $k$-means clustering algorithm as a pre-estimation step \citep{bonhomme2019distributional}, which partitions the $J$ employers in each sample into $G$ groups with similar observable characteristics, increasing the number of movers per group and improving estimation precision. The algorithm requires two inputs: (i) a vector of establishment-level variables on which to classify employers, and (ii) a number of groups $G \leq J$. Our clustering variables capture three characteristics of establishments: wages (the empirical cumulative density function of wages evaluated at deciles of the sample-wide wage distribution), worker flows (job creation and destruction rates, distinguishing direct employer-to-employer transitions from transitions through non-employment), and industry, occupation, and location composition.\footnote{For location, we cut the distribution of municipality employment into deciles and construct indicators for the establishment's municipality. For example, Oslo accounts for more than 10\% of Norwegian employment, so there is a specific indicator for establishments in Oslo.} We set $G = \text{round}(J/50)$ in our baseline to balance heterogeneity against precision. As a result, our baseline model features $1,422$ unique employer clusters. We provide robustness checks using smaller ($G = \text{round}(J/25)=2,843$) and larger ($G = \text{round}(J/100)=711$) clusters.

\subsection{Correlations}

Table \ref{tab:correlation_quality_measures} shows the correlations between the five employer quality measures. Panel A is for the baseline sample (2021-2024, $G = \text{round}(J/50)$). Panels B-D show the same correlations for several alternative samples: 2015-2024 (Panel B), smaller clusters (Panel C), and larger clusters (Panel D). Across samples, we find that the measures of employer quality are imperfectly correlated, with most correlation coefficients falling in the 30--50\% range. There is stronger agreement between the flow-based measures. The correlation between the poaching index and the measures derived from the utility-posting model are consistently above 80\%.

\begin{table}[htbp]
    \caption{Correlations between Employer Quality Measures.}
    \label{tab:correlation_quality_measures}
    \vspace{-0.5cm}
    \begin{center}
        \scalebox{0.45}{\begin{tabular}{l c c c c c}
\toprule \midrule
 & Pay Premium & Sorkin Overall Value & Poaching Index & Sorkin Flow Value & Employment Size \\
\midrule
Panel A. Baseline Sample &  &  &  &  &  \\
\quad Pay Premium & 1.000 &  &  &  &  \\
\quad Sorkin Overall Value & 0.369 & 1.000 &  &  &  \\
\quad Poaching Index & 0.434 & 0.876 & 1.000 &  &  \\
\quad Sorkin Flow Value & 0.329 & 0.981 & 0.849 & 1.000 &  \\
\quad Employment Size & 0.535 & 0.412 & 0.414 & 0.334 & 1.000 \\
\addlinespace
Panel B. 2015--2024 &  &  &  &  &  \\
\quad Pay Premium  & 1.000 &  &  &  &  \\
\quad Sorkin Overall Value  & 0.485 & 1.000 &  &  &  \\
\quad Poaching Index  & 0.531 & 0.898 & 1.000 &  &  \\
\quad Sorkin Flow Value & 0.407 & 0.965 & 0.848 & 1.000 &  \\
\quad Employment Size & 0.540 & 0.336 & 0.357 & 0.191 & 1.000 \\
\addlinespace
Panel C. Smaller Clusters: $G=\text{round}(J/25)$ &  &  &  &  &  \\
\quad Pay Premium  & 1.000 &  &  &  &  \\
\quad Sorkin Overall Value & 0.334 & 1.000 &  &  &  \\
\quad Poaching Index & 0.389 & 0.849 & 1.000 &  &  \\
\quad Sorkin Flow Value & 0.291 & 0.976 & 0.817 & 1.000 &  \\
\quad Employment Size & 0.504 & 0.372 & 0.378 & 0.285 & 1.000 \\
\addlinespace
Panel D. Larger Clusters: $G=\text{round}(J/100)$ &  &  &  &  &  \\
\quad Pay Premium  & 1.000 &  &  &  &  \\
\quad Sorkin Overall Value  & 0.394 & 1.000 &  &  &  \\
\quad Poaching Index & 0.461 & 0.897 & 1.000 &  &  \\
\quad Sorkin Flow Value & 0.348 & 0.985 & 0.874 & 1.000 &  \\
\quad Employment Size & 0.568 & 0.437 & 0.438 & 0.364 & 1.000 \\
\bottomrule  \midrule
\end{tabular}

}
    \end{center}
    {\footnotesize \textit{Notes}: This table shows the correlation coefficients for the five employer quality measures described in Appendix \ref{app:quality_employers}. Panel A: 2021-2024, $G = \text{round}(J/50)$. Panel B: 2015-2024, $G = \text{round}(J/50)$. Panel C: 2021-2024, $G = \text{round}(J/25)$. Panel D: 2021-2024, $G = \text{round}(J/100)$. All correlations are weighted by worker-years.}
\end{table}
\FloatBarrier

\global\long\def\thetable{D.\arabic{table}}%
\setcounter{table}{0}
\global\long\def\thefigure{D.\arabic{figure}}%
\setcounter{figure}{0}
\global\long\def\theequation{D.\arabic{equation}}%
\setcounter{equation}{0}

\newpage
\section{Details on the Monopsony Model}
\label{app:monopsony_model}

\subsection{Estimation Procedure}

\paragraph{Step 1: Sample selection}
We restrict the sample to larger employers for which we can credibly estimate employer-specific components implied by Equations \eqref{eq:indirect_utility}-\eqref{eq:indirect_amenity}. We use establishments as the empirical counterpart to employers (``$j$'') in the model. Specifically, we only retain establishments with at least 30 stayers and 15 movers over the period, where stayers are workers observed at the same employer in the following period, and movers are workers observed at a different employer. As we subsequently recover employer fixed effects from a standard two-sided unobserved heterogeneity wage regression \citep{abowd1999high}, we must restrict the sample to the largest connected set of employers, which is required for identification. This restriction leaves us with 12,973 establishments. Additional details on the composition of this sample can be found in Appendix Table \ref{tab:sample_characteristics_2021_2024}.

\paragraph{Step 2: Clustering}
We reduce the number of parameters to be estimated by grouping employers paying similar wages. Specifically, we use a $k$-means clustering algorithm and input the establishment-level empirical cumulative distribution function (CDF) of hourly wages as the clustering variable \citep{bonhomme2019distributional}. The establishment-level empirical CDF is computed at the deciles of the distribution of hourly wages in the whole sample. Establishments are weighted by their number of worker-year observations over the sample period. We choose the number of employer clusters $G$ with the formula $G = \mathrm{round}(J/50)$, where $J$ is the number of establishments in the estimation sample.

\paragraph{Step 3: Pre-estimated parameters} 
Some of the parameters in the monopsony model are pre-estimated, as described in the main text. The labor supply elasticity ($\sigma$) is derived from the central pass-through estimate for Norway reported in \citet[][Table 2]{bhuller2025wage}, which implies $\widehat{\sigma} = 3.29$. The WTP estimates $\widehat{\gamma}_k$ are taken from Column (4) of Table \ref{tab:experiment_wtp}. Since the survey experiment is restricted to full-time job ads, we construct an estimated WTP for part-time hours from recent evidence by \cite{dube2022power} and \cite{lachowska2025work}.\footnote{We assign a WTP of -0.0785 for jobs that offer part-time hours, which is one minus the average WTP estimate for jobs offering full-time hours reported in \cite{dube2022power} and \cite{lachowska2025work}.} These parameters are listed in Appendix Table \ref{tab:model_preestimated}.

\begin{table}[h!]
    \caption{Monopsony Model: Pre-Estimated Parameters.}
    \label{tab:model_preestimated}
    \vspace{-0.5cm}
    \begin{center}
        \begin{tabular}{l c}
\toprule \midrule
Parameter & Value \\
\midrule

$\sigma$: Labor Supply Elasticity from the Literature & 3.290 \\
\addlinespace
$\gamma_k$: Willingness-to-Pay from the Survey Experiment & \\
\quad Temporary Job & -0.254 \\
\quad Convenient Hours & 0.189 \\
\quad Inconvenient Hours & -0.282 \\
\quad Insurance Scheme & 0.105 \\
\quad Mortgage Possibility & 0.117 \\
\quad Career Opportunities & 0.085 \\
\quad Possibility for Remote Work & 0.065 \\
\quad Good Colleagues & 0.094 \\
\quad Inclusive Work-life Scheme & 0.173 \\
\quad Company Gym or Sports Team & 0.234 \\
\quad Company Cabin & 0.272 \\
\addlinespace
$\gamma_k$: Willingness-to-Pay from the Literature & \\
\quad Part-time Contract & -0.079 \\
\bottomrule \midrule
\end{tabular}
    \end{center}
    {\footnotesize \textit{Notes}: The table shows the pre-estimated parameters used in the model. The labor supply elasticity ($\widehat{\sigma}$) is implied by the pass-through estimates in \citet[][Column (4) of Table 2]{bhuller2025wage}. The willingness-to-pay (WTP) parameters ($\widehat{\gamma}_k$) for the advertised non-pay attributes are taken from Table \ref{tab:experiment_wtp}, Column (4), and estimated using the survey experiment described in Section \ref{subsec:linked_choice_to_ad}. The WTP for jobs with part-time contract is one minus the average WTP for full-time hours reported in \cite{dube2022power} and \cite{lachowska2025work}.}
\end{table}

\paragraph{Step 4: Employer pay premiums}
We recover the employer fixed effects in a standard two-sided unobserved heterogeneity wage regression with worker and employer effects. We use the wage regression model in Equation \eqref{eq:twoway_fe}. The estimated employer fixed effects $\widehat{\psi}_{g(j)}$ are our estimates of employers' pay premiums $\ln w_j$ in Equation \eqref{eq:indirect_utility}.

\paragraph{Step 5: Intrinsic amenity value}
The intrinsic amenity parameters $\ln \tilde{a}_j$ are estimated by maximum likelihood using Equation \eqref{eq:choice_probability}, conditional on the pre-estimated parameters from Step 3 and the employer pay premiums from Step 4. Since the number of parameters to be estimated is large, we rely on standard quasi-Newton methods for optimization.

\subsection{Comparison of Utility Dispersion}

Appendix Table \ref{tab:dispersion_comparison} compares the estimated dispersion of utility to the dispersion of log-wages for several recent papers. We rely on the references discussed in Section 8.5 of the review paper by \cite{mas2025nonwage}. For each of the four studies considered in Appendix Table \ref{tab:dispersion_comparison}, we construct a relative dispersion statistic from the most directly comparable results. Although the underlying studies differ across many dimensions---they all use different models and data---we harmonize these results to make them comparable to our estimates. For example, utility is in wage units in the model introduced by \cite{morchio2024gender}, so we report relative dispersion as the variance of log-utility over the variance of log-wages to make it comparable to the other studies, where utility is in log-wage units.

\begin{table}[htpb]
    \caption{Relative Dispersion of Utility in Selected Studies.}
    \label{tab:dispersion_comparison}
    \vspace{-0.5cm}
    \begin{center}
        \scalebox{0.8}{\begin{tabular}{l l c}
\toprule \midrule
Source & Reported dispersion statistic & Value \\
\midrule

Panel A. Published Studies & & \\
\quad \citet[][Tables VI-VII]{taber2020estimation} & Var(Utility) / Var(Log-Wages) & $0.234/0.104 = 2.250$ \\
\quad \citet[][Table 2]{hall2018wage} & SD(Job Value) / SD(Offered Log-Wages) & $0.38/0.24 = 1.583$ \\
\quad \citet[][Table 10]{morchio2024gender} & Var(Log-Utility) / Var(Log-Wages) [men,women] & $[0.044,0.036]$ \\
\quad \citet[][Table 5]{maestas2023value} & IDR(Log-Compensation) / IDR(Log-Wages) & $1.769/1.664 = 1.063$ \\
\addlinespace
Panel B. Model Estimates & & \\
\quad --- & Var($u_j$) / Var($\ln w_j$) & 2.200 \\
\quad --- & SD($u_j$) / SD($\ln w_j$) & 1.483 \\
\quad --- & IDR($u_j$) / IDR($\ln w_j$) & 1.914 \\
\bottomrule \midrule
\end{tabular}}
    \end{center}
    {\footnotesize \textit{Notes}: This table compares the relative dispersion of workers' indirect utility ($u_j$ in our framework) derived from working at employer $j$ relative to the dispersion of employer log-wages ($\ln w_j$ in our framework) in the selected studies listed in the first column. We include the studies discussed in the review paper by \citet[][Section 8.5]{mas2025nonwage}. For each study, we construct one relative dispersion statistic from the most relevant reported results. ``Var'': Variance. ``SD'': standard deviation. ``IDR'': inter-decile range.}
\end{table}

We find that our estimates broadly align with the numbers reported in \cite{taber2020estimation} (variance ratio: 2.200 vs 2.250) and \cite{hall2018wage} (standard deviation ratio: 1.483 vs 1.583). Our estimates also imply a larger inter-decile range ratio than that reported in \cite{maestas2023value} (1.914 vs 1.063), possibly because their analysis is restricted to a subset of non-pay amenities related to working conditions. Finally, the model developed in \cite{morchio2024gender} implies a much smaller variance ratio than ours (0.040 vs 2.250), most likely because of differences in the model environment.

We see two main takeaways from the comparison of utility dispersion estimates in Table \ref{tab:dispersion_comparison}. First, we provide a range of published estimates for this summary statistic, suggesting that both modeling choices and data sources matter. Second, we confirm that the relative dispersion of employer utility implied by our estimates is reasonable within that range.

\subsection{Distributions in Counterfactual Scenarios}

\begin{figure}[htbp]

    \caption{Relative Contribution of Job Attributes in Counterfactual Scenarios.}
    \label{fig:counterfactual_densities}
    \vspace{-0.5cm}
    \begin{center}
        \begin{subfigure}{0.49\textwidth}
            \centering
            \caption{Employer Pay Premiums: $\ln w_j$}
            \label{fig:counterfactuals_pay_diff}
            \includegraphics[width=\textwidth]{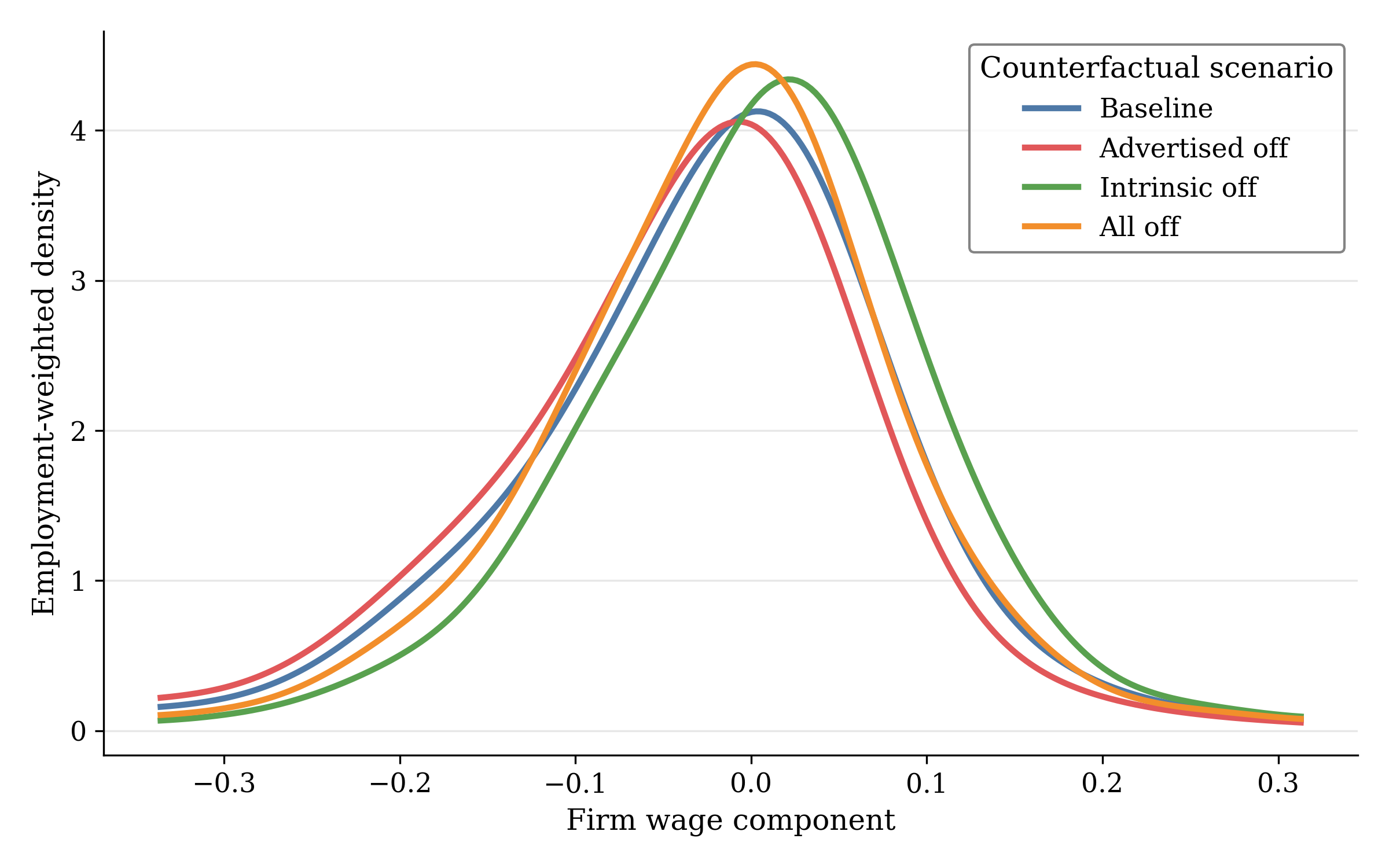}
        \end{subfigure}
        \\
        \begin{subfigure}{0.49\textwidth}
            \centering
            \caption{Employer Utility Differentials: $u_j$}
            \label{fig:counterfactuals_utility_diff}
            \includegraphics[width=\textwidth]{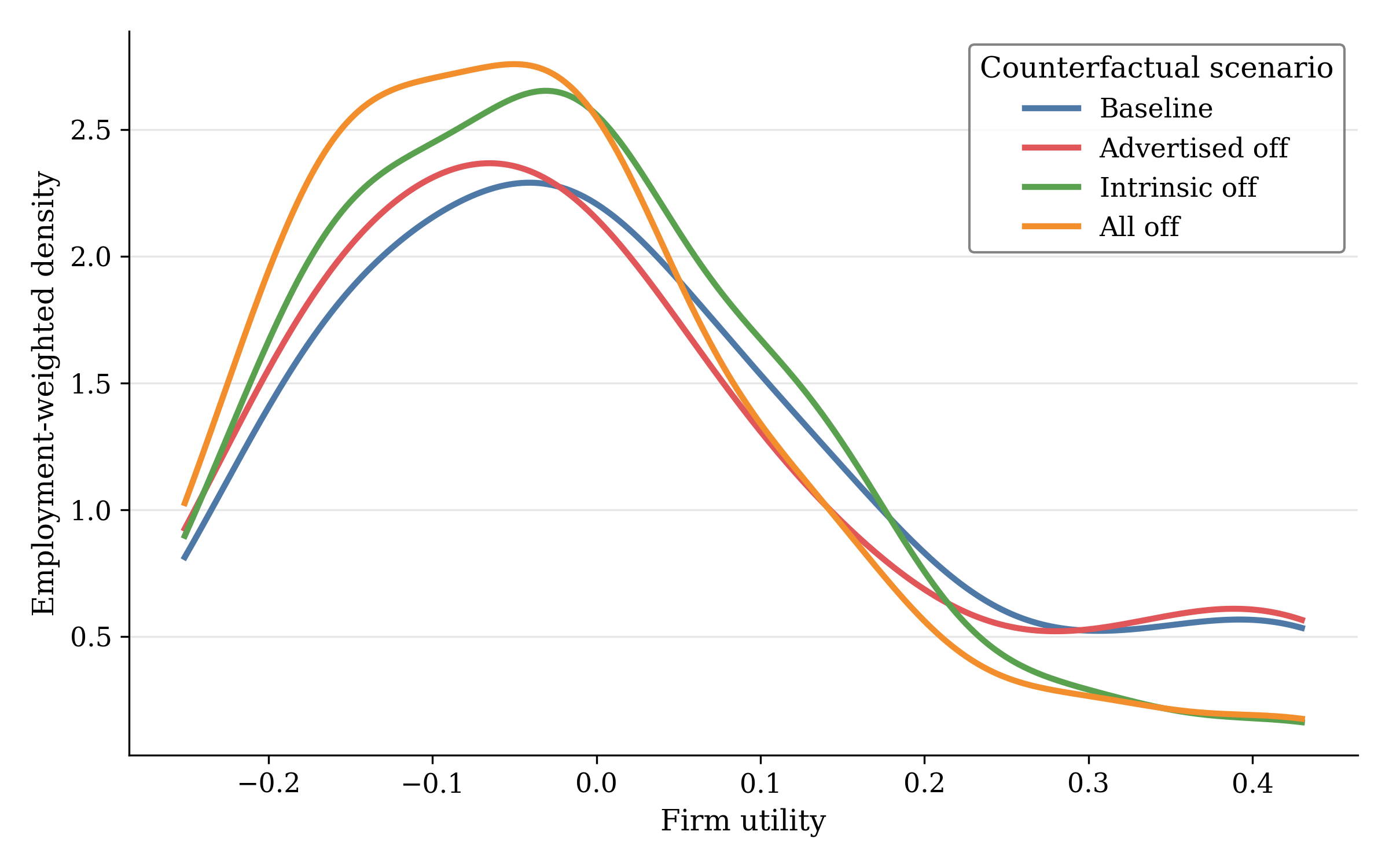}
        \end{subfigure}
        \hfill
        \begin{subfigure}{0.49\textwidth}
            \centering
            \caption{Employment Size}
            \label{fig:counterfactuals_size_diff}
            \includegraphics[width=\textwidth]{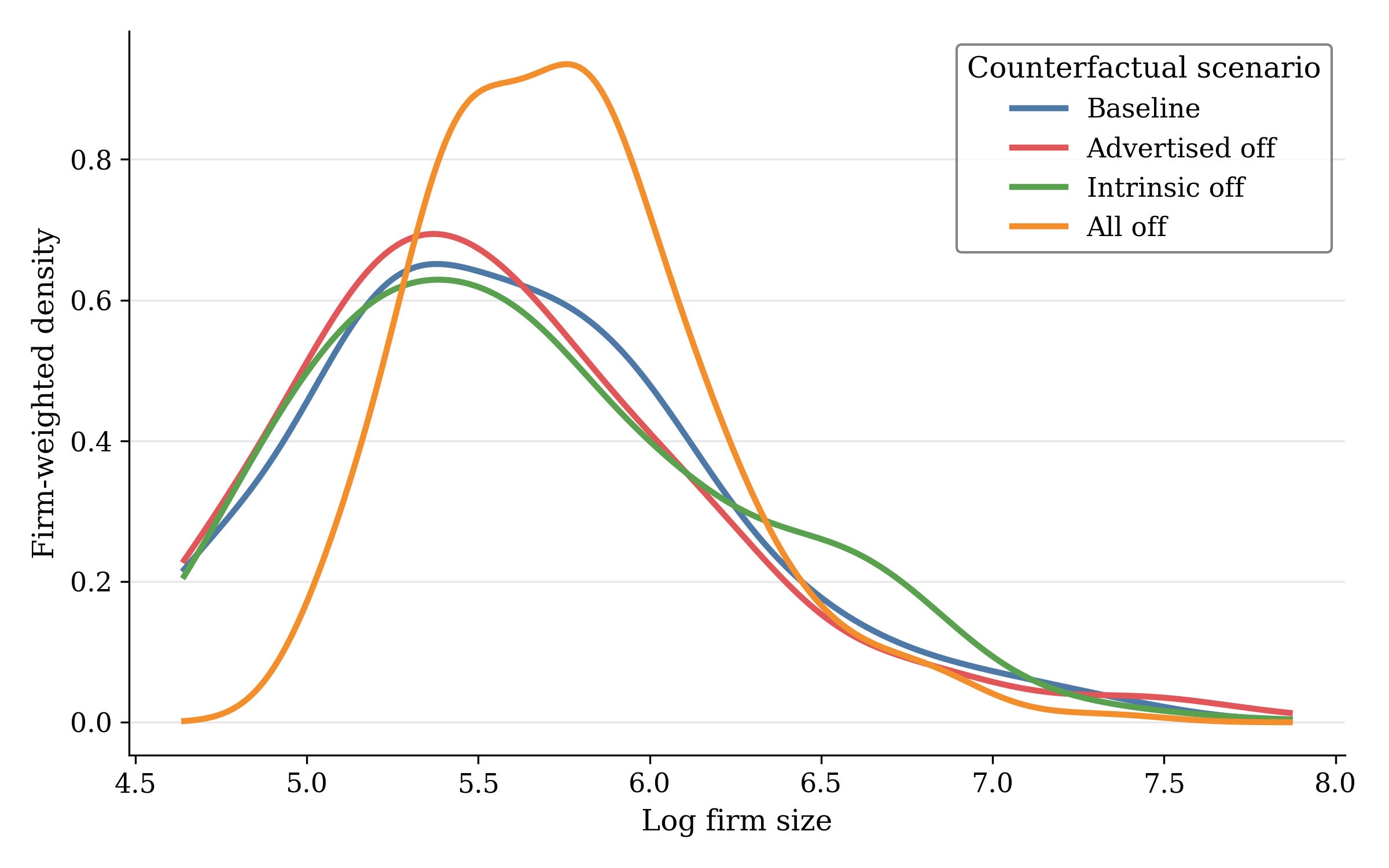}
        \end{subfigure}    
    \end{center}    
    \vspace{-0.25cm}
    {\footnotesize \textit{Notes}: This figure shows kernel density estimates for the distribution of the employer pay premiums (panel \subref{fig:counterfactuals_pay_diff}), the employer utility differentials (panel \subref{fig:counterfactuals_utility_diff}), and the employment size across employers (panel \subref{fig:counterfactuals_size_diff}) in the baseline and each counterfactual scenario. Panels (\subref{fig:counterfactuals_pay_diff}) and (\subref{fig:counterfactuals_utility_diff}) are weighted by employment size. The density estimates in Panel (\subref{fig:counterfactuals_size_diff}) are constructed by computing average employer size in each cluster, taking logs, and weighting each cluster by the corresponding number of establishments. The blue lines show the baseline distributions. The red lines shows the counterfactual distributions when the advertised amenities are turned off ($A_j = 0$, ``advertised off''), while the green lines reflect the scenario when intrinsic amenities are turned off ($\ln \tilde{a}_j = 0$, ``intrinsic off''), and the orange lines reflect the scenario when all amenities are turned off.}
\end{figure}

We complement the summary statistics on the distribution of employer pay, employer utility, and employment size in each scenario by showing kernel density estimates in Figure \ref{fig:counterfactual_densities}. Panel (\subref{fig:counterfactuals_pay_diff}) shows the distribution of employer pay premiums in each scenario, weighted by employment. Panel (\subref{fig:counterfactuals_utility_diff}) shows the distribution of employer utility differentials in each scenario, also weighted by employment. Panel (\subref{fig:counterfactuals_size_diff}) shows the distribution of employment size across employers in each scenario. This last figure is constructed under the assumption that the number of employers is constant in each scenario (i.e., no entry or exit).

\global\long\def\thetable{E.\arabic{table}}%
\setcounter{table}{0}
\global\long\def\thefigure{E.\arabic{figure}}%
\setcounter{figure}{0}
\global\long\def\theequation{E.\arabic{equation}}%
\setcounter{equation}{0}

\FloatBarrier

\global\long\def\thetable{E.\arabic{table}}%
\setcounter{table}{0}
\global\long\def\thefigure{E.\arabic{figure}}%
\setcounter{figure}{0}
\global\long\def\theequation{E.\arabic{equation}}%
\setcounter{equation}{0}

\newpage
\section{Predictive Power of Publicly Posted Job Attributes}
\label{app:restricted_models}

Which job attributes are most predictive of employer quality? In Section \ref{subsec:linked_ad_to_estab}, we decompose the overall $R^2$ by adding \textit{groups} of attributes, which abstracts away from the 47 \textit{individual} attributes. This Appendix focuses on the predictive power of individual attributes directly.

\begin{table}[htbp]
    \caption{Lasso Regressions -- The Most Predictive Publicly Advertised Job Attributes.}  \vspace{-1.5em}
    \label{tab:restricted_top_5_10_15}
    \begin{center}
    \scalebox{.5}{\begin{tabular}{lllll}
\toprule
\multicolumn{1}{c}{Pay Premium} & \multicolumn{1}{c}{Overall Sorkin Value} & \multicolumn{1}{c}{Flow Sorkin Value} & \multicolumn{1}{c}{Poaching Index} & \multicolumn{1}{c}{Employment Size} \\
\multicolumn{1}{c}{(1)} & \multicolumn{1}{c}{(2)} & \multicolumn{1}{c}{(3)} & \multicolumn{1}{c}{(4)} & \multicolumn{1}{c}{(5)} \\
\midrule
\multicolumn{5}{l}{\textbf{Panel A: Lasso Attributes 1-5}} \\
-- Full-time Contract & -- Pension Scheme & -- Part-time Contract & -- Collective Agreement Pay & -- Collective Agreement Pay \\
-- Part-time Contract & -- On-the-Job Training & -- Pension Scheme & -- Pension Scheme & -- On-the-Job Training \\
-- Shift Work & -- Full-time Contract & -- Possibility to Work Flexible Hours & -- On-the-Job Training & -- Full-time Contract \\
-- Temporary Job & -- Permanent Job & -- Social Environment & -- Full-time Contract & -- Challenging Tasks \\
-- Challenging Tasks & -- Social Environment & -- Good Colleagues & -- Social Environment & -- Central Location \\
\midrule
\multicolumn{5}{l}{\textbf{Panel B: Lasso Attributes 6-10}} \\
-- Any Other Mention of Pay & -- Collective Agreement Pay & -- Challenging Tasks & -- Compensation Level & -- Shift Work \\
-- Pension Scheme & -- Part-time Contract & -- Compensation Level & -- Insurance Scheme & -- Weekend/Evening/Night Work \\
-- Insurance Scheme & -- Possibility to Work Flexible Hours & -- On-the-Job Training & -- Part-time Contract & -- Permanent Job \\
-- Good Colleagues & -- Shift Work & -- Work Involves Travelling & -- Weekend/Evening/Night Work & -- Company Gym or Sports Team \\
-- Central Location & -- Weekend/Evening/Night Work & -- Company Vehicle & -- Permanent Job & -- Any Welfare Scheme \\
\midrule
\multicolumn{5}{l}{\textbf{Panel C: Lasso Attributes 11-15}} \\
-- Compensation Level & -- Good Colleagues & -- Any Welfare Scheme & -- Good Colleagues & -- Compensation Level \\
-- Company Gym or Sports Team & -- Challenging Tasks & -- Collective Agreement Pay & -- Challenging Tasks & -- Any Other Mention of Pay \\
-- Company Vehicle & -- Compensation Level & -- Incentive Pay Scheme & -- Involves Leadership Responsibility & -- Insurance Scheme \\
-- Any Welfare Scheme & -- Competitive Pay & -- Any Other Mention of Pay & -- Work Involves Traveling & -- Temporary Job \\
-- Company Cabin & -- Incentive Pay Scheme & -- Shift Work & -- Central Location & -- Company Vehicle \\
\bottomrule
\end{tabular}}
    \end{center}
    {\footnotesize \textit{Notes:} This table shows the variables included in Lasso regressions with penalty parameter set so that the model includes 5, 10, or 15 attributes. Columns (1)-(5) show the first variables selected for each revealed-preference employer quality measure. All regressions include a constant term and other ad characteristics without penalization, including the number of words, the total number of attributes, an indicator for online publication, and an indicator for whether the establishment name was disclosed. As in Section \ref{subsec:linked_ad_to_estab}, we define $A_{g(j)}^k$ is the share of ads from employers in cluster $g$ featuring attribute $k$. The sample corresponds to ``Employer Quality Sample'' in Table \ref{tab:sample_selection}.}
\end{table}

We use the Least Absolute Shrinkage and Selection Operator (Lasso) to select subsets of individual attributes for each employer quality measure. Lasso minimizes residual squared error with an additional penalty on the absolute value of coefficients.\footnote{Specifically, Lasso minimizes $$\sum_{g}\left\{\widehat{Q}_{g(j)}-\sum_k\left\{\beta_kA^k_{g(j)}\right\}\right\}^2+\phi\sum_{k}|\beta_k|,$$ where the first term is the sum of squared errors and the second is the penalty over $K$ penalized parameters. The penalty parameter $\phi$ determines the number of excluded regressors (i.e., variables $k$ with $\beta_k=0$).} For each measure, we start with a penalty high enough to exclude all attributes and reduce it until the Lasso retains 5, 10, or 15 attributes.\footnote{All regressions include a constant and additional ad characteristics (number of words, total number of attributes, an indicator for online posting, and an indicator for employer name disclosure) without penalization, so the selection criteria reflect variation in advertised job attributes rather than general ad characteristics.} Rather than emphasizing the exact order of entry, which may be sensitive when attributes are correlated, we report the sets of attributes selected at each threshold. This provides a parsimonious summary of the attributes with the strongest predictive power for each employer quality measure.

Appendix Table \ref{tab:restricted_top_5_10_15} reports the attributes selected by Lasso for various thresholds. Attributes related to hours of work are important predictors for all quality measures, as found in Section \ref{subsec:linked_ad_to_estab}. For the pay premium (Column 1), the five most predictive attributes relate to hours of work, inconvenient schedules, contract duration, and tasks. For the broader quality measures (Columns 2-4), pension scheme, hours of work, contract duration, convenient or inconvenient hours, collective agreement pay, and social environment are consistently among the most predictive. For employment size (Column 5), collective agreement pay, on-the-job training, and central location are highly predictive. Overall, the attributes flagged as salient in Section \ref{subsec:linked_ad_to_estab} remain highly predictive when we focus on individual attributes directly.

\end{document}